\documentclass{aa}
\usepackage{natbib,twoopt}
\usepackage{graphicx}
\usepackage{titlesec} 

\usepackage{xcolor}
\usepackage{txfonts}
\usepackage{comment}
\usepackage{longtable}
\usepackage{ulem}
\usepackage{comment}
\usepackage{appendix}
\usepackage{pgfplotstable}

\usepackage[T1]{fontenc}
\usepackage{csvsimple}
\usepackage{booktabs}  
\providecommand{\feh}{[Fe/H]}
\usepackage{xltabular}
\usepackage{booktabs}
\usepackage{hyperref}
\usepackage{tabularx}

\begin{document}

   \title{Exploring the dependence of chemical traits on metallicity}

   \subtitle{Chemical trends for red giant stars with asteroseismic ages}

    \author{
S. Vitali\inst{\ref{UDP}, \ref{ESO},\ref{ERIS}}\thanks{analysis of data taken as part of ESO programm ID 108.22DX} \and D. Slumstrup\inst{\ref{ESO}, \ref{UDP}} \and 
P. Jofr\'e\inst{\ref{UDP}, \ref{ERIS}}\and L. Casamiquela\inst{\ref{GEPI}}\and H. Korhonen  \inst{\ref{MPIA}, \ref{ESO}}\and S. Blanco-Cuaresma\inst{\ref{HAR},\ref{CHUV}}\and M.L. Winther\inst{\ref{AA}}\and V.Aguirre B{\o}rsen-Koch \inst{\ref{DARK}}
}

\institute{
Instituto de Estudios Astrof\'isicos, Facultad de Ingenier\'ia y Ciencias, Universidad Diego Portales, Av. Ej\'ercito Libertador 441, Santiago, Chile 
 \label{UDP},
\email{sara.vitali@mail.udp.cl}
\and 
European Southern Observatory, Alonso de Cordova 3107, Vitacura, Chile
 \label{ESO}
 \and
 Millenium Nucleus ERIS \label{ERIS}
\and
GEPI, Observatoire de Paris, PSL Research University, CNRS, Sorbonne Paris Cité, 5 place Jules Janssen, 92190 Meudon, France\label{GEPI}
\and
Max-Planck-Institut f{\"u}r Astronomie, K{\"o}nigstuhl 17, D-69117 Heidelberg, Germany\label{MPIA}
\and
Harvard-Smithsonian Center for Astrophysics, 60 Garden Street, Cambridge, MA 02138, USA\label{HAR}
\and
Laboratoire de Recherche en Neuroimagerie, University Hospital (CHUV) and University of Lausanne (UNIL), Lausanne, Switzerland\label{CHUV}
\and
Stellar Astrophysics Centre, Department of Physics and Astronomy, Aarhus University, Ny Munkegade 120, DK-8000 Aarhus C, Denmark
\label{AA}
\and
DARK, Niels Bohr Institute, University of Copenhagen, Jagtvej 128, DK-2200 Copenhagen, Denmark \label{DARK}
}

 \date{Received \today, accepted }
 
  \abstract
   {Given the massive spectroscopic surveys and the \textit{Gaia} mission, the Milky Way has turned into a unique laboratory to be explored using abundance ratios that show a strong dependency with time. Within this framework, the data provided through asteroseismology serve as a valuable complement. Yet, it has been demonstrated that chemical traits can not be used as universal relations across the Galaxy.}
   {To complete this picture, it is important to investigate the dependence on metallicity of the chemical ratios employed for inferring stellar ages. We aim to explore different combinations of neutron-capture, odd-Z and $\alpha$ elements as a function of age, particularly focusing on their metallicity dependence for a sample of 74 giant field stars.}
   {Using UVES observations, we derive  atmospheric parameters and high-precision line-by-line chemical abundances (<0.04 dex) for the entire set of spectra, which covers a wide spread in age (up to 14 Gyr) and metallicity ($-0.7< \mathrm{[Fe/H]}< +0.1$). Stellar ages are inferred from astereoseismic information. }
   {By fitting chemical-age trends for three different metallicity groups, we estimated their dependence on metallicity. Simultaneously, we identified those exhibiting stronger correlations with time. We found that the stronger chemical-age relations ([Zr/$\alpha$]) are not necessarily the ratios with the smaller dependence on metallicity ([Ce/$\alpha$] and [Ce/Eu]).}
   {We confirm the [n-capture/$\alpha$]-age trends for evolved stars, wherein the most significant correlation is evident in stars with solar-metallicity, gradually diminishing in stars with lower iron content. The lack of homogeneity within the metallicity range highlights the intricate nature of our Galaxy's star formation history and yield production. The dependence on metallicity of the yields involving s-process elements and the influence of radial stellar migration pose challenges to relying solely on chemical abundances for dating stars. These findings contest the feasibility of establishing universally applicable chemical clocks valid across the entire Galaxy and across various metallicity ranges.  }


   \maketitle
%
%
\section{Introduction}
With the massive collection of data by the \textit{Gaia} satellite \citep{2016A&A...595A...1G, 2023A&A...674A...1G}, and the ground-based spectroscopic surveys, such as APOGEE \citep{2017AJ....154...94M}, GALAH \citep{2015MNRAS.449.2604D}, LAMOST \citep{2012RAA....12..735D}, and \textit{Gaia}-ESO \citep{2012Msngr.147...25G}, Galactic astronomy has been significantly propelled forward. The extensiveness of these datasets has enabled to build multi-dimensional maps of the Milky Way which are used to decipher the timeline of events that have shaped its history and evolution. In this framework, chemical abundances of stars are powerful repositories of the composition of the birth cloud from which they originate (see e.g. \citeauthor{2002ARA&A..40..487F} \citeyear{2002ARA&A..40..487F}; \citeauthor{2012ceg..book.....M} \citeyear{2012ceg..book.....M}). The elements studied through stellar spectra carry information of the nuclear reactions and astrophysical processes which have modified the evolution of the birth cloud's chemical composition \citep{2002Freeman,2016Hogg,Jofre17,2020ApJ...900..165R}, and thus are used to unveil the chemical enrichment history of the hosting environment \citep{2012Bovy,2017Minchev,2019Bland-Hawthorn,2021MNRAS.506..150B}.

Among the multiple chemical abundance ratios, it has been found that some specific combinations are particularly informative as they show strong correlations with age (e.g. \citealt{2012A&A...542A..84D, 2015A&A...579A..52N, 2018MNRAS.474.2580S, 2020A&A...640A..81N}). These chemical tracers, also dubbed  as {\it chemical clocks} \citep{2015A&A...579A..52N,2016A&A...590A..32T}, have been proposed as an alternative to determine stellar ages. 

Dating stars is of paramount importance as it provides crucial insights into the timeline of several evolutionary processes, but it is very difficult to do despite the various existing techniques \citep{2015ASSP...39..167N}. Consequently, the prospects of estimated ages inferred from element abundances have been widely explored, especially for single field stars for which the age determination is even more difficult than for stars in groups or clusters \citep{2022MNRAS.517.5325H,2023MNRAS.522.4577L,2023arXiv230408276A}.

Despite the valuable use of these chemical-age relations, there are different factors which can alter these tracers, threatening their reliability and universality as age calibrators \citep{2017MNRAS.465L.109F, 2020A&A...639A.127C}. Indeed, a dependency with metallicity has been found by \cite{2017MNRAS.465L.109F}, \cite{2019A&A...624A..78D}, and \cite{2020A&A...639A.127C}. Furthermore, although the relations might not be affected by the stellar type \citep{2017A&A...604L...8S}, they could be affected by the techniques used for age determination \citep{2022ApJ...936..100B}. 
The lack of homogeneity in these chemical clocks can be attributed to both the diverse nucleosynthetic processes that take place during a star's lifetime and the intricate nature of chemical enrichment and star formation histories (SFHs) which undergo significant changes when examining a considerable spatial volume \citep{1997ApJ...477..765C,2018Feuillet,2021A&A...646L...2M, 2021A&A...652A..25C}.

Another important aspect to be taken into account is that stars move away from their birth places (see for example \citealt{2002MNRAS.336..785S,2010ApJ...722..112M,2020ApJ...896...15F}). Radial gradients and chemical signatures can be significantly weakened by this stellar migration making it difficult to reproduce the expected chemical trends. This dependence on galactocentric position and the impossibility of establishing unique chemical-age trends across the Galaxy correlates with the radial fluctuations in SFH, which in itself also creates a dependence of the stellar yields on the metallicity of the considered environment \citep{2009Magrini,2021A&A...652A..25C,2022A&A...660A.135V,2023arXiv230711159R,2023MNRAS.525.2208R}.

All these effects needed to be considered for establishing reliable chemical-age calibrations. To this purpose, it is indispensable to have high precision abundances to minimize the effects of internal abundance determination errors and to disentangle the different processes which might play a role in altering the relations. In addition, accurate ages determined from a separate reliable method are needed for a solid study of this kind 

\cite{2017A&A...604L...8S} confirmed the tight correlation between [Y/Mg] and age discussed by \cite{2015A&A...579A..52N} for six evolved stars at solar metallicity in four open clusters, extending the validity of the [Y/Mg] chemical clock from dwarfs to red giants. A later study of the red giant $\epsilon$ Tau in the Hyades at super-solar metallicity was found to fall just above the solar twin relation \citep{2019A&A...622A.190A}. This suggests, that the chemical clock for evolved stars could as well depend on \feh\ as solar analogue stars \citep{2017MNRAS.465L.109F}. More recently, \cite{2022ApJ...936..100B} demonstrated that [Y/Mg]-age relation behaves differently for stars other than solar-twins. These evidences may indicate that the picture of chemical tagging is more complicated for giants than otherwise expected, e.g., poorly understood mixing processes happening in their atmospheres might have an impact on their abundances. 

At the moment GALAH is the only large-scale spectroscopic survey that provides yttrium (Y) abundances, but this is focused on main-sequence and sub-giant stars. \cite{2022MNRAS.517.5325H} used GALAH data for main-sequence turn-off stars to show that a precision of $\sim$ 0.05 dex was not sufficient to calibrate different chemical clocks, even if ages were estimated with a precision of 1-2 Gyr \citep[though see the recent work of ][who improved both, abundances and selection of stars]{Walsen23}. 

Asteroseismology is currently one of the most precise ways to determine ages for single field stars \citep{2013ARA&A..51..353C,2014ApJS..210....1C}, especially for red giant (RG) stars \citep{2012ASSP...26...11M,2023arXiv230408276A}, which are bright enough for inferring reliable age information. In synergy with the large-scale spectroscopic surveys, asteroseismology has widely contributed to spectroscopic age determination \citep{2014ApJS..215...19P}. While asteroseismic ages for an abundance of stars have been made available by the CoRoT \citep{2016AN....337..970V,2017A&A...597A..30A}, {\it Kepler} \citep{ 2010PASP..122..131G,2014ApJS..215...19P} and K2 \citep{2014PASP..126..398H,2019MNRAS.490.4465R}, they only surveyed parts of the sky. With the launch of the TESS mission \citep{2018MNRAS.473.2004S, 2020ApJ...889L..34S}, a larger part of the sky has been surveyed, thus allowing for asteroseismic ages to be used in wider studies.

In this work we use a sample of high-resolution UVES spectra of RGB field stars with asteroseismic ages to explore the homogeneity and dependencies of chemical clocks accros metallicity. With our homogeneous high resolution and signal-to-noise spectra, we determine abundances and ages for more than seventy giant stars. Our analysis serves to deepen the understanding and usage of chemical clocks in the case of more evolved stars. 

The paper is structured as follow. In Sect.~\ref{sect:data} we introduce the observational data used for this work and in Sect.~\ref{sect:method} we explain the methods we used for abundance and age determination. In Sect.~\ref{sect:results} we present our results for the relations between ages and chemical abundance rations, and a discussion about their implications with a conclusion can be found in Sect.~\ref{sect:discussion}. 
\begin{figure}[t]
\centering
  \includegraphics[width=0.8\columnwidth]{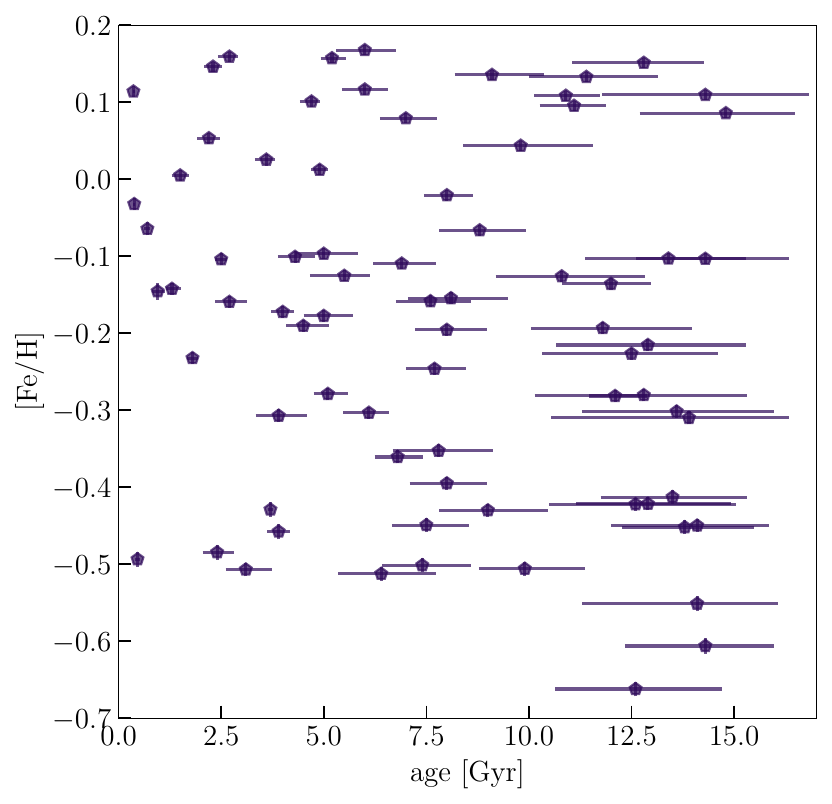} 
 \includegraphics[width=0.95\columnwidth]{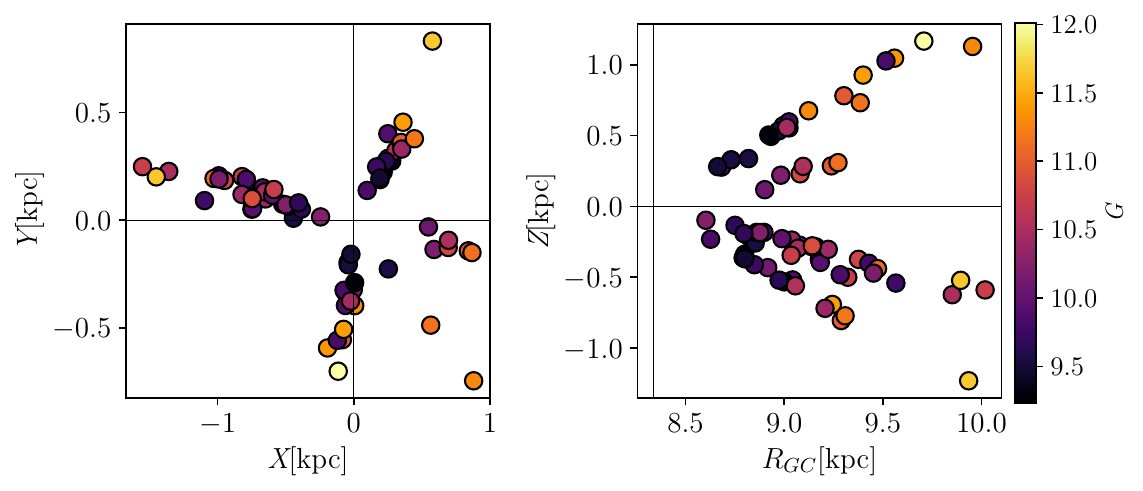} 
 \caption{\textbf{Top:} [Fe/H] vs. age distribution for the proposed sample. The iron values are taken from APOGEE with ages determined in this work. \textbf{Bottom:} Cartesian Galactic coordinates \textit{X},\textit{Y} of the targets. On the right, Galactic \textit{Z} and Galactocentric radii $\mathrm{R_{GAL}}$. The color-coding represents the \textit{G} magnitude of the sample derived from \textit{Gaia} DR3. The black lines indicate the position of the Sun ($\mathrm{X_{\odot}}$; $\mathrm{Y_{\odot}}$; $\mathrm{Z_{\odot}}$ ) = (0; 0; 0) and $\mathrm{R_{GAL}}$ = 8.34 placed at kpc.}
 \label{fig:cord}
\end{figure}
\section{Data}\label{sect:data}

\subsection{Target selection}

We have analyzed a sample of 74 red giant branch (RGB) stars in five K2 fields, from which we extract their Epic identification number, their corresponding Gaia DR3 ID, coordinates, magnitudes and seismic information, summarised in Table~\ref{tab:infotarg}. These stars have been selected from the K2 targets already observed by APOGEE DR17  \citep{2022ApJS..259...35A}, from which an assessment of the metallicity and temperature was available. This allows us to infer the stellar ages with an error of $\leq 30 $\% (average $\sim 20$\%, see section \ref{subsect:ages} for a more detailed discussion). The inferred ages are shown for all targets at the top of Fig. \ref{fig:cord}, showcasing a significant range in both age and metallicity which serves for the purpose of this study. Thanks to the \textit{Gaia} third data release (DR3) \citep{2023A&A...674A...1G} the selection focused on mostly bright disk giants located in the solar region (29 stars with 8.5 < $\mathrm{R_{GAL}}$ < 9 kpc ), and the majority in the outer disk (9 < $\mathrm{R_{GAL}}$ < 10 kpc).

At the bottom of Fig. \ref{fig:cord} we show the Galactic positions of our targets, coloured by their magnitude. In the left hand panel, we plot the cartesian coordinates X and Y, and in the right hand panel we plot the polar coordiantes R and z.  We can see how the stars are located at different positions, and that toward the outer disk, the stars are about 1 kpc above (or below) the Galactic plane. 

\subsection{Spectral observations and data reduction}
We followed up the selected APOGEE DR17 targets with UVES,  a cross-dispersed echelle spectrograph mounted on the Very Large Telescope (VLT), at Paranal Observatory. The data was observed under the ESO programe ID 108.22DX. The spectra were taken between September 2021 and January 2022 and the configuration employed ensures a resolving power of $R \sim 110 000$ covering a wavelength range from 480 to 680 nm. The high quality of the spectra translates to a signal-to-noise ratio (SNR) of $\sim 80-100$ estimated by the ESO pipeline, which is also listed in Tab.~\ref{tab:infotarg}. 

\begin{figure}[]
\centering
 \includegraphics[width=\columnwidth]{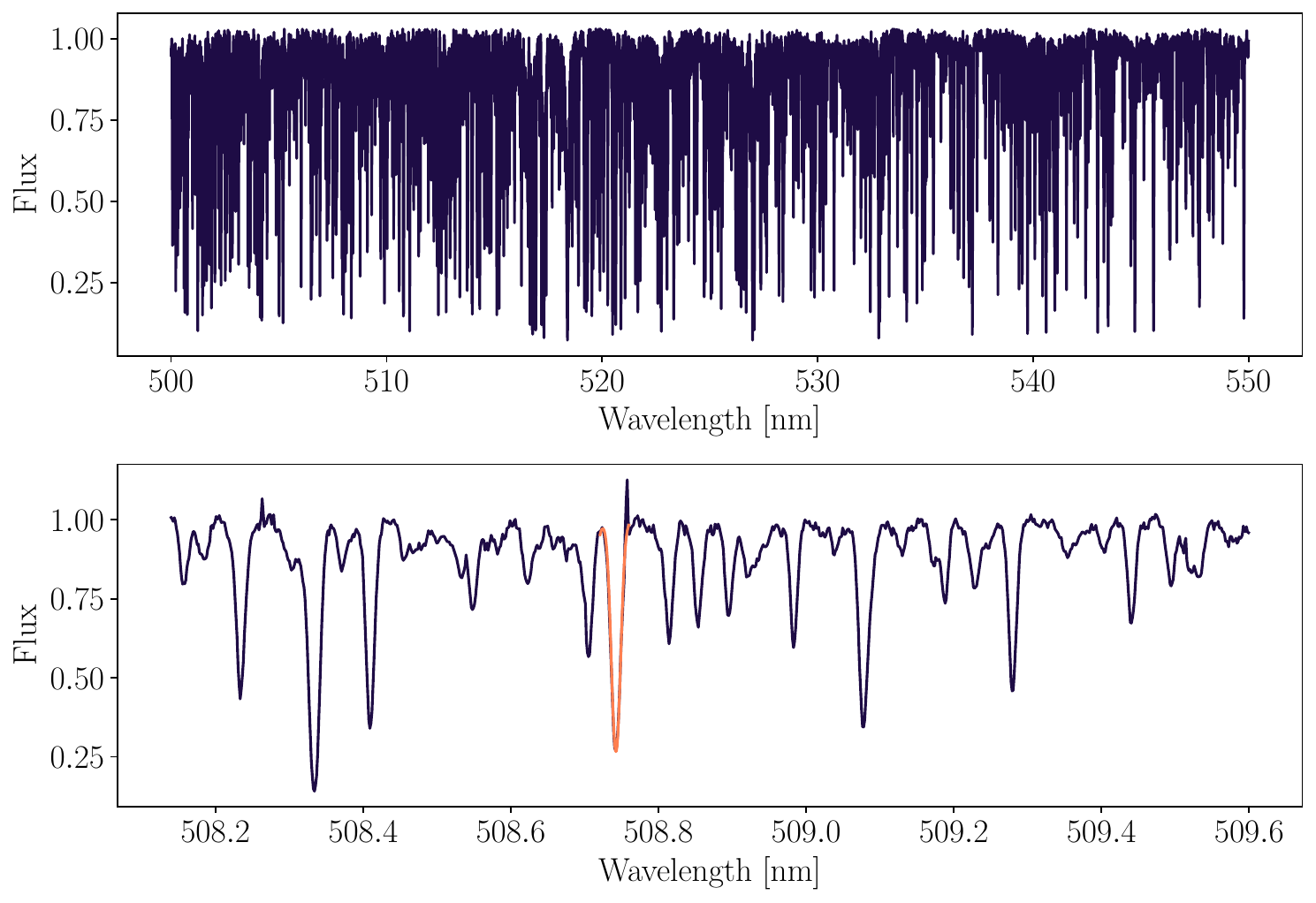} 
 \caption{UVES spectrum of J04034842+1551272 from $\approx 500$ to 550nm. The bottom panel is a zoom around the YII line at 508.74 nm. The SNR is approximately 90.}
 \label{fig:spectrum}
\end{figure}

\section{Method: stellar analysis}\label{sect:method}
We adopted the automatic pipeline presented in the work of \cite{2020A&A...635A...8C} which relies on the public software \texttt{iSpec} \citep{2014A&A...569A.111B,2019MNRAS.486.2075B}. The workflow was personalized for our purpose to perform the correction for the barycentric velocity and the stellar radial velocity (RVs). The RVs were determined using a synthetic template. The sky subtraction and the cleaning from telluric features were applied using an internal telluric line list provided by \texttt{iSpec}. Finally, we normalized the spectra in segments using cubic splines and finding the continuum level with a median-maximum filter. An example of a portion of a normalised and velocity corrected spectrum is reported in Fig.~\ref{fig:spectrum}.

\subsection{Atmospheric parameters} \label{sect:AP}

The atmospheric parameters ($T_{\mathrm{eff}}$, [M/H], [$\alpha$/M] and broadening parameters) were calculated via synthetic spectral fitting. This technique produces synthetic spectra on the fly and compares them with the observed spectrum taking into account the uncertainties on its flux. The parameters are found by $\chi^2$ minimization between the flux and the synthesis. We chose Turbospectrum as the radiative transfer code \citep{1998A&A...330.1109A,2012ascl.soft05004P}, which considers local thermodynamic equilibrium and the one dimensional spherical MARCS model atmospheres \citep{2008A&A...486..951G}. The atomic parameters are contained in the latest version (v.6) of the line list of the \textit{Gaia}-ESO Survey \citep{2021A&A...645A.106H} and the line selection to perform the fitting is taken from \cite{2019MNRAS.486.2075B}. Lines affected by telluric features, blends, or continuum displacement are automatically discarded. It leads to a total of 202 lines for 21 elements. 

Following \cite{2019MNRAS.486.2075B}, we initially performed the synthesis fitting leaving effective temperature $T_{\mathrm{eff}}$, surface gravity $\log g$, metallicity [M/H], alpha-enhancement [$\alpha$/M] and the microturbulence velocity $v_{\rm mic}$ as free parameters. The resolution was fitted for each spectrum for accounting for the broadening effects, while the rotational velocity $v_{\text{sin}i}$ was fixed to 1.6 $\mathrm{kms}^{-1}$ as we expected a rotation velocity less than 2  $\mathrm{kms}^{-1}$  for giant stars. By visual inspection we checked the goodness of the fits. We removed the spectrum with low SNR ($\sim 30$) of the star EPIC 201456500, due to unsatisfactory results obtained from the fit compared to the rest of the 73 targets, with SNR $\sim 70-100$.

\begin{figure}[t]
\centering
  \includegraphics[width=0.95\columnwidth]{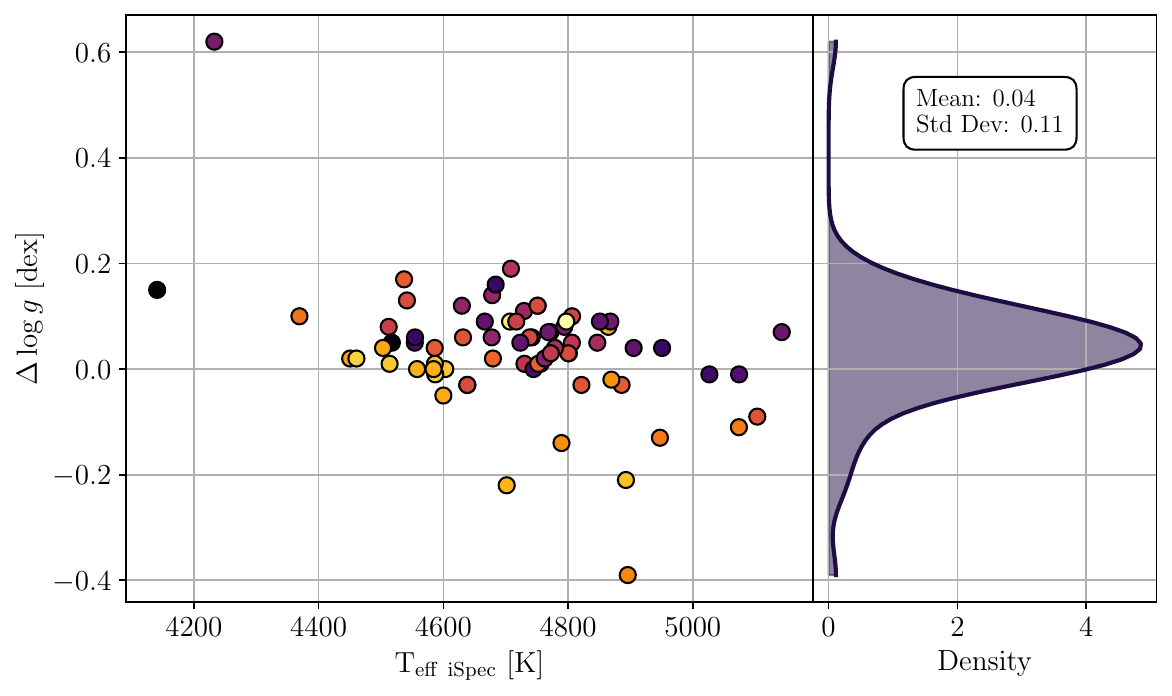} 
  \includegraphics[width=0.95\columnwidth]{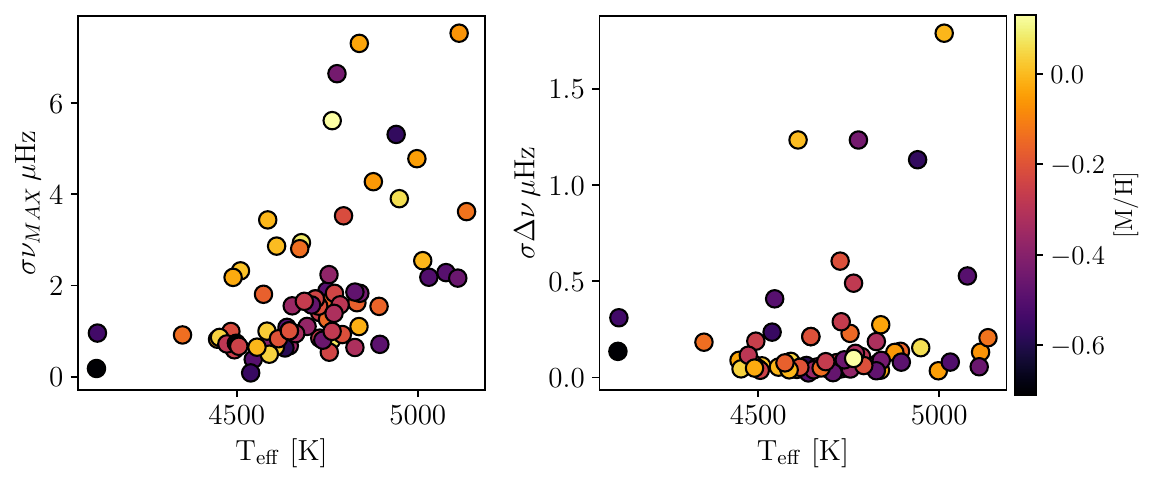} 
 \caption{\textbf{Top:} Difference between spectroscopic $\log g$ and the outcomes derived from the asteroseismic scaling relation of Eq.~\ref{eq:eq1} and the corresponding KDE distribution. Larger deviations between the two methods are observed for higher metallicity and the three coldest stars. \textbf{Bottom:} Uncertainties in the seismic maximum frequency and the separation $\Delta\nu$ are depicted as a function of temperature. In all cases, the color-coding is determined based on the global metallicity calculated using iSpec. }
 \label{fig:seismo-spectro-log}
\end{figure}

All targets have seismic measurements, thus making it possible to validate our results with those obtained considering the seismic information. To do so, for a second run of spectral fitting, we adopted the scaling relation from astroseismology \citep{1991ApJ...368..599B,1995A&A...293...87K} that involves the frequenct of maximum power $\nu_{\rm max}$ from the K2 catalogue to infer $\log g$ as :
\begin{equation} \label{eq:eq1}
    \log g = \log \left(\left(\frac{\nu_{\rm max}}{\nu_{\rm max\,\odot}}\right)\cdot\left(\frac{T_{\rm eff}}{T_{\rm eff\,\odot}}\right)^{1/2}\right)+ \log g_{\odot}.
\end{equation}
where we as \cite{2015Heiter} adopt the solar values $\nu_{\rm max\,\odot} = 3100\,\mathrm{\mu Hz}$, $T_{\rm eff\,\odot} = 5777$ K, and $\log g_{\odot} = 4.44$ dex. To achieve this, we used the temperatures determined in the previous run. We proceeded to compute $\log g$ values using the provided equation. This independent determination of  $\log g$ enables us a further validation of internal consistency between our parameters and the ones that can be derived from astroseismology. 

The relation in equation \ref{eq:eq1} is extrapolated from the Sun. Since we apply it here for giants, deviations should be expected, which have been quantified in the literature, e.g. with studies of binary systems, parallaxes and comparisons to interferometry. Asteroseismic masses have been shown to be accurate to better than 8\% \citep{2016Stello, 2016Miglio,2016Brogaard} and radius better than
4\% \citep{2012Aguirre,2016Huber,2018Sahlholdt}. Assuming a stellar mass of 1.2 solar masses and radius of 10 $\mathrm{R_{\odot}}$, this translates to an accuracy on $\log g$ better than 0.04 dex.

\begin{figure}[t]
\centering
 \includegraphics[width=0.38\textwidth]{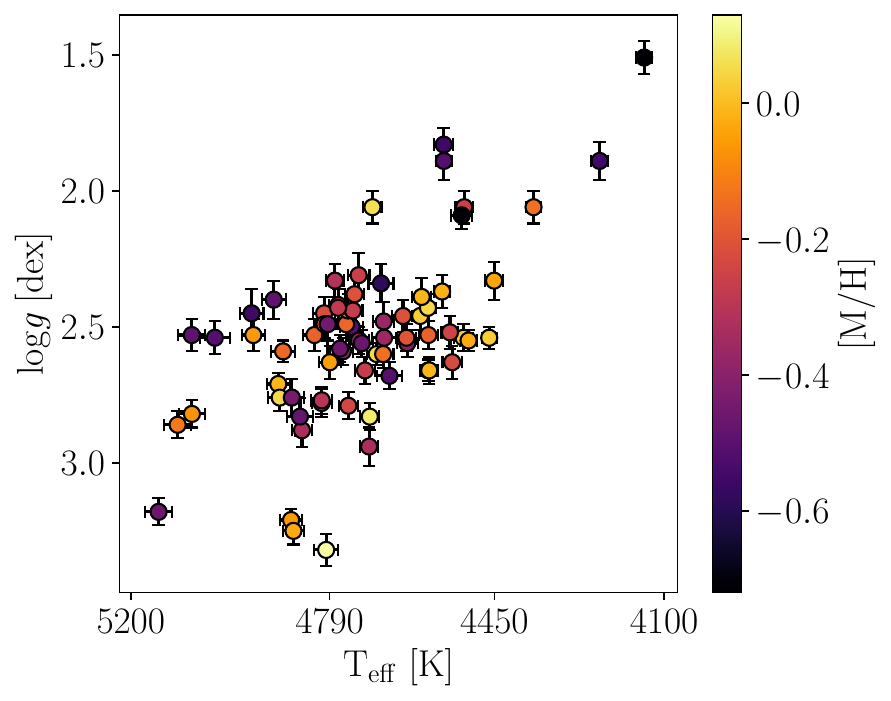} 
 \includegraphics[width=0.48\textwidth]{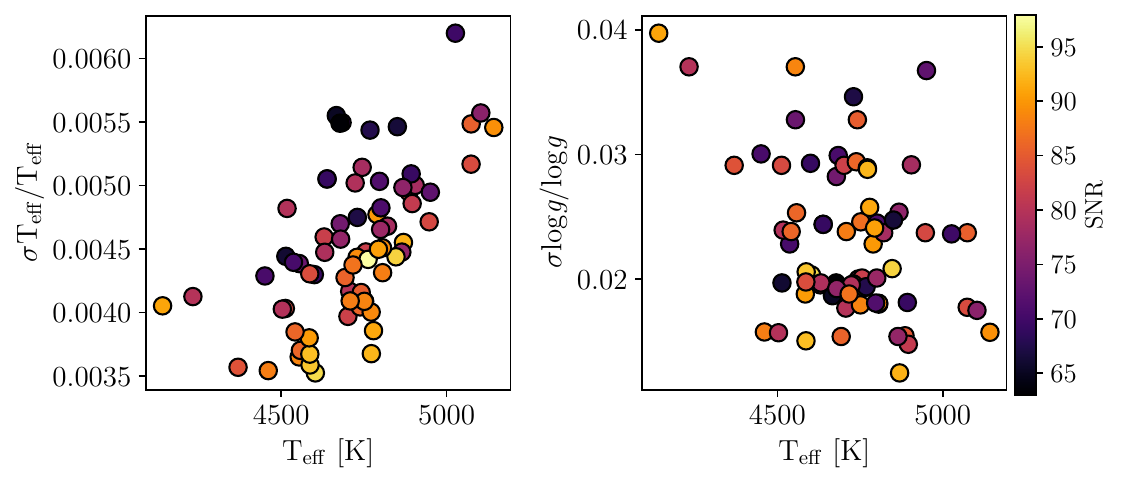} 
\caption{\textbf{Top}: Kiel diagram of the entire spectroscopic sample. The parameters and their errors are derived with spectral synthesis. All spectra are colour-coded by their global metallicity. \textbf{Bottom}: Relative uncertainties on $T_{\mathrm{eff}}$ and $\log g$ as a function of temperature and coloured by the SNR of the targets.}
 \label{fig:kiel_delta}
\end{figure}

A comparison of the spectroscopic and asteroseismic $\log g$ results appears in Fig.
~\ref{fig:seismo-spectro-log}. The stars are sorted according to their temperature, and coloured according to their metallicity. For most of the stars the agreement between both determinations is within 0.2 dex, with few exceptions which differ more. These uncertainties might stem both from increased variability in sigma on $\nu_{\mathrm{max}}$ , 
and $\Delta \nu$ as presented by the two lower subplots of Fig.~\ref{fig:seismo-spectro-log} and from the larger errors on the temperatures of warmer giants (as shown at the bottom left of Fig.~\ref{fig:kiel_delta}). Concerning the two coldest outliers, the $\log g$ values computed with the seismic relation exceed the spectroscopic values by more than 0.2 dex.  Despite conducting a visual inspection, no issues were identified in their fits. Consequently, we have opted not to exclude these cases. However, in light of these findings, caution must be taken when interpreting the results for these three giants, given that accuracy may be compromised for such cold stars.

We then derived again the other atmospheric parameters fixing the asteroseismic $\log g$ iteratively to ensure consistent parameters. Our final parameters thus are those which combine both spectroscopic temperatures and metallicities with seismic $\log g$. These appear in the Kiel diagram in the upper panel of Fig.~\ref{fig:kiel_delta}.

\begin{figure}[t]
\centering
 \includegraphics[width=\columnwidth]{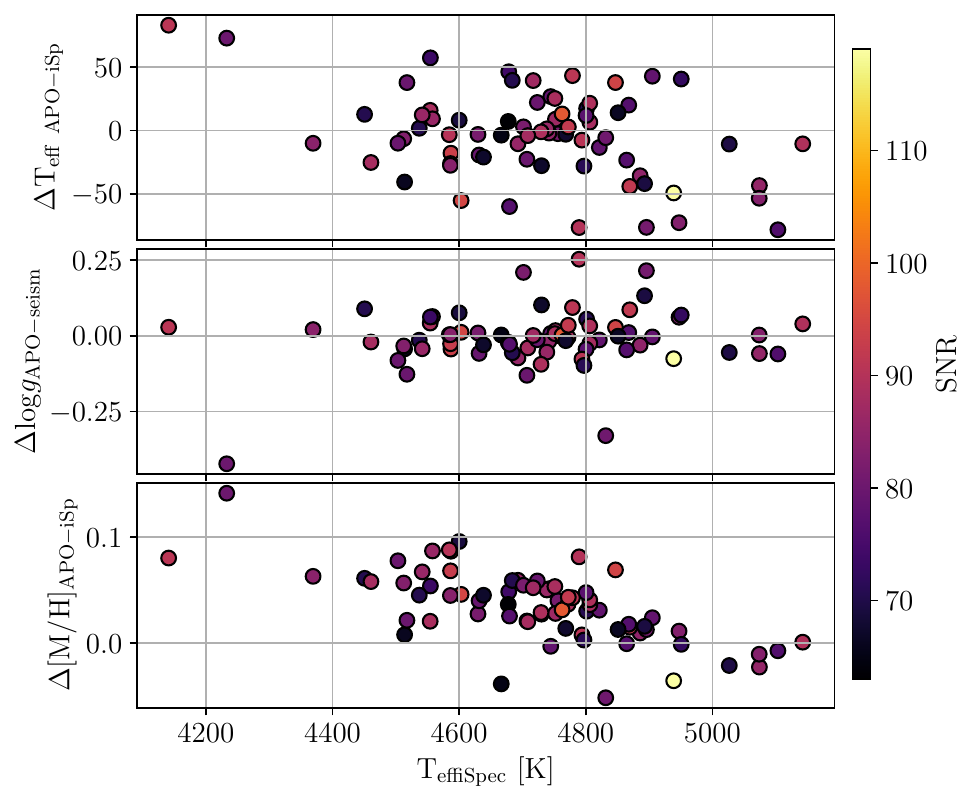} 
 \includegraphics[width=\columnwidth]{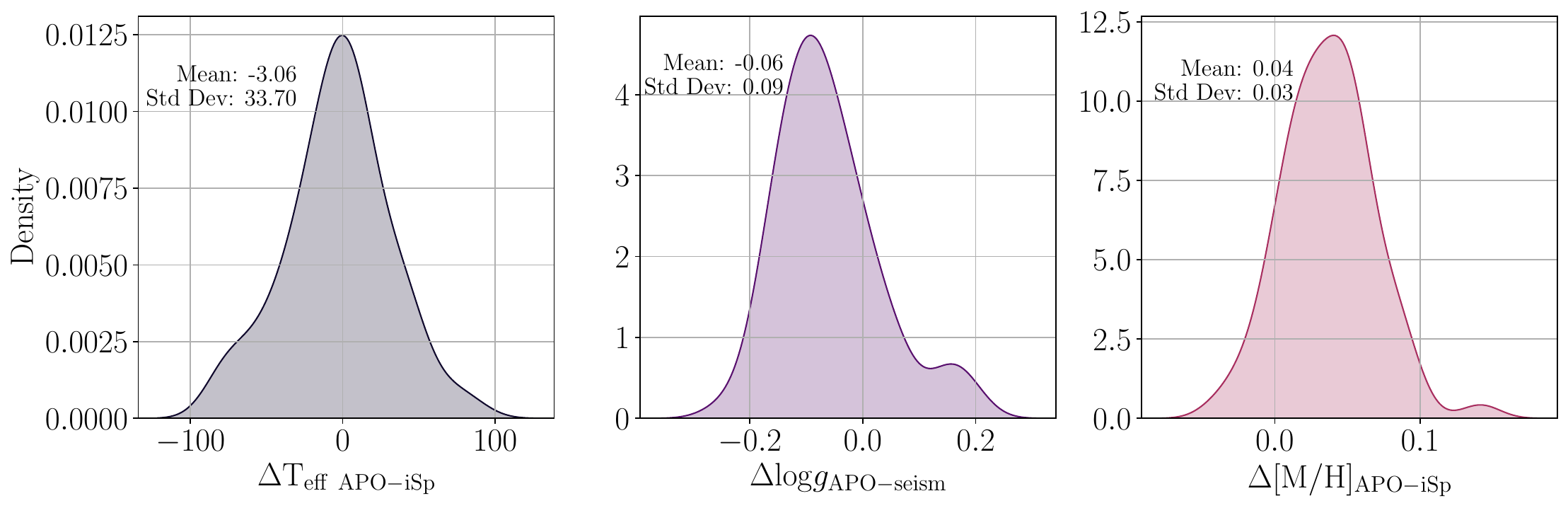} 
\caption{\textbf{Top}: Differences between our computed and adopted atmospheric parameters and the APOGEE results as a function of spectroscopic $\mathrm{T_{eff}}$. The color code expresses the SNR of the targets. Our temperature and metallicities are directly derived from spectroscopy, while the $\log g$ are calculated using the scaling relation of Eq.~\ref{eq:eq1}. \textbf{Bottom}: Kernel density estimation of the three deltas together with their mean differences and dispersions.}
 \label{fig:apo_comp}
\end{figure}

It is possible to notice the extension of the metallicity range covered by the sample, i.e. -0.7 $\lesssim$ [M/H] $\lesssim$ +0.1.  The two coolest stars are the most metal-poor ones. iSpec estimates the uncertainties for the atmospheric parameters from the covariance matrix (the square root of the diagonal elements) computed by the least square algorithm, and they are plotted at the bottom of Fig.~\ref{fig:kiel_delta}. For the temperature the average uncertainty is 21 K and the relative uncertainties increase for hotter temperatures as well as for lower SNR. The seismic $\log g$, however, remains unaffected by such trends, even with a temperature change of 21 K, resulting in a variation of approximately 0.001 dex.
For the spectroscopic $\log g$ the average value of the relative uncertainty is $\sim 0.02$ with a spread around 0.01, but no evident trend is present looking at the right panel at the bottom of Fig.~\ref{fig:kiel_delta}.

Since the stars were targeted as a follow up from APOGEE observations, it is natural to compare our results with those obtained from APOGEE DR17. A comparison for temperature, surface gravity and metallicity is illustrated in in Fig.~\ref{fig:apo_comp}. The differences (APOGEE-ours) are plotted in three different panels starting from the spectroscopic temperature at the top, the seismic gravity in the middle and the metallicity at the bottom. 

In addition, in the same plot we show the distribution of the differences of all three parameters, with the corresponding mean and standard deviation indicated in each panel. It is possible to see that the parameters are generally consistent for the two methods exhibiting larger deviations in the cases of the most metal-poor and most metal-rich and hotter stars. Given that these parameters are correlated, the scatter for the hotter giants can be attributed to the increase of the errors with the temperature. However, for the temperatures and $\log g$ the mean differences reported in the histograms are within the average parameter uncertainties. For the global metallicity [M/H] there is an offset of around 0.05 dex which exceeds the average combined uncertainties $\bar{\sigma}_{\mathrm{[M/H]}} \sim 0.02$ dex. This variation is higher for colder temperature and it decreases for $\mathrm{T_{eff}}$ higher than $\sim 4800$ K. This offset, together with the scatters found, can be explained by the differences in methodologies, analysis and calibrations  \citep[see .e.g][for extensive discussions]{2019Jofre, 2020AJ....160..120J, 2023A&A...670A.107H}. 

\subsection{Age determination} \label{subsect:ages}
All stars in this study are in the giant phase of their evolution. Therefore, we expect them to be low and intermediate mass stars undergoing H-shell burning. Since they are typically bright, this category is a perfect benchmark for studying the ages of field stars located at different distances from the Sun using asteroseismology. To this purpose, we used the public BAyesian STellar Algorithm (\texttt{BASTA}, \citealt{2015MNRAS.452.2127S,2022MNRAS.509.4344A}). It is a python-based code which infers stellar properties by matching observations to grids of stellar models. In our case, we employed the updated isochrones from a Bag of Stellar Tracks and Isochrones (BaSTI, \citealt{2018ApJ...856..125H, 2021ApJ...908..102P}).

As observables, we used the $T_{\mathrm{eff}}$ and [Fe/H] derived by our spectral analysis, and the large frequency separation $\Delta \nu$ and frequency of maximum power $\nu_{max}$ from the K2 catalogue listed in Tab.~\ref{tab:infotarg}. Additionally, we used the magnitudes \textit{J,H} and \textit{K} from 2MASS \citep{2003yCat.2246....0C} together with the parallaxes from \textit{Gaia} DR3 to match to the synthetic magnitudes of the stellar models. With this information we determined ages, final masses and $\log g$ for the entire sample. 

\begin{figure}[t]
\centering
 \includegraphics[width=\columnwidth]{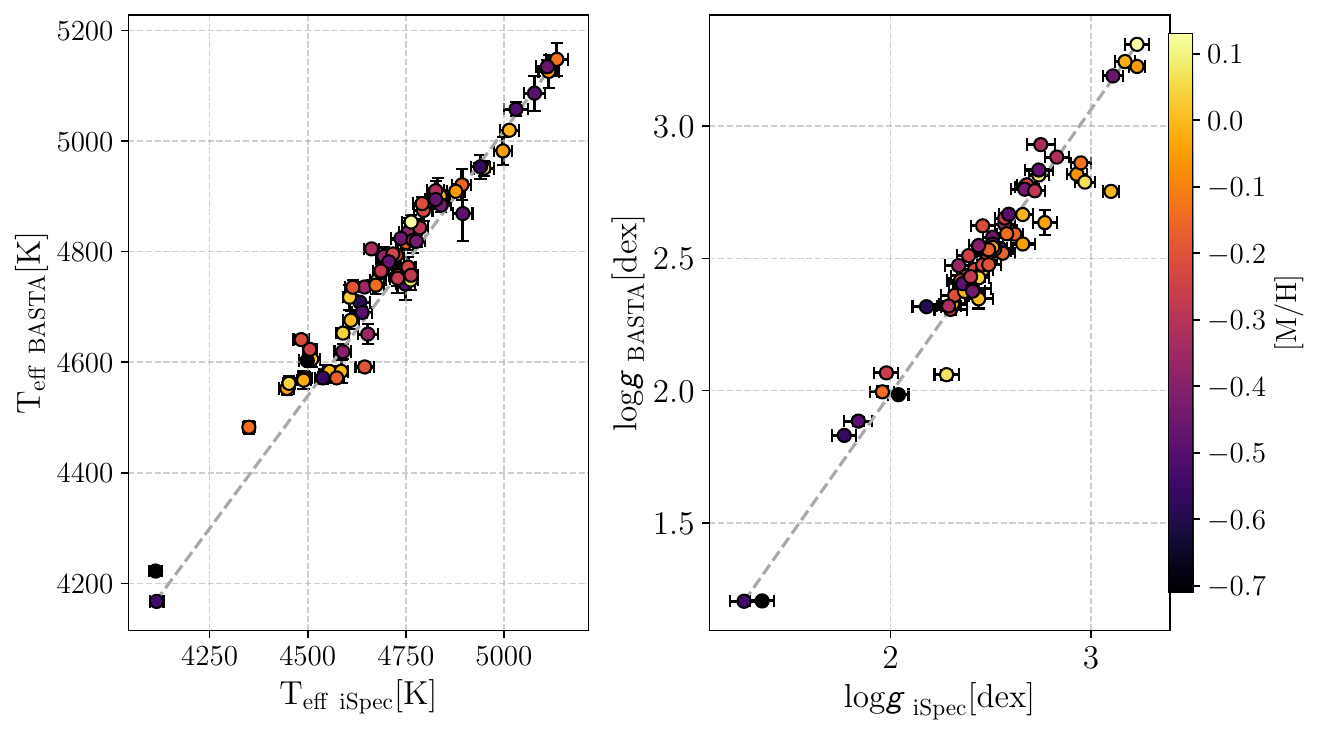} 
 \caption{A comparison between the properties from the spectroscopic analysis (iSpec) and stellar inference (BASTA). \textbf{Left:} Surface temperature with the color code representing the derived spectroscopic metallicities. \textbf{Right:} same comparison but for the $\log g$ parameter.}
 \label{fig:basta_ispec}
\end{figure}

We matched the observations to the set of BaSTI isochrones computed with core overshooting, atomic diffusion and mass loss enabled (case 4 in table 1 of \citealt{2022MNRAS.509.4344A}) deemed appropriate for this sample following \cite{2023MNRAS.524.1634S}. Our results of ages can be checked for consistency by comparing the best BASTA parameters with the spectroscopic ones, which is plotted in Fig.~\ref{fig:basta_ispec}. We can see good agreement having a mean difference of 46 K with a dispersion of 40 K for the temperature, and a 0.02 average difference with 0.09 in dispersion for $\log g$. 
\subsection{Chemical abundances}
We measured chemical abundances for the following families of elements: $\alpha$-capture (Mg, Si, Ca, Ti); odd-Z (Na, Al, Sc, Cu, V); iron-peak (Cr, Mn, Fe, Co, Ni, Zn);  neutron-capture (slow-processed : Sr, Y, Zr, La, Ce, Nd; rapid-process: Eu). We fixed the atmospheric parameters calculated as explained in Sect. \ref{sect:AP}. We proceeded with specral synthesis adopting the same radiative code and atmospheric model as for the stellar parameters. We determined the mean abundances ratios $ \langle \left[\mathrm{\frac{X}{Fe}}\right] \rangle $ using the line-by-line absolute abundances of each element. To be able to compare the chemical relations with the literature (see section \ref{subsect:validation}), we expressed the bracket abundances with respect to the Sun by adding the absolute $\left[\mathrm{\frac{X}{Y}}\right]_{\odot}$ ratios calculated from a solar HARPS spectrum which we took from the library of \cite{2014A&A...566A..98B}. The HARPS spectrum has a similar resolution to our  UVES sample (R $\sim$ 110000). 

We customized the line selection according to the number of lines per element: when dealing with a substantial number of measurable lines, we discarded the lines that exhibited systematically different results. Conversely, for the elements with 1 or 2 lines (Al, Cu, Zn, Sr, Ce, Eu), we relied on the quality flags \textit{synflag} (for the blending properties) and \textit{gf\_flag} introduced by \cite{2021A&A...645A.106H}. The first flag refers to the blending properties of a line, meanwhile the second indicates the quality of the $\mathrm{log} gf$ value of the line. This meant that we excluded lines with potentially inaccurate $\mathrm{log} gf$ measurements from our analysis.

It is worth to stress that our analysis is done in LTE. It has been demonstrated that for the solar metallicity range non-LTE effects have small impact on the computation of abundances (e.g  \citealt{2011Bergemann,2015A&A...579A..52N,2019Mashonkina,2020A&A...642A..62A, 2020Amarsi,2020ApJ...896...64L}) with respect more metal-poor companions. Nevertheless, recent works by \cite{ 2023arXiv230901402A,2023MNRAS.525.3718S} examined the impact on NLTE for classical elements used as chemical clocks, such as yttrium. They demonstrated that the for late type stars, the corrections concerning Y II lines do not exceed $\approx$0.12-0.15 dex close to solar metallicity. The deviations can be more severe for metal-poor red giants, and they exceed 0.5 dex only for [Fe/H] < -3.0, 
a metallicity range that falls out the coverage of our sample.

Furthermore, it is important to consider that the number of suitable lines for abundance determination can considerably differ from one element to the other. Elements with fewer lines (e.g. Mg I, Sr I, Zr II) may be more susceptible to the influence of atomic or molecular data uncertainties and blends due to the lower statistics of the measurements. All these effects systematically contribute to the overall uncertainty in the determination of abundance ratios, and separating these effects can be challenging. 

We decided to compute the uncertainties of the abundance ratios by perturbing the spectra within their flux errors and by repeating the entire analysis (the synthetic spectral fitting both for the atmospheric parameters and the abundances) ten times. By taking the mean and the standard deviation of these "repeated" measurements, we inferred the average dispersion of our abundance measurements  as a response of the SNR of our spectra. We note this represents an internal precision of our abundances, and not a measurement of their accuracy. For the purpose of this paper, which is to test the relations of these abundances with ages as a function of metallicity, we are more interested in the relative difference among our measurements and thus we aim to reach high precision. For all the elements the internal precision is around 0.01-0.02 dex, with a maximum of 0.04 dex for strontium and zirconium.

Complete tables providing the atmospheric parameters, the line selection and the mean abundance ratios with their associated uncertainties are available as online material. 

\begin{figure}[t]
\centering
 \includegraphics[width=\columnwidth]{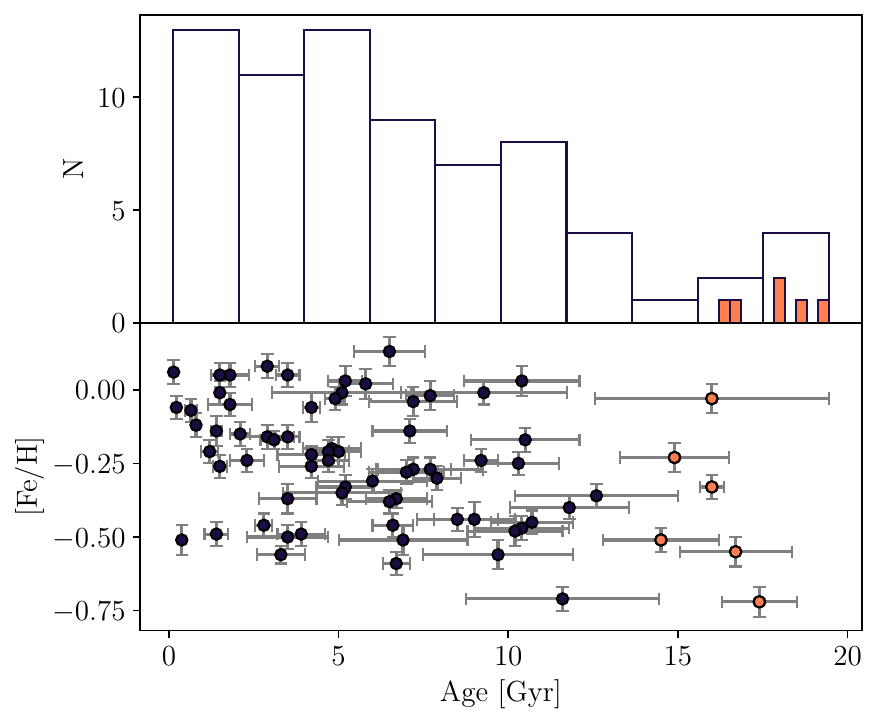} 
 \caption{\textbf{Top:} Histogram of the age distribution of the giant sample. The orange color denotes the stars for which we inferred ages greater than the age of the Universe, i.e. > 14 Gyr. \textbf{Bottom:} the  panel illustrates the age-metallicity distribution following the same color code.}
 \label{fig:feh_age}
\end{figure}

\begin{figure}[]
\centering
 \includegraphics[width=0.99\columnwidth]{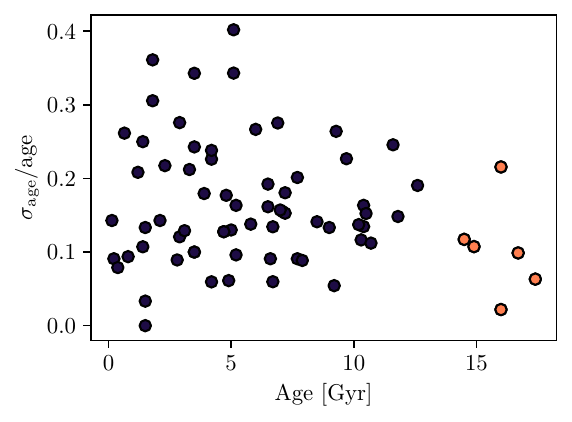} 
 \caption{Relative age uncertainties for the entire sample with stars older than 14 Gyr colored in orange.}
 \label{fig:age_err}
\end{figure}
\section{Results}\label{sect:results}

\subsection{Age-metallicity relation}

\begin{figure*}[]
\centering
 \includegraphics[width=\textwidth]{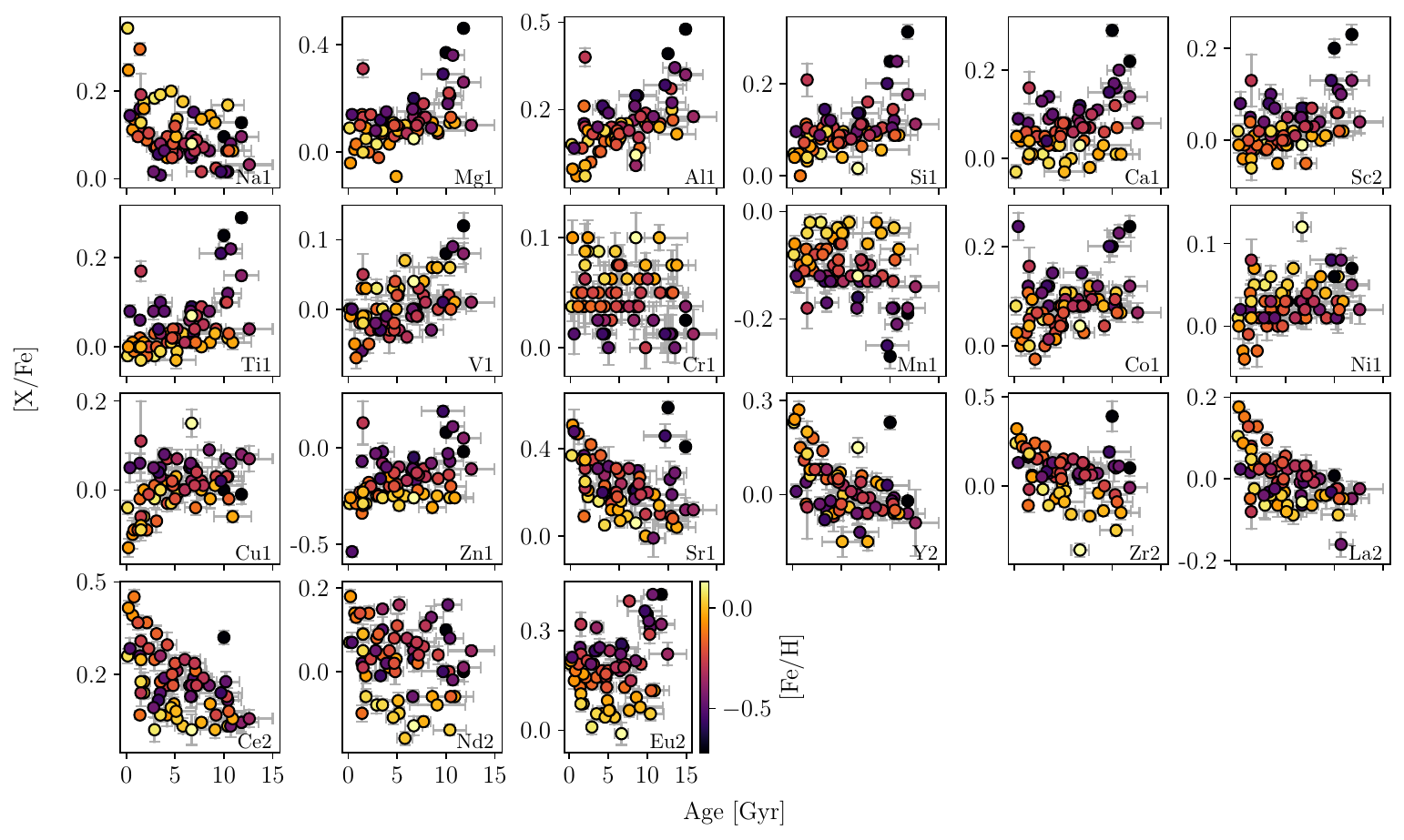} 
 \caption{[X/Fe] as a function of ages, for stars younger than 14 Gyr. The colors are regulated by the iron abundances and the panels are ordered according to the atomic number of each element. }
 \label{fig:el_ages}
\end{figure*}

Figure~\ref{fig:feh_age} depicts the distribution of the resulting ages on the upper panel, and the age-metallicity relation (AMR) on the lower panel. Our sample hence covers ages of the entire range of cosmic history, from very young to very old stars. The range in metallicity is also significant, allowing us to have an AMR to study the chemical evolution of the Milky Way. Similar to previous studies \citep[e.g.,][]{2018MNRAS.475.5487S,2019MNRAS.489.1742F}, the AMR of our stars shows no correlation, thought we lack young stars at low metallicity. Nevertheless, the scatter is very large, in fact much larger than the uncertainties. This underscores the complexity and dynamics of our Galaxy, implying that relying solely on this relationship is insufficient for constraining the formation and the chemical evolution of the Milky Way. 

Despite the improved precision in our abundances and accuracy in our ages, with this sample we are not able to see the dual age-metallicity relation recently discussed by \cite{2020A&A...640A..81N}, \cite{2021ApJ...920...23J} and \cite{Sahlholdt22}. This might be due to the fact that our stars span a wide range of Galactic volume, as well as metallicities. 

In Fig.~\ref{fig:feh_age} it is possible to see that six stars show non-physical ages, i.e. > 14 Gyr. This group includes one of the two most metal-poor and cooler stars within the sample. For the entire group we checked the values of their $\alpha$ abundances (using Mg, Ca, Si and Ti) finding $[\alpha/\mathrm{Fe}] > 0.3$. We attempted to utilize isochrones with $\alpha$ enhancement for this sub-sample, but the results did not improve significantly. We thus decided to not consider these stars in the analysis, reducing the sample from 72 to 66 stars. The challenge of age calculation is evident from Fig.\ref{fig:age_err}, displaying the relative uncertainties, which extend up to 40 \% and a mean uncertainty of $\sim 10\%$.

\subsection{Abundance-age trends}\label{sect:ab-age}

For two stars we measured very high abundances of all s-process elements, hinting towards a pollution from an asymptotic-giant-branch (AGB) star. The abundances thus cannot be used for studying the chemical evolution of the Galaxy and were removed from our analysis. The study of their binary nature will be published in a complementary paper. This cut reduced the sample to 64 stars. 

In Fig.~\ref{fig:el_ages} we show the abundance ratios as a function of stellar age for the stars of our sample that had reliable ages and no signs of binarity. Each panel shows a different element. We note that most metal-poor and old stars are enhanced in the $\alpha$ elements (Mg, Si, Ca, Ti). In fact, there is a general increasing trend of [$\alpha/\mathrm{Fe}$] with stellar ages. This can be seen as the result of chemical evolution which is considerably shaped by the yields of SN type II and SN type Ia (hereafter SN II and SN Ia, respectively) occurring on different timescales\citep{2016Matteucci}. 

Our trends agree with previous chemo-chronological studies conducted on solar-like stars \citep{2019A&A...624A..78D, 2020A&A...640A..81N, 2012A&A...542A..84D} where calcium exhibits a less pronounced correlation while Na abundances lie on a opposite and spread trend. Referring to the models presented by \cite{2020ApJ...900..179K} for sodium (Na), the trend within a comparable metallicity range to ours does not exhibit significant differences. Additionally, it is important to remember that Na abundances are heavily influenced by NLTE effects \citep{2007Andrievsky}, which we are not taking into account in this study. Plus, as we are studying giants, which have likely undergone mixing processes. These factors can potentially blur our chemical-age relation. In the case of calcium (Ca), this observation may be partially attributed to the contribution of SN Ia in its production \citep{2020ApJ...900..179K}. Silicon (Si) displays a slightly weak increasing trend with age, which is in agreement with \cite{2019A&A...624A..78D} and \cite{2020A&A...640A..81N} but conflicting with the results reported by \cite{2012A&A...542A..84D}, who reported no discernible trend.

Generally the behaviour of the $\alpha$ elements is opposite to the one of the s-process elements, which have a lower ratio with respect to iron for older stars. There is no noticeable difference between the lighter strontium (Sr), yttrium (Y) and heavier zirconium (Zr), lanthanum (La), cerium (Ce) n-capture elements, apart from a slightly flatter trend of La for younger stars. As in \cite{2019A&A...624A..78D, 2012A&A...542A..84D} the s-processed element neodymium (Nd) creates a flatter trend especially for older stars, while its production seems to increase at more recent times. 

\begin{figure}[t]
\centering
 \includegraphics[width=\columnwidth]{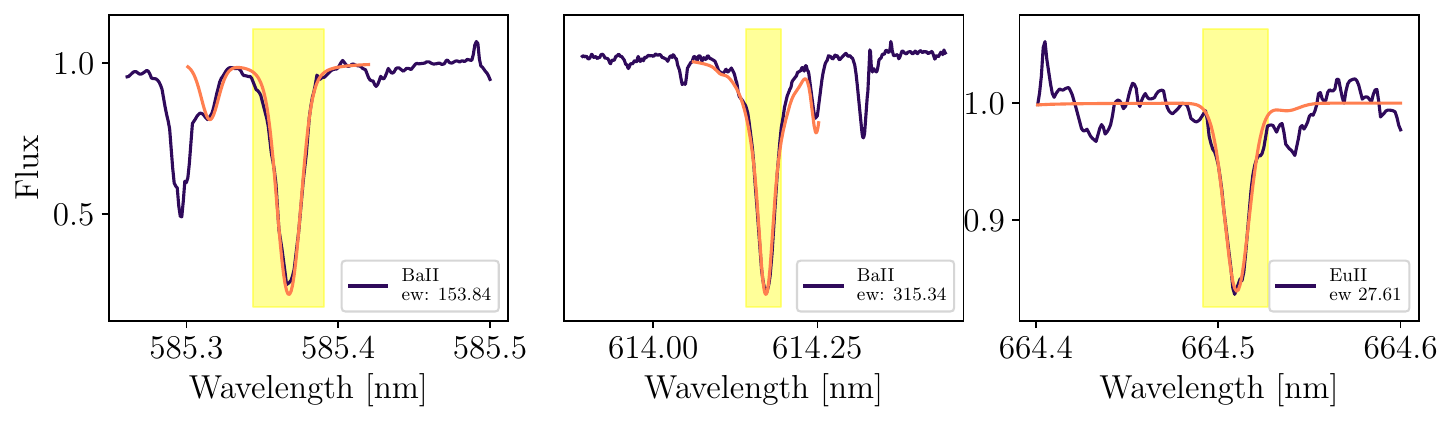} 
 \caption{Profile of two Ba II lines at 585.366 and 614.171 nm and the Eu II at 664.5101 nm for the giant J04034842+1551272. The darker lines represent the observed spectra, the orange color denotes the synthetic fit and the yellow color the area in which the abundances are computed. The values of the equivalent widths of each line are reported in each panel.}
 \label{fig:euba_lines}
\end{figure}

Concerning the r-process element europium (Eu), the observed scatter can be attributed to the challenges in measuring its abundances. Our measurements were limited to only one line ($\lambda$664.5101 nm), located in the reddest part of the spectrum. An example of the Eu line profile for J04034842+1551272 can be seen in the right hand panel of Fig.~\ref{fig:euba_lines}. Nonetheless, the enhancement for the oldest giants agrees with the expected increasing of Eu for lower [Fe/H] values \citep{2019A&A...624A..78D, 2021A&A...649A.126T}. 

Barium (Ba) is a good representative element for the s-process family and the abundance of barium was measured for the entire sample. However, we have opted not to report our results. The two Ba II lines used in our analysis (585.366 and 614.141 nm, which are also shown in Fig.~\ref{fig:euba_lines}) require careful assessment. These lines are known to have a strong correlation with stellar activity \citep{2017ApJ...845..151R}. Furthermore, the intensity of the lines identified in our sample of giants raised concerns about entering a regime influenced by saturation and NLTE effects \citep{2020ApJ...896...64L,2021A&A...653A..67B}. Indeed, for the star J04034842+1551272 plotted in Fig.~\ref{fig:euba_lines}, the Ba lines are saturated. 

Regarding the iron-peak elements, as in other studies, we did not find any specific correlations. Cobalt (Co) and nickel (Ni) show an increasing trend for stars younger than the Sun with the oldest experiencing a flattening trend. Chromium (Cr) and manganese (Mn) exhibit more dispersed trends, possibly indicating the influence of AGB stars in their formation processes.(\citealt{2020ApJ...900..179K}).

Finally, among the odd-Z element, aluminium (Al) shows a strong increasing trend with age, which makes this explosive element a valuable component for chemical tagging \citep{2020A&A...633L...9J,2021A&A...652A..25C, 2015MNRAS.453..758H, 2020MNRAS.493.5195D}. A similar strong trend can be seen for vanadium (V), which is mostly produced by core-collapse supernovae (CCSNe) \citep{2020ApJ...900..106O} with a smaller contribution from SN Ia \citep{2020ApJ...900..179K}. Scandium (Sc), present a moderately smaller decrease over time, but a trend is still visible, being mostly produced by CCSNe. Copper (Cu) is a more complex case, lacking a clear trend with time, being partially formed from AGB stars, while models for Sc attributed its formation to SN Ia \citep{2020ApJ...900..179K}.

\subsection{Comparison with established chemical clocks}\label{subsect:validation}

\begin{table}[]
    \centering
    \begin{tabular}{c c c c} 
    \hline \hline
        Group & [Fe/H] range & Number of stars & Color \\
        \hline \hline
         \texttt{solar} &  [-0.1, 0.1] & 18 & blue \\
         \texttt{mix} & [-0.35, -0.1] & 24 & yellow \\
         \texttt{poor} & [-0.7, -0.35] & 20 & red \\
         \hline
    \end{tabular}
    \caption{Definition of metallicity groups used to divide the clean sample with the respective number of stars for each group. These groups exclude stars eliminated due to age or chemical cuts, in addition to the two potential blue stragglers as as discussed in sect. \ref{subsect:validation}.) }
    \label{tab:groups}
\end{table}

\begin{figure}
\centering
 \includegraphics[width=1\columnwidth]{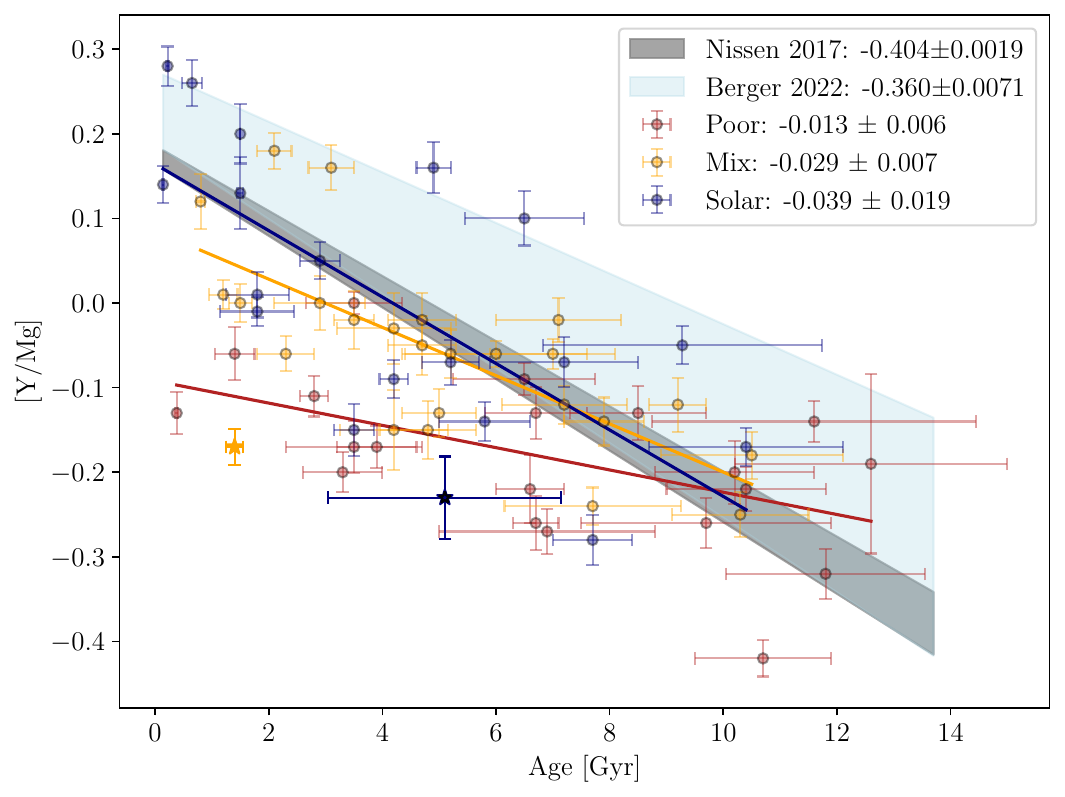}
 \includegraphics[width=1\columnwidth]{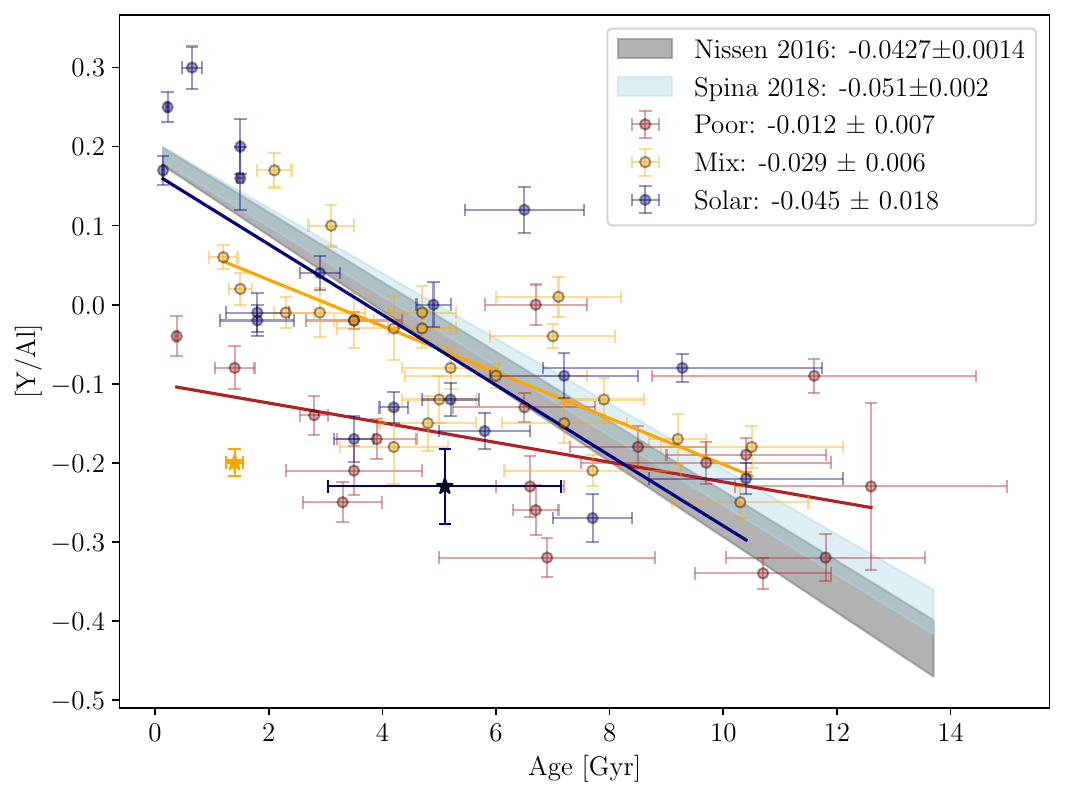}
 \caption{\textbf{Top:} [Y/Mg] vs ages for stars with age < 14 Gyr. \textbf{Bottom:} [Y/Al] vs ages for the same stellar selection. For both panels the three solid lines represent the relation identified for the three distinct metallicity groups outlined in Tab\ref{tab:groups}. The grey and light blue shaded regions display the previously established relationships from the literature derived for solar-twin stars which we compare with the findings for our solar group (in blue). The figure's legends provide the slopes of each relation along with their associated uncertainties. The two giants marked with star symbols are two possible evolved blue stragglers which have been excluded from the fitting procedures, as discussed in section \ref{subsect:validation}.}
 \label{fig:Y_Mg_wo_blue}
\end{figure}

To validate our chemical abundance measurements, we consider the well-studied chemical clocks, which relate [Y/Mg] \citep[for instance,][]{2012A&A...542A..84D, 2015A&A...579A..52N}, [Y/Al] \citep[e.g.,][]{2016Nissen,2016Spina,2018MNRAS.474.2580S,2021A&A...652A..25C} and age. In the case of Y and magnesium (Mg), these two abundances are representative of the families of the n-capture and $\alpha$ elements. Since these two groups of elements are produced through various production channels which operate on different time-scales, their dependencies with time can be seen as a direct consequence of the chemical evolution within our Galaxy. In fact Mg, like other $\alpha$-elements, primarily originates from CCSNe, and it contributed to the enrichment of the interstellar medium (ISM) on earlier and shorter time-scales. On the other hand, Y is an s-processed element that comes from low and intermediate-mass stars (1-8 $\mathrm{M_{\odot}}$) during their AGB phase. These stars release their elemental yields over significantly longer time intervals (for extensive explanations see e.g., \citealt{2006Kobayashi,2012ceg..book.....M, Karakas2014} and references therein). Therefore, the ratios of these element abundances enclose intrinsic galactic time information and can be used as age indicators.

The primary sources of Al are CCSNe \citep{2014Bisterzo,2020Fujimoto}. As stated before, these stars act on significantly shorter timescales compared to AGBs, which are primarily responsible for the production of neutron-capture elements. 

For Y and Mg we compare our results with the relations discussed by \cite{2017A&A...608A.112N} and by \cite{2022ApJ...936..100B}. These relations are based on solar-twin data and were derived in LTE. Ages were calculated with isochrones in \cite{2022ApJ...936..100B}, and using asteroseismic data in \cite{2017A&A...608A.112N}.

To best compare these results with our own, we consider only the stars with measured metallicities around solar ($ -0.1 < [\mathrm{Fe/H}] < 0.1$). We call that group of stars {\tt solar} for better reference and will be represented in blue color throughout the rest of the work. Since our sample covers a wider range in [Fe/H], to study the dependency on metallicity, we divide our sample in more bins, namely, one of intermediate metallicity ($-0.35 < \mathrm{[Fe/H]}<  -0.1$), and one metal poor ($\mathrm{ [Fe/H]} < -0.35$). We call the stars belonging to these divisions {\tt mix} and {\tt poor}. The stars from these groups are plotted in yellow and red, respectively, and their ranges and number of stars can be found in Tab.~\ref{tab:groups} toghether with the {\tt solar} group. 

The abundance-age linear relations were estimated using the {\tt RANSAC} (RANdom SAmple Consensus, e.g. \citealt{fischler1981random}) algorithm for linear regression. {\tt RANSAC} estimates slope and intercept using repeated random sub-sampling to reduce the influence of highly uncertain measurements (or outliers) on the estimated relations. To account for the uncertainties in the data, we perform this regression multiple times, using values from normal distributions for each abundance and age measurement. These distributions are centered around the reported measurements (see results in Tab. \ref{tab:coeff_alpha_n2}) and have standard deviations based on their reported uncertainties. We fit 1000 regressors by sampling abundances and ages in accordance with their uncertainties. The final values of the linear regression coefficients and their uncertainties are computed by determining the mean and variance after these iterations.

In the top panel of Fig.~\ref{fig:Y_Mg_wo_blue} we present the linear relations we fit for the three metallicity groups. The colored shaded areas denote the literature fits along including their respective uncertainties. Grey color represents the relation discovered by \cite{2017A&A...608A.112N} while the blue shade represents the relation of \cite{2022ApJ...936..100B}. Comparing our {\tt solar} results with the literature, we find an agreement with the relations found in the literature for solar metallicity stars. Indeed, our relation is also consistent with other studies that consider red giant branch stars \citep{2017A&A...604L...8S,2020A&A...635A...8C}. 

The bottom of Fig.~\ref{fig:Y_Mg_wo_blue} shows  the same metallicity groups for the [Y/Al] ratio. Here the comparison is performed with respect to the works by \cite{2016Nissen,2018MNRAS.474.2580S} that again analyse solar-twin stars and ages from isochrone models, assuming LTE. The slope found in this work for the solar group fall within the established literature ranges. A slight deviation is noticeable for very young ages, potentially caused by the observed dispersion among younger stars. Additionally, we are comparing our giants with less evolved stars, specifically solar-twin stars. Despite this, the comparison aligns with the findings in the literature, accounting for uncertainties.

\begin{figure*}[t]
\centering
 \includegraphics[width=\textwidth]{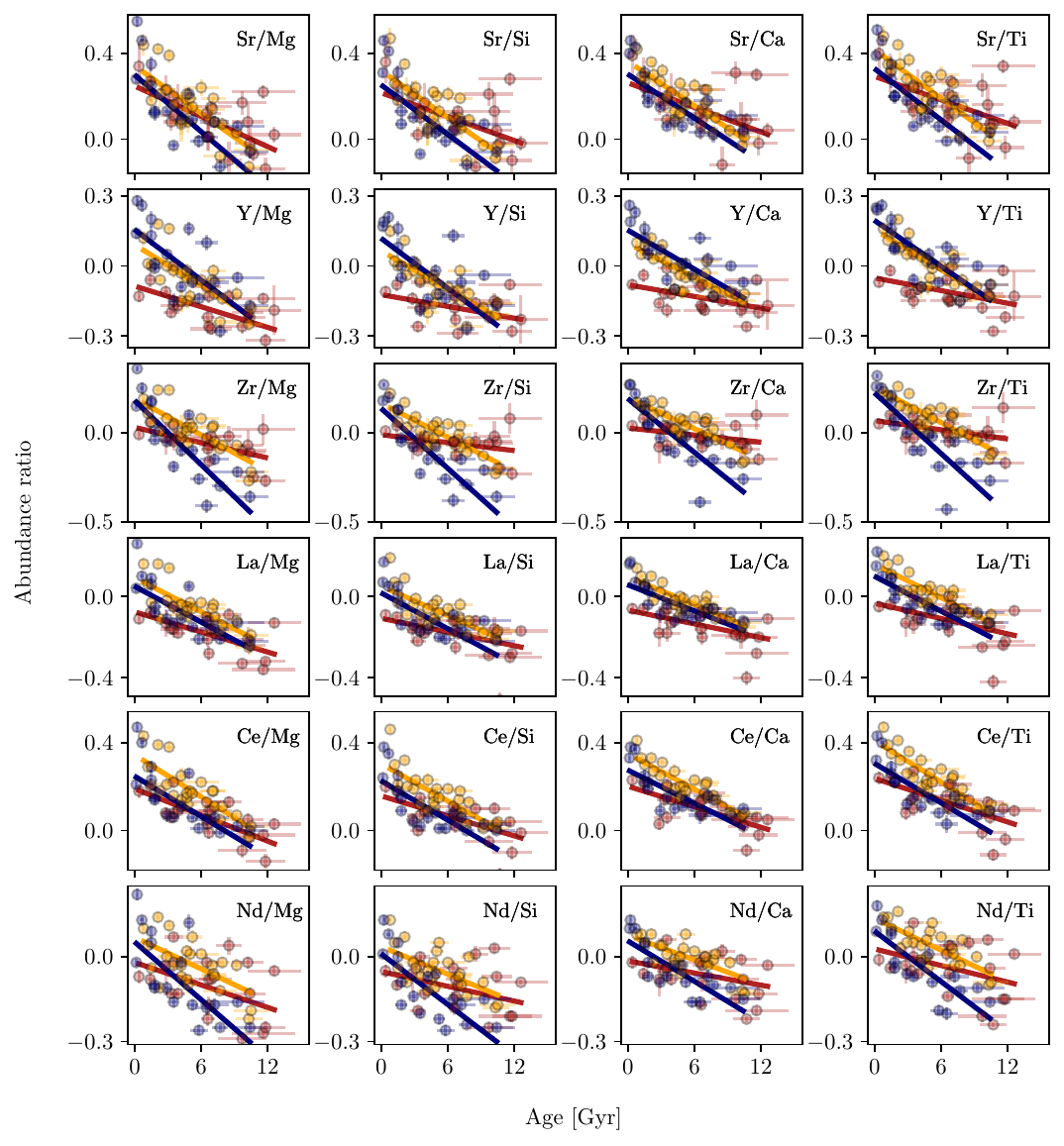} 
 \caption{Chemical traits as a function of age for the 3 metallicity groups with linear regression fits. Blue represents the {\tt solar}, yellow the {\tt mix} and red the {\tt poor} groups (see Tab.~\ref{tab:groups}). The solid lines are the linear fits to the respective metallicity groups. The stellar sample represented do not contain stars older than 14 Gyr and J16081431-2130041, J03573726+242533.}
 \label{fig:nalpha}
\end{figure*}

Similarly to the solar metallicity range, we perform linear regression for the {\tt mix} and {\tt poor} groups. The fits are plotted with a continuous line whose slope is indicated in the legend.  We find for both cases that the chemical-age relation becomes flatter with metallicity, which agrees with previous findings (e.g., \citealt{2017MNRAS.465L.109F,2019A&A...624A..78D,2020A&A...639A.127C,2022ApJ...936..100B}). 

We comment on the stars J16081431-2130041, J03573726+2425332 (marked with blue and yellow star symbols), which have noticeably lower [Y/Mg] and [Y/Al] given their age. By visual inspection of their spectra, we were unable to find anything particular that would make us conclude that the chemical abundances were wrongly determined. The target with a blue star symbol is associated with a relatively large abundance error bar, nevertheless this discrepancy does not account for the low abundance ratio. We further inspect the results of the ages, reaching the same conclusion that the results seem to be well behaved. We believe these stars might be evolved blue stragglers. Recent studies have shown that such stars have indeed very normal chemical abundance patterns and might not necessarily be binaries anymore \citep{Jofre23}, but can still be explained with the merger of two stars \citep{Izzard18}, henceforth being evolved blue stragglers. In the field, stars can have a wide range of ages, blue stragglers tend to be very hidden, as they do not necessarily show oddly young ages, such as blue straggles in clusters \citep{sandage}, blue metal-poor stars in the halo \citep{Preston} or the stars dubbed as young $\alpha$ rich stars \citep{Martig15}. Chemical clocks might offer an interesting way to identify them, because their chemistry predicts an older age than what can be measured with standard evolutionary tracks that consider single stars.

\begin{figure}
\centering
 \includegraphics[width=\columnwidth]{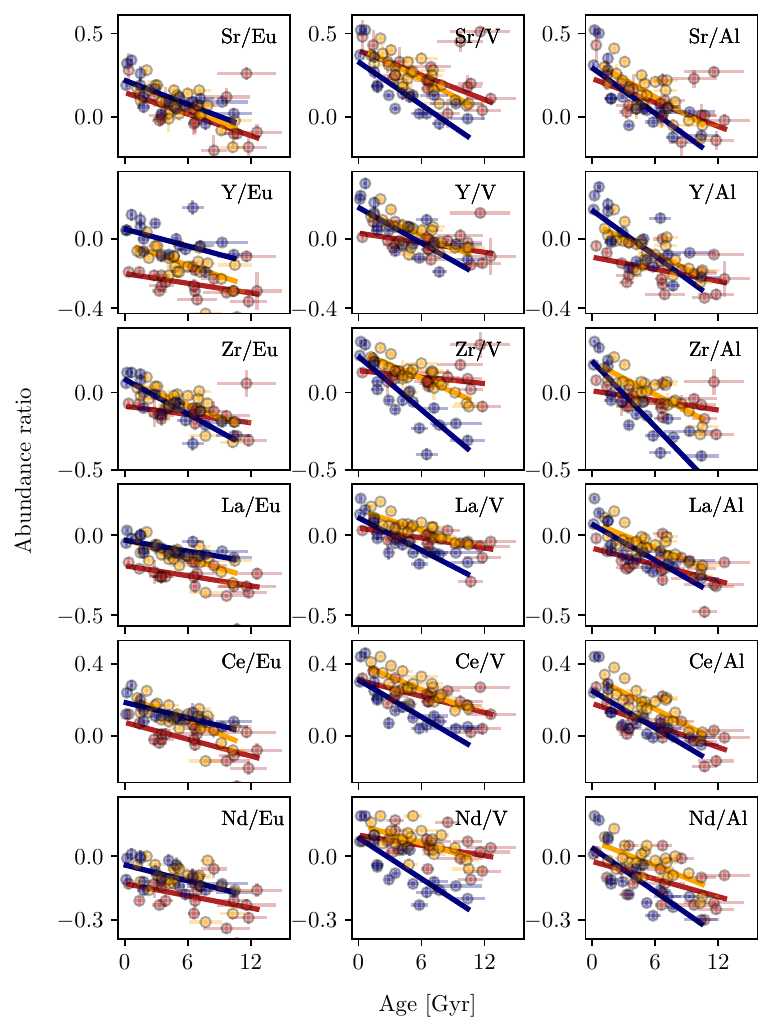} 
 \caption{Ratios of slow (s)/rapid (r) n-capture elements, and s-elements over vanadium and over aluminium as a function of ages. The color scheme for the stars and the regression lines is the same as for Fig.\ref{fig:nalpha}. }
 \label{fig:s_r_al}
\end{figure}

\subsection{Dependency of metallicity in chemical traits} \label{sect:dep_met}

\cite{2020A&A...633L...9J} discussed how many other abundance ratios which combine an s-process element with an $\alpha$-capture element also show a strong dependency with age. They called these abundance ratios chemical traits instead of chemical clocks, arguing that a clock has a universal dependency on time, which might not be the case of these abundance ratios. That work focused on a solar-twin sample, thus it remains to be explored if they all had a similar dependency on metallicity than the more studied [Y/Mg] discussed in the previous section (see also \citealt{2017MNRAS.465L.109F} and \citealt{2020A&A...640A..81N} for discussion). \cite{2017Delgado} explored for a set of FGK dwarf stars, examining various combinations of elements alongside some [s/$\alpha$] ratios. Their study revealed as well variability in abundance-age relationships influenced by factors such as metallicity, stellar structure, and evolution.

Here we have a larger range of metallicity and we can investigate the dependency on these chemical traits with metallicity. To do so, we keep the separation of the 3 metallicity groups listed in Tab.~\ref{tab:groups} and plot the abundance ratios as a function of age in Fig.~\ref{fig:nalpha}, where we respect the color scheme of Fig.~\ref{fig:Y_Mg_wo_blue}. Each panel shows a different trait. Each column is a different $\alpha$-capture element (Mg, Si, Ca and Ti) and each row a different s-process element (Sr, Y, Zr, La, Ce and Nd). Following the analysis discussed in the previous section, we perform linear regression fits to the data  and plot with the corresponding colour a continuous line in Fig.~\ref{fig:nalpha}. From this point of the analysis we exclude the two possible blue stragglers discussed above.  A table reporting the linear regression fit coefficients for the three groups can be found in the Appendix, \ref{tab:coeff_alpha_n2}.

In general, we see that the chemical traits have a negative trend with age, following the same behavior as [Y/Mg], namely that the neutron-capture element over Fe increases with time while the $\alpha$-capture element over Fe decreases. We further find that the {\tt solar} group, which is coloured with blue,  has traits with stronger correlations  than the {\tt poor} group, which is coloured with red. The {\tt mix} group lies in between. Looking at the solar group, Zr, Sr and then Y combined with alphas show the steeper slopes. These combinations usually show the higher correlation (as found as well by \citealt{2020A&A...633L...9J} and \citealt{2021A&A...652A..25C}), which we quantified by computing for each regression fit the Pearson correlation coefficient ($\rho$ coefficient) reported in table \ref{tab:coeff_alpha_n2}.  
The ratios involving Ca in the denominator for Nd, Ce, La and Y do not satisfy the criterion of having a slope greater than a threshold value set at 0.03 dex $\mathrm{Gyr^{-1}}$. This can be partially attributed to the contribution of SN Ia in the formation processes of Ca, resulting in a less pronounced correlation with age compared to the other $\alpha$-elements. The threshold value for the slope is adopted following the same criterion as in \cite{2020A&A...633L...9J}. They derived this criterion from typical observations of the slope in linear fits involving abundance ratios with Fe, which typically remain below 0.03 dex $\mathrm{Gyr^{-1}}$. 

It is interesting to notice that [La/$\alpha$-elements] ratios drive the slopes to be smaller than 0.03 dex $\mathrm{Gyr^{-1}}$. However, our coefficients considering their uncertainties are comparable with the results found by \cite{2020A&A...633L...9J}. In this same literature work Ca does not produce considerable strong correlations with age, except when it is considered alongside Ba, an element we have opted not to include in our analysis. Nd displays less pronounced slopes as expected from the scatter shown by this element in the [Fe/H]-age plane in Fig.\ref{fig:el_ages}.

In addition to s-process versus $\alpha$-capture, we study traits that combine neutron-capture elements of s-process versus two odd-Z elements: Al and V and with the r-process element Eu. Fig.~\ref{fig:el_ages} reveals an interesting behaviour for these elements, which exhibit a strong and neat correlation with ages. This characteristic makes them  noteworthy candidates for chemical stellar age prediction.

Even if it is not clear which kind of sources are responsible for the production of r-process elements (for instance neutron star mergers, magneto-rotational supernovae, see for further details \citealt{2018Cote}), they constitute the main channel for n-capture nucleosynthesis before the upturn of the AGBs, after ignition of helium shell burning. Therefore, the [s/r] ratios are promising indicators of the contribution of intermediate-mass stars compared to high-mass star, and thus valuable tracers of chemical evolution \citep[e.g.][]{ 2018Magrini,2021RBlanco}. 

Our results for these element are shown in the left-hand column of Fig.~\ref{fig:s_r_al} for the [s/r] abundance ratios, [s/V] in the middle column, and in the right-hand column for the [s/Al] ratio. The rows are sorted in the same way as Fig.~\ref{fig:nalpha}. Similar to the $\alpha$-captures, the traits when considering aluminium have a dependency with metallicity such that the {\tt solar} group yields a linear regression fit with age that has a steeper slope than the {\tt poor} group, with the {\tt mix} group in between. This is because Al is also produced in CCSNe, at rates that are comparable to $\alpha$-capture elements. Also for this case, Zr is one of the elements that shows stronger slopes, followed by Y and Sr. La and Ce, while showing slightly weaker correlations, still have ratios with Al that are related with ages with slopes greater than the threshold value. The trends observed in the other odd-Z element V display a comparable pattern, albeit with slightly less pronounced slopes. This behaviour can be linked to the involvement of SNIa in the production of V, in contrast to Al, which is predominantly formed from CCSNe. Again, Zr and Sr exhibit stronger correlations with steeper ratios together with La, while Ce and Nd showcase more moderate trends as for the [n-capture/$\alpha$]-elements.

The [s/r] abundance ratios have different dependency with age and metallicity, meaning that the stars in the solar group do not necessarily display the stronger dependencies with ages. In fact, it is the mixed group for which we observe the steepest slopes when examining the relationship with age. The only exceptions are [Zr/Eu] and [Al/Eu]. These ratios yield a correlation with age that is consistent with the s-process/$\alpha$-capture traits, namely stronger for the {\tt solar} group. For the other traits, the trends are negative, but with comparable trends with age. It is important to bear in mind that the measure of Eu abundances must be approached with caution since they are based on one line, see explanation in section \ref{sect:ab-age}.

Finally, the scatter of the older and more metal-poor targets which blur the age relations can be be ascribed to the challenges in estimating ages, particularly for giants, resulting in greater imprecision. This makes it more difficult to separate the pure [Fe/H]-dependency from the dispersion due to the age variable.

\begin{figure*}[h]
\centering
 \includegraphics[scale=0.55]{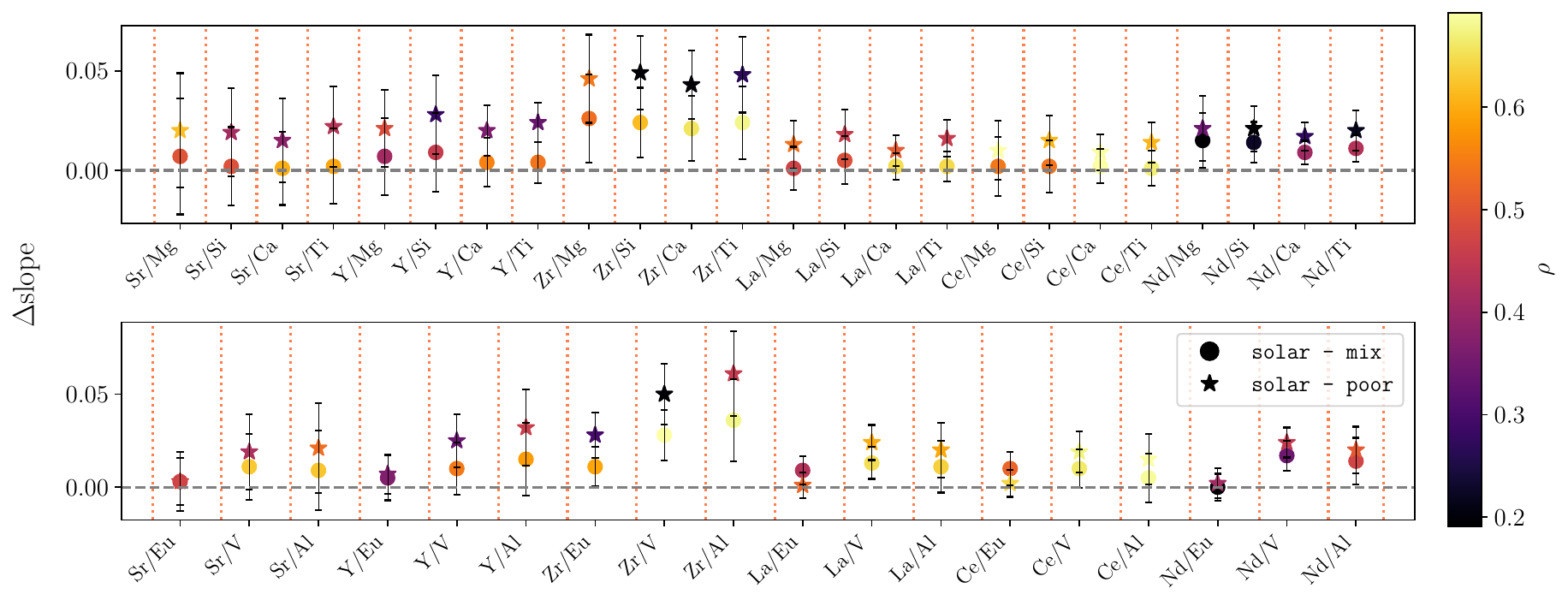} 
 \caption{Slope differences for each element combination of each ratio. The dot and star points specifically highlight variations between the solar and mix groups, and the solar and poor groups, respectively. The color-coding represents the correlation coefficient values. For the dot points, $\rho$ is computed based on the mix group, whereas for the star points, the correlation coefficients are derived from the poor group.}
 \label{fig:delta}
\end{figure*}

\subsection{Comparing slopes}

Generally, from Fig.~\ref{fig:nalpha} and \ref{fig:s_r_al} we find that all chemical traits have a similar dependency with metallicity, namely that the strongest dependency is for solar-metallicity stars, and it weakens towards more metal-poor stars. The exception is when studying only neutron capture elements and the [s/r] abundance ratio, which seems to be more independent of metallicity. 

In Fig.~\ref{fig:delta} we summarize these findings by studying further the change of the slope for each trait.  We plot the difference in slope of our linear regression fits for each group. The circles represent the difference in slope between the {\tt solar} and the {\tt mix} group, while the star symbol represents the difference between the {\tt solar} and the {\tt poor group}. The symbols are color coded by the $\rho$ correlation parameter of the linear regression fit found for the {\tt poor group} (star symbol) and {\tt mix} group (dot marker). The error bars incorporate the combined uncertainties of the differences between the ratios. The top panel of the figure summarizes the differences for the traits shown in Fig.~\ref{fig:nalpha} and the bottom panel summarizes the differences for the traits of Fig.~\ref{fig:s_r_al}. 

From this figure we immediately see that Zr is the element with the strongest dependency on metallicity, followed by Y and Sr. The differences in slope for the correlations of [Zr/$\alpha$-capture] and [Zr/Al] are up to 0.05 dex, twice as high as any other element, while the differences for Sr and Y are lower than 0.03 dex. Similar findings are reported by \cite{2018Magrini} and \cite{2017Delgado}, who have unveiled a pronounced [Zr/Fe] vs. [Fe/H] trend. We note however there is a high intrinsic scatter in these relations, potentially resulting from measurements based on a single absorption line for both Zr and Al, suggesting caution in adopting these elements as possible stellar chemical-clocks.  We further see that the traits that involve La and Ce have a small dependency with metallicity, e.g. the $\Delta$ slope values are close to zero, hence more universal stellar chemical-clock. 

The different metallicity dependence discussed above could be associated with the diverse origins and formation pathways of both light (\textit{ls}) and heavy (\textit{hs}) s-process elements, also known as first-peak (Sr, Y, Zr) and second-peak elements (Ba, La, Ce, Pr, and Nd) respectively. The \textit{ls}-elements in addition to the significant contributions from the s-process channel \citep[65\% for Zr and 69\% for Y,][]{Travaglio2004}, exhibit notable contributions from intermediate-mass AGB stars \citep{2020ApJ...900..179K,2023Goswami} and more massive star \citep{2010ApJ...710.1557P}. On the other hand, the primary sources responsible for these heavy s-process elements are the yields from low-mass AGB stars with masses of $\leq 3 \mathrm{M_{\odot}}$ \citep{2003Lugaro}. For both groups, the complex production chain of s-process elements happening in the thermally-pulsing phase of AGB stars in the burning-shell regions highly depends on metallicity \citep{2014Bisterzo,Karakas2014,2015Cristallo,2021A&A...646L...2M}. This significantly challenges the uniformity of these ratios in conveying age-related information. 

Another notable difference between the first and second-peak elements is the percentage of s-process contribution and the neutron flux within their respective sources. For the second-peak elements, both the s-process contribution and neutron flux are high \citep{Travaglio2004,2016ApJ...825...26K}. Earlier studies have demonstrated that the decline in metallicity has a greater impact on elements from the first peak compared to those from the second peak. This is because the neutron flux rises as metallicity decreases, meaning that first peak elements are expected to be more abundant at higher metallicity \citep{2006Gallino, 2015Cristallo}.
Consequently, it is plausible that within the range of metallicities being studied, the production mechanisms of \textit{ls}-elements  may undergo alterations.
 
In this context, the study of \cite{2018MNRAS.474.2580S} demonstrated that the slopes of [X/Fe]–age relationships exhibit an increase with the s-process contribution, that is elements such as Zr, Y, and Sr present higher values of slopes, i.e. steeper, with only barium displaying even greater values. In their work, the \textit{hs}-elements, La and Ce exhibit less steep slopes compared to the \textit{ls}-companions (as found in our analysis) even when their s-process contributions are similar.

Moving to Nd, one of the heaviest element, we observe relatively weak correlations with age. Its production pathway shows resemblances to Ce, primarily originating from low-mass AGBs \citep{2003Lugaro}. Consequently, one might anticipate strong age correlations when Nd is examined in conjunction with $\alpha$-elements, similar to Ce. Referring to \cite{2014Bisterzo} the s-contribution of Nd is $\approx 58 \%$, while for Ce is $\approx 84 \%$. These different contributions might justify the different age trends detected for the two elements.

Regarding the behavior of the $\alpha$-capture elements, there are not any prominent features that emerge. We can see that when Ca is considered the difference in slope is slightly smaller. Indeed, the [Y/Mg] trait has a stronger dependency with metallicity than its equivalent [Y/Ca]. As mentioned in Sect.~\ref{sect:dep_met} this may originate from the yields of SN Ia, which contribute to the release of calcium \citep{2023Zenati}. 

Regarding the behavior of the r-process element Eu, we see that in general the dependency on metallicity is very small, and that the slopes are flatter compared to the ratios with the same n-capture elements but with $\alpha$ or odd-Z elements in the numerators. Similar conclusions are reported in the work of \cite{2018MNRAS.474.2580S,2019A&A...624A..78D}. We find that most traits have negligible differences in the slope of the regression fits, except [Zr/Eu], but we have already found that Zr might be driving the metallicity dependency. The distinct behaviours of these element ratios which result in more shallow slopes can be attributed to the debated origin of r-processed elements. Apart from CCSNe, more exotic scenarios have been proposed as potential sources for r-processed elements, including neutron star mergers \citep{2018Cote}.

The odd-Z elements V and Al, when combined with neutron-capture elements, once again exhibit stronger correlations with age within the solar metallicity group. However, there is a noticeable flattening trend in the cases of Y, Zr, and La when they are combined with vanadium for the metal-poor stars. Overall, Al appears to have a lesser effect on flattening the age-metallicity relationships for lower metallicities. For both odd-Z elements, their dependence on metallicity, when considering the same specific n-capture, is more significant than that of Eu and it is more similar to the $\alpha$-elements.

Finally, we focus on the correlation coefficient. The traits with the tighter correlation for both {\tt mix} and {\tt poor} are [Ce/Ca], [Ce/Al] and [Ce/Ti], which interestingly have a very small dependency with metallicity. Generally, the [Ce/$\alpha$] ratios show this small dependence, with the majority having slopes around 0.03 dex $\mathrm{Gyr}^-1$. The study conducted by \cite{2020A&A...633L...9J} yielded comparable values, finding an equivalent slope to ours concerning [Ce/Si], along with a minor variation of 0.007 dex $\mathrm{Gyr}^-1$ for [Ce/Mg]. 

The traits that involve Zr, which have the highest metallicity dependency, have quite weak correlation factor for the most metal-poor group. It is thus possible that our Zr measurements are simply very uncertain at low [Fe/H]. A similar effect might be causing the poor correlation found for Nd. We know that Nd lines are strong and heavily affected by hyperfine structure splitting, causing a large line-to-line scatter in abundance measurements \citep{2020A&A...635A...8C}. It is interesting that the classical [Y/Mg] trait is not the element with the strongest correlation coefficient, or the least dependency in metallicity. Indeed, the trait that is the best candidate for chemical clock seems to be [Ce/Ca], followed by [Ce/Al] (see Fig. \ref{fig:nalpha} and \ref{fig:s_r_al} respectively) because of their strength of the slope and weak metallicity dependency. Another ratio displaying similar characteristics is [Ce/Eu], even tough from Fig.~\ref{fig:s_r_al} we can see that its overall trend with age is very flat, thus not making this element ratio vary much with time. \cite{2022Ssilva} tested the [Ce/$\alpha$] ratios as stellar chemical clocks by examining the evolution of Ce in 42 open clusters. Their conclusion suggests that these ratios exhibit variability across the disk owing to the influence of AGB yields, which are dependent on metallicity. It needs however to be remarked that the clusters analyzed in their study encompass a wider Galactic radius ($6 \lesssim \mathrm{R_{GC}} \lesssim 20$ kpc) compared to our work. At the same time, their analysis revealed a comparable age-dependence pattern within two distinct metallicity groups (namely -0.6 < [Fe/H] < -0.1 and -0.1 < [Fe/H] < +0.4), which is a similar behaviour observed in our relations. The study on 47 open clusters conducted by \cite{2021A&A...652A..25C} also revealed an increasing scatter in the age-abundance relations across various radii outside the solar neighborhood ($d > 1$ kpc), demonstrating a dependency on the spatial volume analysed. However, they did not consider in particular ratios involving Ce. 

Accounting for all the astrophysical explanations, deriving robust conclusions solely based on pure metallicity dependence becomes challenging due to the dispersion in abundance-age trends. Factors beyond chemical evolution arguments can significantly impact these relations, and complicate the comprehension of our chemical trends. These variables might include uncertainties associated with determining atmospheric parameters (partly due to the level of the SNR of the measurements), inaccuracies in estimating the ages for giant stars \citep{2022MNRAS.517.5325H,2022Sharma}. 

Another aspect to consider concerns the use of linear fittings which can potentially constrain the interpretation of the data, particularly when dealing with older stars. In fact, due to their increased dispersion, linear regressions may no longer accurately capture their trends. To best reproduce the different chemical-age features, as turnovers or varying slopes across age ranges, \cite{2016Spina} explored hyperbolic and two-segmented line fits in their study, while \cite{2020A&A...639A.127C} used a multivariate linear regression, incorporating metallicity as a dimension. On the same line, \cite{2017Delgado} fit different multi-dimension relations to the abundance-age trends analysed in their work. These attempts highlight that employing fitting models beyond linear regression could offer an alternative to replicate the observed tagging trends. Nevertheless, employing higher-order polynomials might also introduce additional complexity, making it more challenging to interpret the results and to disentangle the various dependencies. 

\section{Discussion}\label{sect:discussion}

Despite the extensive exploration of chemical tagging in recent years, there are still open questions to be addressed. For instance, which traits show the strongest dependence on metallicity and how do we explain these  dependences from the point of view of Galactic evolution. Ultimately, in the complex scenario of chemical trends, is it feasible to pinpoint "golden" candidates for chemical tagging?

\subsection{Assumption and implication in Galactic evolution}

Given the extended metallicity and age ranges covered by our sample, we were able to explore various abundance ratios at different metallicities. As outlined in other  studies (e.g. \citealt{2017MNRAS.465L.109F,2019A&A...624A..78D, 2020A&A...639A.127C,2021A&A...646L...2M}), we observed a non-uniform pattern in the relationship between abundance and age, attributed to variations in metallicity. Specifically we found a decline of the slopes with the [Fe/H] of the stars, meaning a decline in age predictability. 

This behaviour needs to be considered when pursuing chemical tagging, particularly for individual field stars. The dependence on metallicity poses a constraint on the possibility of using the chemical-age relation within the context of Galactic Archaeology. Leveraging observed abundance patterns to constrain chemical enrichment history and infer stellar ages can serve as a potent tool but requires consideration from various perspectives.

Firstly, if we assume that stars of a specific age retain the characteristics of their original gas, which should remain chemically distinct, then the chemical space enables dating the star and its hosting environment. However, this notion becomes less straightforward when considering that stars posses different birth radii, being born in different places. 
  
Focusing on the metallicity dependence, it is essential to acknowledge that despite assuming that the heritability of stars descendent from the same gas cloud can be retrieved by chemical abundances, the complexity of the combination of production sites of the elements and the different Galactic locations where the stars are born can render this picture less straightforward to interpret. Indeed, the abundances analysed in this work come from a variety of sources which act on different timescales according to the theory of Galactic evolution \citep{2002Freeman,2013Rix,2016Matteucci}. As discussed in Sect.~ \ref{sect:dep_met}, certain nucleosynthesis processes and their associated yields, particularly for AGBs, are subject to the influence of the source's metallicity. This dependence will inevitably impact the production of various elements, leading in the final production to the prevalence of certain elements over others,  for example in the first-peak and second-peak n-capture elements. 

Alongside the diversity of production channels, it is important to add to the picture the intricate diversity of chemical substructures within our Galaxy, which suggests the presence of a complex mixture of stellar populations. Indeed, the ISM will be enriched differently according to the star formation rate and star formation history characterising the Galactic volume considered. This stochastic enrichment \citep{2018Krumholz} grows more intricate when adding the role of stellar migration in blurring chemical gradients and signatures as stars from different sites mix together. The variations in the chemical clock relations are notably influenced by the diverse birth radii of stars \citep{2002Sellwood,2012Minchev}, which in their abundances carry the fingerprints of distinct local star formation histories, as demonstrated by the recent work by \cite{2023arXiv230711159R}.  

In our case the stellar sample is composed of field stars. Although their positions place them in the Solar neighborhood (see Fig.~\ref{fig:cord}), it can not be ruled out that they might have travelled across the Galactic disk. The spread in the age-metallicity relation (Fig. \ref{fig:feh_age}) and the presence of solar metallicity stars with ages $\sim 10$ Gyr  suggests that those older stars might have formed in the inner Galaxy and then  migrated radially towards the Solar neighborhood. In such instances, the dispersion in the chemical-age relations could potentially be attributed to differences in birth radii (consequently different local metallicity), that trace the composition of the original ISM. A star formed in an ISM with lower iron content undergoes a diminished supply of iron for neutron capture processes \citep{2016ApJ...825...26K, 2022Univ....8..173C}, hence diminishing the production of n-capture elements and weakening the [n-capture/$\alpha$]-age trends.

Secondly, through the analysis with two known chemical clocks mostly employed for solar analogous stars (see sect. \ref{subsect:validation}), we noted that these ratios for the \texttt{solar} group continue to serve as reliable indicators of chemical enrichment also in the case of our more evolved targets, i.e. giant stars. This aligns with the results presented by \cite{2017A&A...604L...8S}, who confirmed the tight correlation between [Y/Mg] and age for six evolved stars. \cite{2022ApJ...936..100B} recovered significant trend for non-solar FGK stars, even if less precise than the relations inferred for solar analogues. Moreover, most of the coefficients of the \texttt{solar} group (see Table~\ref{tab:coeff_alpha_n2}) are in good agreement with the ones found by \cite{2020A&A...633L...9J}, with the exception of Zr and Al which show greater slopes in our work.

Hence, by confining the analysis to the solar metallicity range, the chemical-age relations appear relatively consistent between RGB and MS stars, thus not dependent on detailed stellar evolution. While we acknowledge the possibility that the poorly understood mixing processes occurring in evolved stars \citep{2014Mosser} can alter the stellar compositions, these mechanisms do not appear to erase the correlation between ages and $\alpha$, n-capture elements for giant stars making them valuable for Galactic archaeology studies. 

\subsection{Limitation and caveats}

When using chemical stellar dating, other variables might cause the variations in the different [X/Y]-age planes. 
Disentangling the scatter due to astrophysical origin from the scatter caused by method uncertainties is not trivial. The situation becomes more severe in the low metallicity regime, where the age predictability of the chemical clocks decreases. 

The systematic average internal uncertainties of $\sim$ 0.03 dex for [Y/Mg] and [Y/Al] (see Fig. \ref{fig:Y_Mg_wo_blue}) is smaller than the scatter around the linear fit for each metallicty group, ranging from 0.05 to 0.15 dex. This means that the observed dispersion around the linear fit can not be accounted for only by the internal precision. Other factors that affect the abundance measurements can be  errors from atomic diffusion, molecular data, omission of NLTE corrections and 3D effects, \citep{2015Adibekyan,2023MNRAS.525.3718S,2023Alexeeva}, as well as the reliance on the chosen lines for abundance measurement \citep{2018Nissenrew,2019Jofre,2019Slumstrup}. All these external factors are partially responsible in the spread seen in the chemical-age trends. The situation described above stress the importance of having accurate and precise age and abundance estimations when studying chemical trends in the context of Galactic evolution.

\subsection{Exploring the best chemical traits}

Despite the aforementioned factors, through our analysis we have once again unveiled the age-related information conveyed by the s-process over $\alpha$ elements. Owing to their different nucleosynthesis processes and production timescales, the change with time is strongly correlated to their element ratios, as can be seen in Fig.~\ref{fig:el_ages}. We confirm that the most significant correlation is evident in stars
with solar-metallicity, gradually weakening in stars with lower [Fe/H].

When examining the metallicity dependence of the various combinations, we see that some n-capture elements such as Zr, Sr, Y display considerable differences in the slopes, as can be seen in Fig.~\ref{fig:delta}. Therefore, we believe that, even when these elements display steep slopes and high correlation coefficients with ages, their sensitivity to stellar metallicity implies their lack of universality. Nevertheless they can be used for specific metallicity ranges, or their measurements could potentially be improved by relying on a greater number of lines for their abundances. Other abundance ratios display low dependencies in metallicity but also small $\rho$ values, such as ratios involving Nd, thus  failing to encompass age-related information within their trends. The combinations [s/r]-processes element also exhibit low dependencies on [Fe/H] but looking at the values of the slopes in Table \ref{tab:coeff_alpha_n2} only the slope for [Zr/Eu] is greater than the threshold value, that is 0.03 dex $\mathrm{Gyr^-1}$. On the other hand, ratios such as [Ce/Al] and [Ce/Ti], despite having small dependence with metallicty, are related with age with slopes greater than 0.03 dex $\mathrm{Gyr^-1}$ showing higher correlation coefficients. On the whole, ratios comprising Ce at numerators have generally lower metallicity dependence, making them potentially reliable and consistent chemical clocks.

From these results it is not obvious to identify an ideal candidate for chemical-clocks. Although our analysis leans toward ratios exhibiting slopes steep enough to retrieve age information, but at the same time with small dependence on metallicity. Even so, the selection of one element combination over another relies on the precision attainable for each specific element. This depends on factors such as the spectral type of the star, the precision at which we can measure atmospheric parameters and abundance, the number of measurable lines. We note that in this analysis, we excluded iron-peak elements as we chose to prioritize examining the dependence on iron, hence not exploring all the possible combinations of chemical traits.

\section{Conclusion}

No universally optimal abundance ratios for chemically dating stars have emerged. This lack of universality spans across different metallicities \citep[as discovered here and in various studies such as ][]{2017MNRAS.465L.109F,2017Delgado} and position \citep{2019A&A...622A..59T,2020A&A...639A.127C}. We suggest that the selection criteria should be contingent upon the metallicity range of the stars. Within this context, our analysis offers a method to dentangle the dependence on metallicity by examining the efficacy of various element ratios in retrieving age information within specific metallicity ranges. Indeed, it has been demonstrated that certain ratios are applicable at solar metallicity, exhibiting more pronounced age-related patterns, but they lose their age correlations when applied to stars with lower iron content. Others can be regarded as "golden" ratios because they consistently hold true across all metallicities. As an example the combinations of Zr over $\alpha$ elements can be used for solar metallicity stars but proves less effective for most metal-poor stars, as their slopes in the [X/Y]-age trend  weaken. On the other hand, ratios with Ce and La appear to be easily usable at different ranges of metallicities, being more independent on the stellar iron content. Additionally, when selecting chemical candidates, factors such as measurement precision, the type of the star (which significantly impacts age determination), and the spatial coverage of the sample should be kept in mind.

Despite not having found a consistency of the various chemical clocks for the different metallicity groups, we recovered [n-capture/$\alpha$]-age trends with an internal abundance precision of < 0.03 dex and a mean relative age uncertainties around  20\%. 

Our exploration of different chemical ratios highlighted that the potential of chemical clocks extends beyond stellar dating. They can serve as valuable tools in identifying discrepancies in age estimation, especially in cases where stars might have previously been part of binary systems and subsequently accreted mass, as we speculate for two stars in our sample. This potential can be harnessed to detect such stars, which might otherwise remain hidden by a chemical analysis alone or within single evolutionary tracks that do not account for binary systems. 

Notwithstanding the non-universality of chemical traits, abundance ratios remain a source of key information in Galactic Archaeology. Thanks to the recent massive spectroscopic surveys and the revolutionary data from \textit{Gaia} DR3 they enable us to build  a chronological map of the Milky Way by linking chemistry and stellar ages. 

\begin{acknowledgements}
SV thanks ANID (Beca Doctorado Nacional) and Universidad Diego Portales for the financial support provided. SV and PJ acknowledges the Millennium Nucleus ERIS (ERIS NCN2021017) and FONDECYT for the funding. I express my gratitude to Theosamuele Signor for the valuable insights and the helpful discussions.

This work has made use of data from the European Space Agency (ESA) mission
{\it Gaia} (\url{https://www.cosmos.esa.int/gaia}), processed by the {\it Gaia}
Data Processing and Analysis Consortium (DPAC,
\url{https://www.cosmos.esa.int/web/gaia/dpac/consortium}). Funding for the DPAC
has been provided by national institutions, in particular the institutions
participating in the {\it Gaia} Multilateral Agreement.
\end{acknowledgements}

%
%
\bibliographystyle{aa}
\bibliography{bib} 

\begin{thebibliography}{140}
\expandafter\ifx\csname natexlab\endcsname\relax\def\natexlab#1{#1}\fi

\bibitem[{{Abdurro'uf} {et~al.}(2022){Abdurro'uf}, {Accetta}, {Aerts}, {Silva
  Aguirre}, {Ahumada}, {Ajgaonkar}, {Filiz Ak}, {Alam}, {Allende Prieto},
  {Almeida}, {Anders}, {Anderson}, {Andrews}, {Anguiano}, {Aquino-Ort{\'\i}z},
  {Arag{\'o}n-Salamanca}, {Argudo-Fern{\'a}ndez}, {Ata}, {Aubert},
  {Avila-Reese}, {Badenes}, {Barb{\'a}}, {Barger}, {Barrera-Ballesteros},
  {Beaton}, {Beers}, {Belfiore}, {Bender}, {Bernardi}, {Bershady}, {Beutler},
  {Bidin}, {Bird}, {Bizyaev}, {Blanc}, {Blanton}, {Boardman}, {Bolton},
  {Boquien}, {Borissova}, {Bovy}, {Brandt}, {Brown}, {Brownstein}, {Brusa},
  {Buchner}, {Bundy}, {Burchett}, {Bureau}, {Burgasser}, {Cabang}, {Campbell},
  {Cappellari}, {Carlberg}, {Wanderley}, {Carrera}, {Cash}, {Chen}, {Chen},
  {Cherinka}, {Chiappini}, {Choi}, {Chojnowski}, {Chung}, {Clerc}, {Cohen},
  {Comerford}, {Comparat}, {da Costa}, {Covey}, {Crane}, {Cruz-Gonzalez},
  {Culhane}, {Cunha}, {Dai}, {Damke}, {Darling}, {Davidson}, {Davies},
  {Dawson}, {De Lee}, {Diamond-Stanic}, {Cano-D{\'\i}az}, {S{\'a}nchez},
  {Donor}, {Duckworth}, {Dwelly}, {Eisenstein}, {Elsworth}, {Emsellem},
  {Eracleous}, {Escoffier}, {Fan}, {Farr}, {Feng}, {Fern{\'a}ndez-Trincado},
  {Feuillet}, {Filipp}, {Fillingham}, {Frinchaboy}, {Fromenteau}, {Galbany},
  {Garc{\'\i}a}, {Garc{\'\i}a-Hern{\'a}ndez}, {Ge}, {Geisler}, {Gelfand},
  {G{\'e}ron}, {Gibson}, {Goddy}, {Godoy-Rivera}, {Grabowski}, {Green},
  {Greener}, {Grier}, {Griffith}, {Guo}, {Guy}, {Hadjara}, {Harding},
  {Hasselquist}, {Hayes}, {Hearty}, {Hern{\'a}ndez}, {Hill}, {Hogg},
  {Holtzman}, {Horta}, {Hsieh}, {Hsu}, {Hsu}, {Huber}, {Huertas-Company},
  {Hutchinson}, {Hwang}, {Ibarra-Medel}, {Chitham}, {Ilha}, {Imig}, {Jaekle},
  {Jayasinghe}, {Ji}, {Johnson}, {Jones}, {J{\"o}nsson}, {Katkov}, {Khalatyan},
  {Kinemuchi}, {Kisku}, {Knapen}, {Kneib}, {Kollmeier}, {Kong}, {Kounkel},
  {Kreckel}, {Krishnarao}, {Lacerna}, {Lane}, {Langgin}, {Lavender}, {Law},
  {Lazarz}, {Leung}, {Leung}, {Lewis}, {Li}, {Li}, {Lian}, {Liang}, {Lin},
  {Lin}, {Lin}, {Lintott}, {Long}, {Longa-Pe{\~n}a}, {L{\'o}pez-Cob{\'a}},
  {Lu}, {Lundgren}, {Luo}, {Mackereth}, {de la Macorra}, {Mahadevan},
  {Majewski}, {Manchado}, {Mandeville}, {Maraston}, {Margalef-Bentabol},
  {Masseron}, {Masters}, {Mathur}, {McDermid}, {Mckay}, {Merloni},
  {Merrifield}, {Meszaros}, {Miglio}, {Di Mille}, {Minniti}, {Minsley},
  {Monachesi}, {Moon}, {Mosser}, {Mulchaey}, {Muna}, {Mu{\~n}oz}, {Myers},
  {Myers}, {Nadathur}, {Nair}, {Nandra}, {Neumann}, {Newman}, {Nidever},
  {Nikakhtar}, {Nitschelm}, {O'Connell}, {Garma-Oehmichen}, {Luan Souza de
  Oliveira}, {Olney}, {Oravetz}, {Ortigoza-Urdaneta}, {Osorio}, {Otter},
  {Pace}, {Padilla}, {Pan}, {Pan}, {Parikh}, {Parker}, {Peirani}, {Pe{\~n}a
  Ram{\'\i}rez}, {Penny}, {Percival}, {Perez-Fournon}, {Pinsonneault},
  {Poidevin}, {Poovelil}, {Price-Whelan}, {B{\'a}rbara de Andrade Queiroz},
  {Raddick}, {Ray}, {Rembold}, {Riddle}, {Riffel}, {Riffel}, {Rix}, {Robin},
  {Rodr{\'\i}guez-Puebla}, {Roman-Lopes}, {Rom{\'a}n-Z{\'u}{\~n}iga}, {Rose},
  {Ross}, {Rossi}, {Rubin}, {Salvato}, {S{\'a}nchez}, {S{\'a}nchez-Gallego},
  {Sanderson}, {Santana Rojas}, {Sarceno}, {Sarmiento}, {Sayres}, {Sazonova},
  {Schaefer}, {Schiavon}, {Schlegel}, {Schneider}, {Schultheis}, {Schwope},
  {Serenelli}, {Serna}, {Shao}, {Shapiro}, {Sharma}, {Shen}, {Shetrone}, {Shu},
  {Simon}, {Skrutskie}, {Smethurst}, {Smith}, {Sobeck}, {Spoo}, {Sprague},
  {Stark}, {Stassun}, {Steinmetz}, {Stello}, {Stone-Martinez},
  {Storchi-Bergmann}, {Stringfellow}, {Stutz}, {Su}, {Taghizadeh-Popp},
  {Talbot}, {Tayar}, {Telles}, {Teske}, {Thakar}, {Theissen}, {Tkachenko},
  {Thomas}, {Tojeiro}, {Hernandez Toledo}, {Troup}, {Trump}, {Trussler},
  {Turner}, {Tuttle}, {Unda-Sanzana}, {V{\'a}zquez-Mata}, {Valentini},
  {Valenzuela}, {Vargas-Gonz{\'a}lez}, {Vargas-Maga{\~n}a}, {Alfaro},
  {Villanova}, {Vincenzo}, {Wake}, {Warfield}, {Washington}, {Weaver},
  {Weijmans}, {Weinberg}, {Weiss}, {Westfall}, {Wild}, {Wilde}, {Wilson},
  {Wilson}, {Wilson}, {Wolf}, {Wood-Vasey}, {Yan}, {Zamora}, {Zasowski},
  {Zhang}, {Zhao}, {Zheng}, {Zheng}, \& {Zhu}}]{2022ApJS..259...35A}
{Abdurro'uf}, {Accetta}, K., {Aerts}, C., {et~al.} 2022, \apjs, 259, 35

\bibitem[{{Adibekyan} {et~al.}(2015){Adibekyan}, {Figueira}, {Santos}, {Sousa},
  {Faria}, {Delgado-Mena}, {Oshagh}, {Tsantaki}, {Hakobyan}, {Gonz{\'a}lez
  Hern{\'a}ndez}, {Su{\'a}rez-Andr{\'e}s}, \& {Israelian}}]{2015Adibekyan}
{Adibekyan}, V., {Figueira}, P., {Santos}, N.~C., {et~al.} 2015, \aap, 583, A94

\bibitem[{{Aguirre B{\o}rsen-Koch} {et~al.}(2022){Aguirre B{\o}rsen-Koch},
  {R{\o}rsted}, {Justesen}, {Stokholm}, {Verma}, {Winther}, {Knudstrup},
  {Nielsen}, {Sahlholdt}, {Larsen}, {Cassisi}, {Serenelli}, {Casagrande},
  {Christensen-Dalsgaard}, {Davies}, {Ferguson}, {Lund}, {Weiss}, \&
  {White}}]{2022MNRAS.509.4344A}
{Aguirre B{\o}rsen-Koch}, V., {R{\o}rsted}, J.~L., {Justesen}, A.~B., {et~al.}
  2022, \mnras, 509, 4344

\bibitem[{{Alexeeva} {et~al.}(2023{\natexlab{a}}){Alexeeva}, {Wang}, {Zhao},
  {Wang}, {Wu}, {Wang}, {Yan}, \& {Shi}}]{2023arXiv230901402A}
{Alexeeva}, S., {Wang}, Y., {Zhao}, G., {et~al.} 2023{\natexlab{a}}, arXiv
  e-prints, arXiv:2309.01402

\bibitem[{{Alexeeva} {et~al.}(2023{\natexlab{b}}){Alexeeva}, {Wang}, {Zhao},
  {Wang}, {Wu}, {Wang}, {Yan}, \& {Shi}}]{2023Alexeeva}
{Alexeeva}, S., {Wang}, Y., {Zhao}, G., {et~al.} 2023{\natexlab{b}}, \apj, 957,
  10

\bibitem[{{Alvarez} \& {Plez}(1998)}]{1998A&A...330.1109A}
{Alvarez}, R. \& {Plez}, B. 1998, \aap, 330, 1109

\bibitem[{{Amarsi} {et~al.}(2020{\natexlab{a}}){Amarsi}, {Lind}, {Osorio},
  {Nordlander}, {Bergemann}, {Reggiani}, {Wang}, {Buder}, {Asplund}, {Barklem},
  {Wehrhahn}, {Sk{\'u}lad{\'o}ttir}, {Kobayashi}, {Karakas}, {Gao},
  {Bland-Hawthorn}, {de Silva}, {Kos}, {Lewis}, {Martell}, {Sharma}, {Simpson},
  {Zucker}, {{\v{C}}otar}, {Horner}, \& {GALAH
  Collaboration}}]{2020A&A...642A..62A}
{Amarsi}, A.~M., {Lind}, K., {Osorio}, Y., {et~al.} 2020{\natexlab{a}}, \aap,
  642, A62

\bibitem[{{Amarsi} {et~al.}(2020{\natexlab{b}}){Amarsi}, {Lind}, {Osorio},
  {Nordlander}, {Bergemann}, {Reggiani}, {Wang}, {Buder}, {Asplund}, {Barklem},
  {Wehrhahn}, {Sk{\'u}lad{\'o}ttir}, {Kobayashi}, {Karakas}, {Gao},
  {Bland-Hawthorn}, {de Silva}, {Kos}, {Lewis}, {Martell}, {Sharma}, {Simpson},
  {Zucker}, {{\v{C}}otar}, {Horner}, \& {GALAH Collaboration}}]{2020Amarsi}
{Amarsi}, A.~M., {Lind}, K., {Osorio}, Y., {et~al.} 2020{\natexlab{b}}, \aap,
  642, A62

\bibitem[{{Anders} {et~al.}(2017){Anders}, {Chiappini}, {Rodrigues}, {Miglio},
  {Montalb{\'a}n}, {Mosser}, {Girardi}, {Valentini}, {Noels}, {Morel},
  {Johnson}, {Schultheis}, {Baudin}, {de Assis Peralta}, {Hekker},
  {Theme{\ss}l}, {Kallinger}, {Garc{\'\i}a}, {Mathur}, {Baglin}, {Santiago},
  {Martig}, {Minchev}, {Steinmetz}, {da Costa}, {Maia}, {Allende Prieto},
  {Cunha}, {Beers}, {Epstein}, {Garc{\'\i}a P{\'e}rez},
  {Garc{\'\i}a-Hern{\'a}ndez}, {Harding}, {Holtzman}, {Majewski},
  {M{\'e}sz{\'a}ros}, {Nidever}, {Pan}, {Pinsonneault}, {Schiavon},
  {Schneider}, {Shetrone}, {Stassun}, {Zamora}, \&
  {Zasowski}}]{2017A&A...597A..30A}
{Anders}, F., {Chiappini}, C., {Rodrigues}, T.~S., {et~al.} 2017, \aap, 597,
  A30

\bibitem[{{Anders} {et~al.}(2023){Anders}, {Gispert}, {Ratcliffe}, {Chiappini},
  {Minchev}, {Nepal}, {Queiroz}, {Amarante}, {Antoja}, {Casali}, {Casamiquela},
  {Khalatyan}, {Miglio}, {Perottoni}, \& {Schultheis}}]{2023arXiv230408276A}
{Anders}, F., {Gispert}, P., {Ratcliffe}, B., {et~al.} 2023, arXiv e-prints,
  arXiv:2304.08276

\bibitem[{{Andrievsky} {et~al.}(2007){Andrievsky}, {Spite}, {Korotin}, {Spite},
  {Bonifacio}, {Cayrel}, {Hill}, \& {Fran{\c{c}}ois}}]{2007Andrievsky}
{Andrievsky}, S.~M., {Spite}, M., {Korotin}, S.~A., {et~al.} 2007, \aap, 464,
  1081

\bibitem[{{Arentoft} {et~al.}(2019){Arentoft}, {Grundahl}, {White},
  {Slumstrup}, {Handberg}, {Lund}, {Brogaard}, {Andersen}, {Silva Aguirre},
  {Zhang}, {Chen}, {Yan}, {Pope}, {Huber}, {Kjeldsen}, {Christensen-Dalsgaard},
  {Jessen-Hansen}, {Antoci}, {Frandsen}, {Bedding}, {Pall{\'e}}, {Garcia},
  {Deng}, {Hon}, {Stello}, \& {J{\o}rgensen}}]{2019A&A...622A.190A}
{Arentoft}, T., {Grundahl}, F., {White}, T.~R., {et~al.} 2019, \aap, 622, A190

\bibitem[{{Baratella} {et~al.}(2021){Baratella}, {D'Orazi}, {Sheminova},
  {Spina}, {Carraro}, {Gratton}, {Magrini}, {Randich}, {Lugaro}, {Pignatari},
  {Romano}, {Biazzo}, {Bragaglia}, {Casali}, {Desidera}, {Frasca}, {de Silva},
  {Melo}, {Van der Swaelmen}, {Tautvai{\v{s}}ien{\.{e}}},
  {Jim{\'e}nez-Esteban}, {Gilmore}, {Bensby}, {Smiljanic}, {Bayo},
  {Franciosini}, {Gonneau}, {Hourihane}, {Jofr{\'e}}, {Monaco}, {Morbidelli},
  {Sacco}, {Sbordone}, {Worley}, \& {Zaggia}}]{2021A&A...653A..67B}
{Baratella}, M., {D'Orazi}, V., {Sheminova}, V., {et~al.} 2021, \aap, 653, A67

\bibitem[{{Bergemann} {et~al.}(2011){Bergemann}, {Lind}, {Collet}, \&
  {Asplund}}]{2011Bergemann}
{Bergemann}, M., {Lind}, K., {Collet}, R., \& {Asplund}, M. 2011, in Journal of
  Physics Conference Series, Vol. 328, Journal of Physics Conference Series,
  012002

\bibitem[{{Berger} {et~al.}(2022){Berger}, {van Saders}, {Huber}, {Gaidos},
  {Schlieder}, \& {Claytor}}]{2022ApJ...936..100B}
{Berger}, T.~A., {van Saders}, J.~L., {Huber}, D., {et~al.} 2022, \apj, 936,
  100

\bibitem[{{Bisterzo} {et~al.}(2014){Bisterzo}, {Travaglio}, {Gallino},
  {Wiescher}, \& {K{\"a}ppeler}}]{2014Bisterzo}
{Bisterzo}, S., {Travaglio}, C., {Gallino}, R., {Wiescher}, M., \&
  {K{\"a}ppeler}, F. 2014, \apj, 787, 10

\bibitem[{{Blanco-Cuaresma}(2019)}]{2019MNRAS.486.2075B}
{Blanco-Cuaresma}, S. 2019, \mnras, 486, 2075

\bibitem[{{Blanco-Cuaresma} {et~al.}(2014{\natexlab{a}}){Blanco-Cuaresma},
  {Soubiran}, {Heiter}, \& {Jofr{\'e}}}]{2014A&A...569A.111B}
{Blanco-Cuaresma}, S., {Soubiran}, C., {Heiter}, U., \& {Jofr{\'e}}, P.
  2014{\natexlab{a}}, \aap, 569, A111

\bibitem[{{Blanco-Cuaresma} {et~al.}(2014{\natexlab{b}}){Blanco-Cuaresma},
  {Soubiran}, {Jofr{\'e}}, \& {Heiter}}]{2014A&A...566A..98B}
{Blanco-Cuaresma}, S., {Soubiran}, C., {Jofr{\'e}}, P., \& {Heiter}, U.
  2014{\natexlab{b}}, \aap, 566, A98

\bibitem[{{Bland-Hawthorn} {et~al.}(2019){Bland-Hawthorn}, {Sharma},
  {Tepper-Garcia}, {Binney}, {Freeman}, {Hayden}, {Kos}, {De Silva}, {Ellis},
  {Lewis}, {Asplund}, {Buder}, {Casey}, {D'Orazi}, {Duong}, {Khanna}, {Lin},
  {Lind}, {Martell}, {Ness}, {Simpson}, {Zucker}, {Zwitter}, {Kafle},
  {Quillen}, {Ting}, \& {Wyse}}]{2019Bland-Hawthorn}
{Bland-Hawthorn}, J., {Sharma}, S., {Tepper-Garcia}, T., {et~al.} 2019, \mnras,
  486, 1167

\bibitem[{{Bovy} {et~al.}(2012){Bovy}, {Rix}, {Liu}, {Hogg}, {Beers}, \&
  {Lee}}]{2012Bovy}
{Bovy}, J., {Rix}, H.-W., {Liu}, C., {et~al.} 2012, \apj, 753, 148

\bibitem[{{Brogaard} {et~al.}(2016){Brogaard}, {Jessen-Hansen}, {Handberg},
  {Arentoft}, {Frandsen}, {Grundahl}, {Bruntt}, {Sandquist}, {Miglio}, {Beck},
  {Thygesen}, {Kj{\ae}rgaard}, \& {Haugaard}}]{2016Brogaard}
{Brogaard}, K., {Jessen-Hansen}, J., {Handberg}, R., {et~al.} 2016,
  Astronomische Nachrichten, 337, 793

\bibitem[{{Brown} {et~al.}(1991){Brown}, {Gilliland}, {Noyes}, \&
  {Ramsey}}]{1991ApJ...368..599B}
{Brown}, T.~M., {Gilliland}, R.~L., {Noyes}, R.~W., \& {Ramsey}, L.~W. 1991,
  \apj, 368, 599

\bibitem[{{Buder} {et~al.}(2021){Buder}, {Sharma}, {Kos}, {Amarsi},
  {Nordlander}, {Lind}, {Martell}, {Asplund}, {Bland-Hawthorn}, {Casey}, {de
  Silva}, {D'Orazi}, {Freeman}, {Hayden}, {Lewis}, {Lin}, {Schlesinger},
  {Simpson}, {Stello}, {Zucker}, {Zwitter}, {Beeson}, {Buck}, {Casagrande},
  {Clark}, {{\v{C}}otar}, {da Costa}, {de Grijs}, {Feuillet}, {Horner},
  {Kafle}, {Khanna}, {Kobayashi}, {Liu}, {Montet}, {Nandakumar}, {Nataf},
  {Ness}, {Spina}, {Tepper-Garc{\'\i}a}, {Ting}, {Traven},
  {Vogrin{\v{c}}i{\v{c}}}, {Wittenmyer}, {Wyse}, {{\v{Z}}erjal}, \& {Galah
  Collaboration}}]{2021MNRAS.506..150B}
{Buder}, S., {Sharma}, S., {Kos}, J., {et~al.} 2021, \mnras, 506, 150

\bibitem[{{Casali} {et~al.}(2020){Casali}, {Spina}, {Magrini}, {Karakas},
  {Kobayashi}, {Casey}, {Feltzing}, {Van der Swaelmen}, {Tsantaki},
  {Jofr{\'e}}, {Bragaglia}, {Feuillet}, {Bensby}, {Biazzo}, {Gonneau},
  {Tautvai{\v{s}}ien{\.{e}}}, {Baratella}, {Roccatagliata}, {Pancino}, {Sousa},
  {Adibekyan}, {Martell}, {Bayo}, {Jackson}, {Jeffries}, {Gilmore}, {Randich},
  {Alfaro}, {Koposov}, {Korn}, {Recio-Blanco}, {Smiljanic}, {Franciosini},
  {Hourihane}, {Monaco}, {Morbidelli}, {Sacco}, {Worley}, \&
  {Zaggia}}]{2020A&A...639A.127C}
{Casali}, G., {Spina}, L., {Magrini}, L., {et~al.} 2020, \aap, 639, A127

\bibitem[{{Casamiquela} {et~al.}(2021){Casamiquela}, {Soubiran}, {Jofr{\'e}},
  {Chiappini}, {Lagarde}, {Tarricq}, {Carrera}, {Jordi},
  {Balaguer-N{\'u}{\~n}ez}, {Carbajo-Hijarrubia}, \&
  {Blanco-Cuaresma}}]{2021A&A...652A..25C}
{Casamiquela}, L., {Soubiran}, C., {Jofr{\'e}}, P., {et~al.} 2021, \aap, 652,
  A25

\bibitem[{{Casamiquela} {et~al.}(2020){Casamiquela}, {Tarricq}, {Soubiran},
  {Blanco-Cuaresma}, {Jofr{\'e}}, {Heiter}, \& {Tucci
  Maia}}]{2020A&A...635A...8C}
{Casamiquela}, L., {Tarricq}, Y., {Soubiran}, C., {et~al.} 2020, \aap, 635, A8

\bibitem[{{Cescutti} \& {Matteucci}(2022)}]{2022Univ....8..173C}
{Cescutti}, G. \& {Matteucci}, F. 2022, Universe, 8, 173

\bibitem[{{Chaplin} {et~al.}(2014){Chaplin}, {Basu}, {Huber}, {Serenelli},
  {Casagrande}, {Silva Aguirre}, {Ball}, {Creevey}, {Gizon}, {Handberg},
  {Karoff}, {Lutz}, {Marques}, {Miglio}, {Stello}, {Suran}, {Pricopi},
  {Metcalfe}, {Monteiro}, {Molenda-{\.Z}akowicz}, {Appourchaux},
  {Christensen-Dalsgaard}, {Elsworth}, {Garc{\'\i}a}, {Houdek}, {Kjeldsen},
  {Bonanno}, {Campante}, {Corsaro}, {Gaulme}, {Hekker}, {Mathur}, {Mosser},
  {R{\'e}gulo}, \& {Salabert}}]{2014ApJS..210....1C}
{Chaplin}, W.~J., {Basu}, S., {Huber}, D., {et~al.} 2014, \apjs, 210, 1

\bibitem[{{Chaplin} \& {Miglio}(2013)}]{2013ARA&A..51..353C}
{Chaplin}, W.~J. \& {Miglio}, A. 2013, \araa, 51, 353

\bibitem[{{Chiappini} {et~al.}(1997){Chiappini}, {Matteucci}, \&
  {Gratton}}]{1997ApJ...477..765C}
{Chiappini}, C., {Matteucci}, F., \& {Gratton}, R. 1997, \apj, 477, 765

\bibitem[{{C{\^o}t{\'e}} {et~al.}(2018){C{\^o}t{\'e}}, {Fryer}, {Belczynski},
  {Korobkin}, {Chru{\'s}li{\'n}ska}, {Vassh}, {Mumpower}, {Lippuner},
  {Sprouse}, {Surman}, \& {Wollaeger}}]{2018Cote}
{C{\^o}t{\'e}}, B., {Fryer}, C.~L., {Belczynski}, K., {et~al.} 2018, \apj, 855,
  99

\bibitem[{{Cristallo} {et~al.}(2015){Cristallo}, {Straniero}, {Piersanti}, \&
  {Gobrecht}}]{2015Cristallo}
{Cristallo}, S., {Straniero}, O., {Piersanti}, L., \& {Gobrecht}, D. 2015,
  \apjs, 219, 40

\bibitem[{{Cutri} {et~al.}(2003){Cutri}, {Skrutskie}, {van Dyk}, {Beichman},
  {Carpenter}, {Chester}, {Cambresy}, {Evans}, {Fowler}, {Gizis}, {Howard},
  {Huchra}, {Jarrett}, {Kopan}, {Kirkpatrick}, {Light}, {Marsh}, {McCallon},
  {Schneider}, {Stiening}, {Sykes}, {Weinberg}, {Wheaton}, {Wheelock}, \&
  {Zacarias}}]{2003yCat.2246....0C}
{Cutri}, R.~M., {Skrutskie}, M.~F., {van Dyk}, S., {et~al.} 2003, VizieR Online
  Data Catalog, II/246

\bibitem[{{da Silva} {et~al.}(2012){da Silva}, {Porto de Mello}, {Milone}, {da
  Silva}, {Ribeiro}, \& {Rocha-Pinto}}]{2012A&A...542A..84D}
{da Silva}, R., {Porto de Mello}, G.~F., {Milone}, A.~C., {et~al.} 2012, \aap,
  542, A84

\bibitem[{{Das} {et~al.}(2020){Das}, {Hawkins}, \&
  {Jofr{\'e}}}]{2020MNRAS.493.5195D}
{Das}, P., {Hawkins}, K., \& {Jofr{\'e}}, P. 2020, \mnras, 493, 5195

\bibitem[{{De Silva} {et~al.}(2015){De Silva}, {Freeman}, {Bland-Hawthorn},
  {Martell}, {de Boer}, {Asplund}, {Keller}, {Sharma}, {Zucker}, {Zwitter},
  {Anguiano}, {Bacigalupo}, {Bayliss}, {Beavis}, {Bergemann}, {Campbell},
  {Cannon}, {Carollo}, {Casagrande}, {Casey}, {Da Costa}, {D'Orazi}, {Dotter},
  {Duong}, {Heger}, {Ireland}, {Kafle}, {Kos}, {Lattanzio}, {Lewis}, {Lin},
  {Lind}, {Munari}, {Nataf}, {O'Toole}, {Parker}, {Reid}, {Schlesinger},
  {Sheinis}, {Simpson}, {Stello}, {Ting}, {Traven}, {Watson}, {Wittenmyer},
  {Yong}, \& {{\v{Z}}erjal}}]{2015MNRAS.449.2604D}
{De Silva}, G.~M., {Freeman}, K.~C., {Bland-Hawthorn}, J., {et~al.} 2015,
  \mnras, 449, 2604

\bibitem[{{Delgado Mena} {et~al.}(2019){Delgado Mena}, {Moya}, {Adibekyan},
  {Tsantaki}, {Gonz{\'a}lez Hern{\'a}ndez}, {Israelian}, {Davies}, {Chaplin},
  {Sousa}, {Ferreira}, \& {Santos}}]{2019A&A...624A..78D}
{Delgado Mena}, E., {Moya}, A., {Adibekyan}, V., {et~al.} 2019, \aap, 624, A78

\bibitem[{{Delgado Mena} {et~al.}(2017){Delgado Mena}, {Tsantaki}, {Adibekyan},
  {Sousa}, {Santos}, {Gonz{\'a}lez Hern{\'a}ndez}, \&
  {Israelian}}]{2017Delgado}
{Delgado Mena}, E., {Tsantaki}, M., {Adibekyan}, V.~Z., {et~al.} 2017, \aap,
  606, A94

\bibitem[{{Deng} {et~al.}(2012){Deng}, {Newberg}, {Liu}, {Carlin}, {Beers},
  {Chen}, {Chen}, {Christlieb}, {Grillmair}, {Guhathakurta}, {Han}, {Hou},
  {Lee}, {L{\'e}pine}, {Li}, {Liu}, {Pan}, {Sellwood}, {Wang}, {Wang}, {Yang},
  {Yanny}, {Zhang}, {Zhang}, {Zheng}, \& {Zhu}}]{2012RAA....12..735D}
{Deng}, L.-C., {Newberg}, H.~J., {Liu}, C., {et~al.} 2012, Research in
  Astronomy and Astrophysics, 12, 735

\bibitem[{{Feltzing} {et~al.}(2017){Feltzing}, {Howes}, {McMillan}, \&
  {Stonkut{\.{e}}}}]{2017MNRAS.465L.109F}
{Feltzing}, S., {Howes}, L.~M., {McMillan}, P.~J., \& {Stonkut{\.{e}}}, E.
  2017, \mnras, 465, L109

\bibitem[{{Feuillet} {et~al.}(2018){Feuillet}, {Bovy}, {Holtzman}, {Weinberg},
  {Garc{\'\i}a-Hern{\'a}ndez}, {Hearty}, {Majewski}, {Roman-Lopes}, {Rybizki},
  \& {Zamora}}]{2018Feuillet}
{Feuillet}, D.~K., {Bovy}, J., {Holtzman}, J., {et~al.} 2018, \mnras, 477, 2326

\bibitem[{{Feuillet} {et~al.}(2019){Feuillet}, {Frankel}, {Lind}, {Frinchaboy},
  {Garc{\'\i}a-Hern{\'a}ndez}, {Lane}, {Nitschelm}, \&
  {Roman-Lopes}}]{2019MNRAS.489.1742F}
{Feuillet}, D.~K., {Frankel}, N., {Lind}, K., {et~al.} 2019, \mnras, 489, 1742

\bibitem[{Fischler \& Bolles(1981)}]{fischler1981random}
Fischler, M.~A. \& Bolles, R.~C. 1981, Communications of the ACM, 24, 381

\bibitem[{{Frankel} {et~al.}(2020){Frankel}, {Sanders}, {Ting}, \&
  {Rix}}]{2020ApJ...896...15F}
{Frankel}, N., {Sanders}, J., {Ting}, Y.-S., \& {Rix}, H.-W. 2020, \apj, 896,
  15

\bibitem[{{Freeman} \&
  {Bland-Hawthorn}(2002{\natexlab{a}})}]{2002ARA&A..40..487F}
{Freeman}, K. \& {Bland-Hawthorn}, J. 2002{\natexlab{a}}, \araa, 40, 487

\bibitem[{{Freeman} \& {Bland-Hawthorn}(2002{\natexlab{b}})}]{2002Freeman}
{Freeman}, K. \& {Bland-Hawthorn}, J. 2002{\natexlab{b}}, \araa, 40, 487

\bibitem[{{Fujimoto} {et~al.}(2020){Fujimoto}, {Krumholz}, \&
  {Inutsuka}}]{2020Fujimoto}
{Fujimoto}, Y., {Krumholz}, M.~R., \& {Inutsuka}, S.-i. 2020, \mnras, 497, 2442

\bibitem[{{Gaia Collaboration} {et~al.}(2016){Gaia Collaboration}, {Prusti},
  {de Bruijne}, {Brown}, {Vallenari}, {Babusiaux}, {Bailer-Jones}, {Bastian},
  {Biermann}, {Evans}, {Eyer}, {Jansen}, {Jordi}, {Klioner}, {Lammers},
  {Lindegren}, {Luri}, {Mignard}, {Milligan}, {Panem}, {Poinsignon},
  {Pourbaix}, {Randich}, {Sarri}, {Sartoretti}, {Siddiqui}, {Soubiran},
  {Valette}, {van Leeuwen}, {Walton}, {Aerts}, {Arenou}, {Cropper}, {Drimmel},
  {H{\o}g}, {Katz}, {Lattanzi}, {O'Mullane}, {Grebel}, {Holland}, {Huc},
  {Passot}, {Bramante}, {Cacciari}, {Casta{\~n}eda}, {Chaoul}, {Cheek}, {De
  Angeli}, {Fabricius}, {Guerra}, {Hern{\'a}ndez}, {Jean-Antoine-Piccolo},
  {Masana}, {Messineo}, {Mowlavi}, {Nienartowicz}, {Ord{\'o}{\~n}ez-Blanco},
  {Panuzzo}, {Portell}, {Richards}, {Riello}, {Seabroke}, {Tanga},
  {Th{\'e}venin}, {Torra}, {Els}, {Gracia-Abril}, {Comoretto},
  {Garcia-Reinaldos}, {Lock}, {Mercier}, {Altmann}, {Andrae}, {Astraatmadja},
  {Bellas-Velidis}, {Benson}, {Berthier}, {Blomme}, {Busso}, {Carry},
  {Cellino}, {Clementini}, {Cowell}, {Creevey}, {Cuypers}, {Davidson}, {De
  Ridder}, {de Torres}, {Delchambre}, {Dell'Oro}, {Ducourant}, {Fr{\'e}mat},
  {Garc{\'\i}a-Torres}, {Gosset}, {Halbwachs}, {Hambly}, {Harrison}, {Hauser},
  {Hestroffer}, {Hodgkin}, {Huckle}, {Hutton}, {Jasniewicz}, {Jordan},
  {Kontizas}, {Korn}, {Lanzafame}, {Manteiga}, {Moitinho}, {Muinonen},
  {Osinde}, {Pancino}, {Pauwels}, {Petit}, {Recio-Blanco}, {Robin}, {Sarro},
  {Siopis}, {Smith}, {Smith}, {Sozzetti}, {Thuillot}, {van Reeven}, {Viala},
  {Abbas}, {Abreu Aramburu}, {Accart}, {Aguado}, {Allan}, {Allasia},
  {Altavilla}, {{\'A}lvarez}, {Alves}, {Anderson}, {Andrei}, {Anglada Varela},
  {Antiche}, {Antoja}, {Ant{\'o}n}, {Arcay}, {Atzei}, {Ayache}, {Bach},
  {Baker}, {Balaguer-N{\'u}{\~n}ez}, {Barache}, {Barata}, {Barbier}, {Barblan},
  {Baroni}, {Barrado y Navascu{\'e}s}, {Barros}, {Barstow}, {Becciani},
  {Bellazzini}, {Bellei}, {Bello Garc{\'\i}a}, {Belokurov}, {Bendjoya},
  {Berihuete}, {Bianchi}, {Bienaym{\'e}}, {Billebaud}, {Blagorodnova},
  {Blanco-Cuaresma}, {Boch}, {Bombrun}, {Borrachero}, {Bouquillon}, {Bourda},
  {Bouy}, {Bragaglia}, {Breddels}, {Brouillet}, {Br{\"u}semeister},
  {Bucciarelli}, {Budnik}, {Burgess}, {Burgon}, {Burlacu}, {Busonero}, {Buzzi},
  {Caffau}, {Cambras}, {Campbell}, {Cancelliere}, {Cantat-Gaudin}, {Carlucci},
  {Carrasco}, {Castellani}, {Charlot}, {Charnas}, {Charvet}, {Chassat},
  {Chiavassa}, {Clotet}, {Cocozza}, {Collins}, {Collins}, {Costigan}, {Crifo},
  {Cross}, {Crosta}, {Crowley}, {Dafonte}, {Damerdji}, {Dapergolas}, {David},
  {David}, {De Cat}, {de Felice}, {de Laverny}, {De Luise}, {De March}, {de
  Martino}, {de Souza}, {Debosscher}, {del Pozo}, {Delbo}, {Delgado},
  {Delgado}, {di Marco}, {Di Matteo}, {Diakite}, {Distefano}, {Dolding}, {Dos
  Anjos}, {Drazinos}, {Dur{\'a}n}, {Dzigan}, {Ecale}, {Edvardsson}, {Enke},
  {Erdmann}, {Escolar}, {Espina}, {Evans}, {Eynard Bontemps}, {Fabre},
  {Fabrizio}, {Faigler}, {Falc{\~a}o}, {Farr{\`a}s Casas}, {Faye}, {Federici},
  {Fedorets}, {Fern{\'a}ndez-Hern{\'a}ndez}, {Fernique}, {Fienga}, {Figueras},
  {Filippi}, {Findeisen}, {Fonti}, {Fouesneau}, {Fraile}, {Fraser}, {Fuchs},
  {Furnell}, {Gai}, {Galleti}, {Galluccio}, {Garabato}, {Garc{\'\i}a-Sedano},
  {Gar{\'e}}, {Garofalo}, {Garralda}, {Gavras}, {Gerssen}, {Geyer}, {Gilmore},
  {Girona}, {Giuffrida}, {Gomes}, {Gonz{\'a}lez-Marcos},
  {Gonz{\'a}lez-N{\'u}{\~n}ez}, {Gonz{\'a}lez-Vidal}, {Granvik}, {Guerrier},
  {Guillout}, {Guiraud}, {G{\'u}rpide}, {Guti{\'e}rrez-S{\'a}nchez}, {Guy},
  {Haigron}, {Hatzidimitriou}, {Haywood}, {Heiter}, {Helmi}, {Hobbs},
  {Hofmann}, {Holl}, {Holland}, {Hunt}, {Hypki}, {Icardi}, {Irwin}, {Jevardat
  de Fombelle}, {Jofr{\'e}}, {Jonker}, {Jorissen}, {Julbe}, {Karampelas},
  {Kochoska}, {Kohley}, {Kolenberg}, {Kontizas}, {Koposov}, {Kordopatis},
  {Koubsky}, {Kowalczyk}, {Krone-Martins}, {Kudryashova}, {Kull}, {Bachchan},
  {Lacoste-Seris}, {Lanza}, {Lavigne}, {Le Poncin-Lafitte}, {Lebreton},
  {Lebzelter}, {Leccia}, {Leclerc}, {Lecoeur-Taibi}, {Lemaitre}, {Lenhardt},
  {Leroux}, {Liao}, {Licata}, {Lindstr{\o}m}, {Lister}, {Livanou}, {Lobel},
  {L{\"o}ffler}, {L{\'o}pez}, {Lopez-Lozano}, {Lorenz}, {Loureiro},
  {MacDonald}, {Magalh{\~a}es Fernandes}, {Managau}, {Mann}, {Mantelet},
  {Marchal}, {Marchant}, {Marconi}, {Marie}, {Marinoni}, {Marrese},
  {Marschalk{\'o}}, {Marshall}, {Mart{\'\i}n-Fleitas}, {Martino}, {Mary},
  {Matijevi{\v{c}}}, {Mazeh}, {McMillan}, {Messina}, {Mestre}, {Michalik},
  {Millar}, {Miranda}, {Molina}, {Molinaro}, {Molinaro}, {Moln{\'a}r},
  {Moniez}, {Montegriffo}, {Monteiro}, {Mor}, {Mora}, {Morbidelli}, {Morel},
  {Morgenthaler}, {Morley}, {Morris}, {Mulone}, {Muraveva}, {Musella},
  {Narbonne}, {Nelemans}, {Nicastro}, {Noval}, {Ord{\'e}novic},
  {Ordieres-Mer{\'e}}, {Osborne}, {Pagani}, {Pagano}, {Pailler}, {Palacin},
  {Palaversa}, {Parsons}, {Paulsen}, {Pecoraro}, {Pedrosa}, {Pentik{\"a}inen},
  {Pereira}, {Pichon}, {Piersimoni}, {Pineau}, {Plachy}, {Plum}, {Poujoulet},
  {Pr{\v{s}}a}, {Pulone}, {Ragaini}, {Rago}, {Rambaux}, {Ramos-Lerate},
  {Ranalli}, {Rauw}, {Read}, {Regibo}, {Renk}, {Reyl{\'e}}, {Ribeiro},
  {Rimoldini}, {Ripepi}, {Riva}, {Rixon}, {Roelens}, {Romero-G{\'o}mez},
  {Rowell}, {Royer}, {Rudolph}, {Ruiz-Dern}, {Sadowski}, {Sagrist{\`a}
  Sell{\'e}s}, {Sahlmann}, {Salgado}, {Salguero}, {Sarasso}, {Savietto},
  {Schnorhk}, {Schultheis}, {Sciacca}, {Segol}, {Segovia}, {Segransan},
  {Serpell}, {Shih}, {Smareglia}, {Smart}, {Smith}, {Solano}, {Solitro},
  {Sordo}, {Soria Nieto}, {Souchay}, {Spagna}, {Spoto}, {Stampa}, {Steele},
  {Steidelm{\"u}ller}, {Stephenson}, {Stoev}, {Suess}, {S{\"u}veges}, {Surdej},
  {Szabados}, {Szegedi-Elek}, {Tapiador}, {Taris}, {Tauran}, {Taylor},
  {Teixeira}, {Terrett}, {Tingley}, {Trager}, {Turon}, {Ulla}, {Utrilla},
  {Valentini}, {van Elteren}, {Van Hemelryck}, {van Leeuwen}, {Varadi},
  {Vecchiato}, {Veljanoski}, {Via}, {Vicente}, {Vogt}, {Voss}, {Votruba},
  {Voutsinas}, {Walmsley}, {Weiler}, {Weingrill}, {Werner}, {Wevers},
  {Whitehead}, {Wyrzykowski}, {Yoldas}, {{\v{Z}}erjal}, {Zucker}, {Zurbach},
  {Zwitter}, {Alecu}, {Allen}, {Allende Prieto}, {Amorim},
  {Anglada-Escud{\'e}}, {Arsenijevic}, {Azaz}, {Balm}, {Beck}, {Bernstein},
  {Bigot}, {Bijaoui}, {Blasco}, {Bonfigli}, {Bono}, {Boudreault}, {Bressan},
  {Brown}, {Brunet}, {Bunclark}, {Buonanno}, {Butkevich}, {Carret}, {Carrion},
  {Chemin}, {Ch{\'e}reau}, {Corcione}, {Darmigny}, {de Boer}, {de Teodoro}, {de
  Zeeuw}, {Delle Luche}, {Domingues}, {Dubath}, {Fodor}, {Fr{\'e}zouls},
  {Fries}, {Fustes}, {Fyfe}, {Gallardo}, {Gallegos}, {Gardiol}, {Gebran},
  {Gomboc}, {G{\'o}mez}, {Grux}, {Gueguen}, {Heyrovsky}, {Hoar}, {Iannicola},
  {Isasi Parache}, {Janotto}, {Joliet}, {Jonckheere}, {Keil}, {Kim},
  {Klagyivik}, {Klar}, {Knude}, {Kochukhov}, {Kolka}, {Kos}, {Kutka}, {Lainey},
  {LeBouquin}, {Liu}, {Loreggia}, {Makarov}, {Marseille}, {Martayan},
  {Martinez-Rubi}, {Massart}, {Meynadier}, {Mignot}, {Munari}, {Nguyen},
  {Nordlander}, {Ocvirk}, {O'Flaherty}, {Olias Sanz}, {Ortiz}, {Osorio},
  {Oszkiewicz}, {Ouzounis}, {Palmer}, {Park}, {Pasquato}, {Peltzer}, {Peralta},
  {P{\'e}turaud}, {Pieniluoma}, {Pigozzi}, {Poels}, {Prat}, {Prod'homme},
  {Raison}, {Rebordao}, {Risquez}, {Rocca-Volmerange}, {Rosen}, {Ruiz-Fuertes},
  {Russo}, {Sembay}, {Serraller Vizcaino}, {Short}, {Siebert}, {Silva},
  {Sinachopoulos}, {Slezak}, {Soffel}, {Sosnowska}, {Strai{\v{z}}ys}, {ter
  Linden}, {Terrell}, {Theil}, {Tiede}, {Troisi}, {Tsalmantza}, {Tur},
  {Vaccari}, {Vachier}, {Valles}, {Van Hamme}, {Veltz}, {Virtanen}, {Wallut},
  {Wichmann}, {Wilkinson}, {Ziaeepour}, \& {Zschocke}}]{2016A&A...595A...1G}
{Gaia Collaboration}, {Prusti}, T., {de Bruijne}, J.~H.~J., {et~al.} 2016,
  \aap, 595, A1

\bibitem[{{Gaia Collaboration} {et~al.}(2023){Gaia Collaboration}, {Vallenari},
  {Brown}, {Prusti}, {de Bruijne}, {Arenou}, {Babusiaux}, {Biermann},
  {Creevey}, {Ducourant}, {Evans}, {Eyer}, {Guerra}, {Hutton}, {Jordi},
  {Klioner}, {Lammers}, {Lindegren}, {Luri}, {Mignard}, {Panem}, {Pourbaix},
  {Randich}, {Sartoretti}, {Soubiran}, {Tanga}, {Walton}, {Bailer-Jones},
  {Bastian}, {Drimmel}, {Jansen}, {Katz}, {Lattanzi}, {van Leeuwen}, {Bakker},
  {Cacciari}, {Casta{\~n}eda}, {De Angeli}, {Fabricius}, {Fouesneau},
  {Fr{\'e}mat}, {Galluccio}, {Guerrier}, {Heiter}, {Masana}, {Messineo},
  {Mowlavi}, {Nicolas}, {Nienartowicz}, {Pailler}, {Panuzzo}, {Riclet}, {Roux},
  {Seabroke}, {Sordo}, {Th{\'e}venin}, {Gracia-Abril}, {Portell}, {Teyssier},
  {Altmann}, {Andrae}, {Audard}, {Bellas-Velidis}, {Benson}, {Berthier},
  {Blomme}, {Burgess}, {Busonero}, {Busso}, {C{\'a}novas}, {Carry}, {Cellino},
  {Cheek}, {Clementini}, {Damerdji}, {Davidson}, {de Teodoro}, {Nu{\~n}ez
  Campos}, {Delchambre}, {Dell'Oro}, {Esquej}, {Fern{\'a}ndez-Hern{\'a}ndez},
  {Fraile}, {Garabato}, {Garc{\'\i}a-Lario}, {Gosset}, {Haigron}, {Halbwachs},
  {Hambly}, {Harrison}, {Hern{\'a}ndez}, {Hestroffer}, {Hodgkin}, {Holl},
  {Jan{\ss}en}, {Jevardat de Fombelle}, {Jordan}, {Krone-Martins}, {Lanzafame},
  {L{\"o}ffler}, {Marchal}, {Marrese}, {Moitinho}, {Muinonen}, {Osborne},
  {Pancino}, {Pauwels}, {Recio-Blanco}, {Reyl{\'e}}, {Riello}, {Rimoldini},
  {Roegiers}, {Rybizki}, {Sarro}, {Siopis}, {Smith}, {Sozzetti}, {Utrilla},
  {van Leeuwen}, {Abbas}, {{\'A}brah{\'a}m}, {Abreu Aramburu}, {Aerts},
  {Aguado}, {Ajaj}, {Aldea-Montero}, {Altavilla}, {{\'A}lvarez}, {Alves},
  {Anders}, {Anderson}, {Anglada Varela}, {Antoja}, {Baines}, {Baker},
  {Balaguer-N{\'u}{\~n}ez}, {Balbinot}, {Balog}, {Barache}, {Barbato},
  {Barros}, {Barstow}, {Bartolom{\'e}}, {Bassilana}, {Bauchet}, {Becciani},
  {Bellazzini}, {Berihuete}, {Bernet}, {Bertone}, {Bianchi}, {Binnenfeld},
  {Blanco-Cuaresma}, {Blazere}, {Boch}, {Bombrun}, {Bossini}, {Bouquillon},
  {Bragaglia}, {Bramante}, {Breedt}, {Bressan}, {Brouillet}, {Brugaletta},
  {Bucciarelli}, {Burlacu}, {Butkevich}, {Buzzi}, {Caffau}, {Cancelliere},
  {Cantat-Gaudin}, {Carballo}, {Carlucci}, {Carnerero}, {Carrasco},
  {Casamiquela}, {Castellani}, {Castro-Ginard}, {Chaoul}, {Charlot}, {Chemin},
  {Chiaramida}, {Chiavassa}, {Chornay}, {Comoretto}, {Contursi}, {Cooper},
  {Cornez}, {Cowell}, {Crifo}, {Cropper}, {Crosta}, {Crowley}, {Dafonte},
  {Dapergolas}, {David}, {David}, {de Laverny}, {De Luise}, {De March}, {De
  Ridder}, {de Souza}, {de Torres}, {del Peloso}, {del Pozo}, {Delbo},
  {Delgado}, {Delisle}, {Demouchy}, {Dharmawardena}, {Di Matteo}, {Diakite},
  {Diener}, {Distefano}, {Dolding}, {Edvardsson}, {Enke}, {Fabre}, {Fabrizio},
  {Faigler}, {Fedorets}, {Fernique}, {Fienga}, {Figueras}, {Fournier},
  {Fouron}, {Fragkoudi}, {Gai}, {Garcia-Gutierrez}, {Garcia-Reinaldos},
  {Garc{\'\i}a-Torres}, {Garofalo}, {Gavel}, {Gavras}, {Gerlach}, {Geyer},
  {Giacobbe}, {Gilmore}, {Girona}, {Giuffrida}, {Gomel}, {Gomez},
  {Gonz{\'a}lez-N{\'u}{\~n}ez}, {Gonz{\'a}lez-Santamar{\'\i}a},
  {Gonz{\'a}lez-Vidal}, {Granvik}, {Guillout}, {Guiraud},
  {Guti{\'e}rrez-S{\'a}nchez}, {Guy}, {Hatzidimitriou}, {Hauser}, {Haywood},
  {Helmer}, {Helmi}, {Sarmiento}, {Hidalgo}, {Hilger}, {H{\l}adczuk}, {Hobbs},
  {Holland}, {Huckle}, {Jardine}, {Jasniewicz}, {Jean-Antoine Piccolo},
  {Jim{\'e}nez-Arranz}, {Jorissen}, {Juaristi Campillo}, {Julbe}, {Karbevska},
  {Kervella}, {Khanna}, {Kontizas}, {Kordopatis}, {Korn}, {K{\'o}sp{\'a}l},
  {Kostrzewa-Rutkowska}, {Kruszy{\'n}ska}, {Kun}, {Laizeau}, {Lambert},
  {Lanza}, {Lasne}, {Le Campion}, {Lebreton}, {Lebzelter}, {Leccia}, {Leclerc},
  {Lecoeur-Taibi}, {Liao}, {Licata}, {Lindstr{\o}m}, {Lister}, {Livanou},
  {Lobel}, {Lorca}, {Loup}, {Madrero Pardo}, {Magdaleno Romeo}, {Managau},
  {Mann}, {Manteiga}, {Marchant}, {Marconi}, {Marcos}, {Marcos Santos},
  {Mar{\'\i}n Pina}, {Marinoni}, {Marocco}, {Marshall}, {Martin Polo},
  {Mart{\'\i}n-Fleitas}, {Marton}, {Mary}, {Masip}, {Massari},
  {Mastrobuono-Battisti}, {Mazeh}, {McMillan}, {Messina}, {Michalik}, {Millar},
  {Mints}, {Molina}, {Molinaro}, {Moln{\'a}r}, {Monari}, {Mongui{\'o}},
  {Montegriffo}, {Montero}, {Mor}, {Mora}, {Morbidelli}, {Morel}, {Morris},
  {Muraveva}, {Murphy}, {Musella}, {Nagy}, {Noval}, {Oca{\~n}a}, {Ogden},
  {Ordenovic}, {Osinde}, {Pagani}, {Pagano}, {Palaversa}, {Palicio},
  {Pallas-Quintela}, {Panahi}, {Payne-Wardenaar}, {Pe{\~n}alosa Esteller},
  {Penttil{\"a}}, {Pichon}, {Piersimoni}, {Pineau}, {Plachy}, {Plum}, {Poggio},
  {Pr{\v{s}}a}, {Pulone}, {Racero}, {Ragaini}, {Rainer}, {Raiteri}, {Rambaux},
  {Ramos}, {Ramos-Lerate}, {Re Fiorentin}, {Regibo}, {Richards}, {Rios Diaz},
  {Ripepi}, {Riva}, {Rix}, {Rixon}, {Robichon}, {Robin}, {Robin}, {Roelens},
  {Rogues}, {Rohrbasser}, {Romero-G{\'o}mez}, {Rowell}, {Royer}, {Ruz Mieres},
  {Rybicki}, {Sadowski}, {S{\'a}ez N{\'u}{\~n}ez}, {Sagrist{\`a} Sell{\'e}s},
  {Sahlmann}, {Salguero}, {Samaras}, {Sanchez Gimenez}, {Sanna},
  {Santove{\~n}a}, {Sarasso}, {Schultheis}, {Sciacca}, {Segol}, {Segovia},
  {S{\'e}gransan}, {Semeux}, {Shahaf}, {Siddiqui}, {Siebert}, {Siltala},
  {Silvelo}, {Slezak}, {Slezak}, {Smart}, {Snaith}, {Solano}, {Solitro},
  {Souami}, {Souchay}, {Spagna}, {Spina}, {Spoto}, {Steele},
  {Steidelm{\"u}ller}, {Stephenson}, {S{\"u}veges}, {Surdej}, {Szabados},
  {Szegedi-Elek}, {Taris}, {Taylor}, {Teixeira}, {Tolomei}, {Tonello}, {Torra},
  {Torra}, {Torralba Elipe}, {Trabucchi}, {Tsounis}, {Turon}, {Ulla}, {Unger},
  {Vaillant}, {van Dillen}, {van Reeven}, {Vanel}, {Vecchiato}, {Viala},
  {Vicente}, {Voutsinas}, {Weiler}, {Wevers}, {Wyrzykowski}, {Yoldas}, {Yvard},
  {Zhao}, {Zorec}, {Zucker}, \& {Zwitter}}]{2023A&A...674A...1G}
{Gaia Collaboration}, {Vallenari}, A., {Brown}, A.~G.~A., {et~al.} 2023, \aap,
  674, A1

\bibitem[{{Gallino} {et~al.}(2006){Gallino}, {Bisterzo}, {Straniero}, {Ivans},
  \& {K{\"a}ppeler}}]{2006Gallino}
{Gallino}, R., {Bisterzo}, S., {Straniero}, O., {Ivans}, I.~I., \&
  {K{\"a}ppeler}, F. 2006, \memsai, 77, 786

\bibitem[{{Gilliland} {et~al.}(2010){Gilliland}, {Brown},
  {Christensen-Dalsgaard}, {Kjeldsen}, {Aerts}, {Appourchaux}, {Basu},
  {Bedding}, {Chaplin}, {Cunha}, {De Cat}, {De Ridder}, {Guzik}, {Handler},
  {Kawaler}, {Kiss}, {Kolenberg}, {Kurtz}, {Metcalfe}, {Monteiro}, {Szab{\'o}},
  {Arentoft}, {Balona}, {Debosscher}, {Elsworth}, {Quirion}, {Stello},
  {Su{\'a}rez}, {Borucki}, {Jenkins}, {Koch}, {Kondo}, {Latham}, {Rowe}, \&
  {Steffen}}]{2010PASP..122..131G}
{Gilliland}, R.~L., {Brown}, T.~M., {Christensen-Dalsgaard}, J., {et~al.} 2010,
  \pasp, 122, 131

\bibitem[{{Gilmore} {et~al.}(2012){Gilmore}, {Randich}, {Asplund}, {Binney},
  {Bonifacio}, {Drew}, {Feltzing}, {Ferguson}, {Jeffries}, {Micela},
  {Negueruela}, {Prusti}, {Rix}, {Vallenari}, {Alfaro}, {Allende-Prieto},
  {Babusiaux}, {Bensby}, {Blomme}, {Bragaglia}, {Flaccomio}, {Fran{\c{c}}ois},
  {Irwin}, {Koposov}, {Korn}, {Lanzafame}, {Pancino}, {Paunzen},
  {Recio-Blanco}, {Sacco}, {Smiljanic}, {Van Eck}, {Walton}, {Aden}, {Aerts},
  {Affer}, {Alcala}, {Altavilla}, {Alves}, {Antoja}, {Arenou}, {Argiroffi},
  {Asensio Ramos}, {Bailer-Jones}, {Balaguer-Nunez}, {Bayo}, {Barbuy},
  {Barisevicius}, {Barrado y Navascues}, {Battistini}, {Bellas Velidis},
  {Bellazzini}, {Belokurov}, {Bergemann}, {Bertelli}, {Biazzo}, {Bienayme},
  {Bland-Hawthorn}, {Boeche}, {Bonito}, {Boudreault}, {Bouvier}, {Brandao},
  {Brown}, {de Bruijne}, {Burleigh}, {Caballero}, {Caffau}, {Calura},
  {Capuzzo-Dolcetta}, {Caramazza}, {Carraro}, {Casagrande}, {Casewell},
  {Chapman}, {Chiappini}, {Chorniy}, {Christlieb}, {Cignoni}, {Cocozza},
  {Colless}, {Collet}, {Collins}, {Correnti}, {Covino}, {Crnojevic}, {Cropper},
  {Cunha}, {Damiani}, {David}, {Delgado}, {Duffau}, {Edvardsson}, {Eldridge},
  {Enke}, {Eriksson}, {Evans}, {Eyer}, {Famaey}, {Fellhauer}, {Ferreras},
  {Figueras}, {Fiorentino}, {Flynn}, {Folha}, {Franciosini}, {Frasca},
  {Freeman}, {Fremat}, {Friel}, {Gaensicke}, {Gameiro}, {Garzon}, {Geier},
  {Geisler}, {Gerhard}, {Gibson}, {Gomboc}, {Gomez}, {Gonzalez-Fernandez},
  {Gonzalez Hernandez}, {Gosset}, {Grebel}, {Greimel}, {Groenewegen},
  {Grundahl}, {Guarcello}, {Gustafsson}, {Hadrava}, {Hatzidimitriou}, {Hambly},
  {Hammersley}, {Hansen}, {Haywood}, {Heber}, {Heiter}, {Held}, {Helmi},
  {Hensler}, {Herrero}, {Hill}, {Hodgkin}, {Huelamo}, {Huxor}, {Ibata},
  {Jackson}, {de Jong}, {Jonker}, {Jordan}, {Jordi}, {Jorissen}, {Katz},
  {Kawata}, {Keller}, {Kharchenko}, {Klement}, {Klutsch}, {Knude}, {Koch},
  {Kochukhov}, {Kontizas}, {Koubsky}, {Lallement}, {de Laverny}, {van Leeuwen},
  {Lemasle}, {Lewis}, {Lind}, {Lindstrom}, {Lobel}, {Lopez Santiago}, {Lucas},
  {Ludwig}, {Lueftinger}, {Magrini}, {Maiz Apellaniz}, {Maldonado}, {Marconi},
  {Marino}, {Martayan}, {Martinez-Valpuesta}, {Matijevic}, {McMahon},
  {Messina}, {Meyer}, {Miglio}, {Mikolaitis}, {Minchev}, {Minniti}, {Moitinho},
  {Momany}, {Monaco}, {Montalto}, {Monteiro}, {Monier}, {Montes}, {Mora},
  {Moraux}, {Morel}, {Mowlavi}, {Mucciarelli}, {Munari}, {Napiwotzki},
  {Nardetto}, {Naylor}, {Naze}, {Nelemans}, {Okamoto}, {Ortolani}, {Pace},
  {Palla}, {Palous}, {Parker}, {Penarrubia}, {Pillitteri}, {Piotto}, {Posbic},
  {Prisinzano}, {Puzeras}, {Quirrenbach}, {Ragaini}, {Read}, {Read}, {Reyle},
  {De Ridder}, {Robichon}, {Robin}, {Roeser}, {Romano}, {Royer}, {Ruchti},
  {Ruzicka}, {Ryan}, {Ryde}, {Santos}, {Sanz Forcada}, {Sarro Baro},
  {Sbordone}, {Schilbach}, {Schmeja}, {Schnurr}, {Schoenrich}, {Scholz},
  {Seabroke}, {Sharma}, {De Silva}, {Smith}, {Solano}, {Sordo}, {Soubiran},
  {Sousa}, {Spagna}, {Steffen}, {Steinmetz}, {Stelzer}, {Stempels},
  {Tabernero}, {Tautvaisiene}, {Thevenin}, {Torra}, {Tosi}, {Tolstoy}, {Turon},
  {Walker}, {Wambsganss}, {Worley}, {Venn}, {Vink}, {Wyse}, {Zaggia},
  {Zeilinger}, {Zoccali}, {Zorec}, {Zucker}, {Zwitter}, \& {Gaia-ESO Survey
  Team}}]{2012Msngr.147...25G}
{Gilmore}, G., {Randich}, S., {Asplund}, M., {et~al.} 2012, The Messenger, 147,
  25

\bibitem[{{Goswami} {et~al.}(2023){Goswami}, {Shejeelammal}, {Goswami}, \&
  {Purandardas}}]{2023Goswami}
{Goswami}, A., {Shejeelammal}, J., {Goswami}, P.~P., \& {Purandardas}, M. 2023,
  arXiv e-prints, arXiv:2311.10043

\bibitem[{{Gustafsson} {et~al.}(2008){Gustafsson}, {Edvardsson}, {Eriksson},
  {J{\o}rgensen}, {Nordlund}, \& {Plez}}]{2008A&A...486..951G}
{Gustafsson}, B., {Edvardsson}, B., {Eriksson}, K., {et~al.} 2008, \aap, 486,
  951

\bibitem[{{Hawkins} {et~al.}(2015){Hawkins}, {Jofr{\'e}}, {Masseron}, \&
  {Gilmore}}]{2015MNRAS.453..758H}
{Hawkins}, K., {Jofr{\'e}}, P., {Masseron}, T., \& {Gilmore}, G. 2015, \mnras,
  453, 758

\bibitem[{{Hayden} {et~al.}(2022){Hayden}, {Sharma}, {Bland-Hawthorn}, {Spina},
  {Buder}, {Ciuc{\u{a}}}, {Asplund}, {Casey}, {De Silva}, {D'Orazi}, {Freeman},
  {Kos}, {Lewis}, {Lin}, {Lind}, {Martell}, {Schlesinger}, {Simpson}, {Zucker},
  {Zwitter}, {Chen}, {{\v{C}}otar}, {Feuillet}, {Horner}, {Joyce},
  {Nordlander}, {Stello}, {Tepper-Garcia}, {Ting}, {Wang}, {Wittenmyer}, \&
  {Wyse}}]{2022MNRAS.517.5325H}
{Hayden}, M.~R., {Sharma}, S., {Bland-Hawthorn}, J., {et~al.} 2022, \mnras,
  517, 5325

\bibitem[{{Heged{\H{u}}s} {et~al.}(2023){Heged{\H{u}}s}, {M{\'e}sz{\'a}ros},
  {Jofr{\'e}}, {Stringfellow}, {Feuillet}, {Garc{\'\i}a-Hern{\'a}ndez},
  {Nitschelm}, \& {Zamora}}]{2023A&A...670A.107H}
{Heged{\H{u}}s}, V., {M{\'e}sz{\'a}ros}, S., {Jofr{\'e}}, P., {et~al.} 2023,
  \aap, 670, A107

\bibitem[{{Heiter} {et~al.}(2015){Heiter}, {Jofr{\'e}}, {Gustafsson}, {Korn},
  {Soubiran}, \& {Th{\'e}venin}}]{2015Heiter}
{Heiter}, U., {Jofr{\'e}}, P., {Gustafsson}, B., {et~al.} 2015, \aap, 582, A49

\bibitem[{{Heiter} {et~al.}(2021){Heiter}, {Lind}, {Bergemann}, {Asplund},
  {Mikolaitis}, {Barklem}, {Masseron}, {de Laverny}, {Magrini}, {Edvardsson},
  {J{\"o}nsson}, {Pickering}, {Ryde}, {Bayo Ar{\'a}n}, {Bensby}, {Casey},
  {Feltzing}, {Jofr{\'e}}, {Korn}, {Pancino}, {Damiani}, {Lanzafame}, {Lardo},
  {Monaco}, {Morbidelli}, {Smiljanic}, {Worley}, {Zaggia}, {Randich}, \&
  {Gilmore}}]{2021A&A...645A.106H}
{Heiter}, U., {Lind}, K., {Bergemann}, M., {et~al.} 2021, \aap, 645, A106

\bibitem[{{Hidalgo} {et~al.}(2018){Hidalgo}, {Pietrinferni}, {Cassisi},
  {Salaris}, {Mucciarelli}, {Savino}, {Aparicio}, {Silva Aguirre}, \&
  {Verma}}]{2018ApJ...856..125H}
{Hidalgo}, S.~L., {Pietrinferni}, A., {Cassisi}, S., {et~al.} 2018, \apj, 856,
  125

\bibitem[{{Hogg} {et~al.}(2016){Hogg}, {Casey}, {Ness}, {Rix},
  {Foreman-Mackey}, {Hasselquist}, {Ho}, {Holtzman}, {Majewski}, {Martell},
  {M{\'e}sz{\'a}ros}, {Nidever}, \& {Shetrone}}]{2016Hogg}
{Hogg}, D.~W., {Casey}, A.~R., {Ness}, M., {et~al.} 2016, \apj, 833, 262

\bibitem[{{Howell} {et~al.}(2014){Howell}, {Sobeck}, {Haas}, {Still},
  {Barclay}, {Mullally}, {Troeltzsch}, {Aigrain}, {Bryson}, {Caldwell},
  {Chaplin}, {Cochran}, {Huber}, {Marcy}, {Miglio}, {Najita}, {Smith},
  {Twicken}, \& {Fortney}}]{2014PASP..126..398H}
{Howell}, S.~B., {Sobeck}, C., {Haas}, M., {et~al.} 2014, \pasp, 126, 398

\bibitem[{{Huber} {et~al.}(2016){Huber}, {Bryson}, {Haas}, {Barclay},
  {Barentsen}, {Howell}, {Sharma}, {Stello}, \& {Thompson}}]{2016Huber}
{Huber}, D., {Bryson}, S.~T., {Haas}, M.~R., {et~al.} 2016, \apjs, 224, 2

\bibitem[{{Izzard} {et~al.}(2018){Izzard}, {Preece}, {Jofre}, {Halabi},
  {Masseron}, \& {Tout}}]{Izzard18}
{Izzard}, R.~G., {Preece}, H., {Jofre}, P., {et~al.} 2018, \mnras, 473, 2984

\bibitem[{{Jofr{\'e}}(2021)}]{2021ApJ...920...23J}
{Jofr{\'e}}, P. 2021, \apj, 920, 23

\bibitem[{{Jofr{\'e}} {et~al.}(2017){Jofr{\'e}}, {Das}, {Bertranpetit}, \&
  {Foley}}]{Jofre17}
{Jofr{\'e}}, P., {Das}, P., {Bertranpetit}, J., \& {Foley}, R. 2017, \mnras,
  467, 1140

\bibitem[{{Jofr{\'e}} {et~al.}(2019){Jofr{\'e}}, {Heiter}, \&
  {Soubiran}}]{2019Jofre}
{Jofr{\'e}}, P., {Heiter}, U., \& {Soubiran}, C. 2019, \araa, 57, 571

\bibitem[{{Jofr{\'e}} {et~al.}(2020){Jofr{\'e}}, {Jackson}, \& {Tucci
  Maia}}]{2020A&A...633L...9J}
{Jofr{\'e}}, P., {Jackson}, H., \& {Tucci Maia}, M. 2020, \aap, 633, L9

\bibitem[{{Jofr{\'e}} {et~al.}(2023){Jofr{\'e}}, {Jorissen},
  {Aguilera-G{\'o}mez}, {Van Eck}, {Tayar}, {Pinsonneault}, {Zinn}, {Goriely},
  \& {Van Winckel}}]{Jofre23}
{Jofr{\'e}}, P., {Jorissen}, A., {Aguilera-G{\'o}mez}, C., {et~al.} 2023, \aap,
  671, A21

\bibitem[{{J{\"o}nsson} {et~al.}(2020){J{\"o}nsson}, {Holtzman}, {Allende
  Prieto}, {Cunha}, {Garc{\'\i}a-Hern{\'a}ndez}, {Hasselquist}, {Masseron},
  {Osorio}, {Shetrone}, {Smith}, {Stringfellow}, {Bizyaev}, {Edvardsson},
  {Majewski}, {M{\'e}sz{\'a}ros}, {Souto}, {Zamora}, {Beaton}, {Bovy}, {Donor},
  {Pinsonneault}, {Poovelil}, \& {Sobeck}}]{2020AJ....160..120J}
{J{\"o}nsson}, H., {Holtzman}, J.~A., {Allende Prieto}, C., {et~al.} 2020, \aj,
  160, 120

\bibitem[{{Karakas} \& {Lattanzio}(2014)}]{Karakas2014}
{Karakas}, A.~I. \& {Lattanzio}, J.~C. 2014, \pasa, 31, e030

\bibitem[{{Karakas} \& {Lugaro}(2016)}]{2016ApJ...825...26K}
{Karakas}, A.~I. \& {Lugaro}, M. 2016, \apj, 825, 26

\bibitem[{{Kjeldsen} \& {Bedding}(1995)}]{1995A&A...293...87K}
{Kjeldsen}, H. \& {Bedding}, T.~R. 1995, \aap, 293, 87

\bibitem[{{Kobayashi} {et~al.}(2020){Kobayashi}, {Karakas}, \&
  {Lugaro}}]{2020ApJ...900..179K}
{Kobayashi}, C., {Karakas}, A.~I., \& {Lugaro}, M. 2020, \apj, 900, 179

\bibitem[{{Kobayashi} {et~al.}(2006){Kobayashi}, {Umeda}, {Nomoto}, {Tominaga},
  \& {Ohkubo}}]{2006Kobayashi}
{Kobayashi}, C., {Umeda}, H., {Nomoto}, K., {Tominaga}, N., \& {Ohkubo}, T.
  2006, \apj, 653, 1145

\bibitem[{{Krumholz} \& {Ting}(2018)}]{2018Krumholz}
{Krumholz}, M.~R. \& {Ting}, Y.-S. 2018, \mnras, 475, 2236

\bibitem[{{Leung} {et~al.}(2023){Leung}, {Bovy}, {Mackereth}, \&
  {Miglio}}]{2023MNRAS.522.4577L}
{Leung}, H.~W., {Bovy}, J., {Mackereth}, J.~T., \& {Miglio}, A. 2023, \mnras,
  522, 4577

\bibitem[{{Liu} {et~al.}(2020){Liu}, {Shi}, \& {Wu}}]{2020ApJ...896...64L}
{Liu}, S., {Shi}, J., \& {Wu}, Z. 2020, \apj, 896, 64

\bibitem[{{Lugaro} {et~al.}(2003){Lugaro}, {Herwig}, {Lattanzio}, {Gallino}, \&
  {Straniero}}]{2003Lugaro}
{Lugaro}, M., {Herwig}, F., {Lattanzio}, J.~C., {Gallino}, R., \& {Straniero},
  O. 2003, \apj, 586, 1305

\bibitem[{{Magrini} {et~al.}(2009){Magrini}, {Sestito}, {Randich}, \&
  {Galli}}]{2009Magrini}
{Magrini}, L., {Sestito}, P., {Randich}, S., \& {Galli}, D. 2009, \aap, 494, 95

\bibitem[{{Magrini} {et~al.}(2018){Magrini}, {Spina}, {Randich}, {Friel},
  {Kordopatis}, {Worley}, {Pancino}, {Bragaglia}, {Donati},
  {Tautvai{\v{s}}ien{\.{e}}}, {Bagdonas}, {Delgado-Mena}, {Adibekyan}, {Sousa},
  {Jim{\'e}nez-Esteban}, {Sanna}, {Roccatagliata}, {Bonito}, {Sbordone},
  {Duffau}, {Gilmore}, {Feltzing}, {Jeffries}, {Vallenari}, {Alfaro}, {Bensby},
  {Francois}, {Koposov}, {Korn}, {Recio-Blanco}, {Smiljanic}, {Bayo},
  {Carraro}, {Casey}, {Costado}, {Damiani}, {Franciosini}, {Frasca},
  {Hourihane}, {Jofr{\'e}}, {de Laverny}, {Lewis}, {Masseron}, {Monaco},
  {Morbidelli}, {Prisinzano}, {Sacco}, \& {Zaggia}}]{2018Magrini}
{Magrini}, L., {Spina}, L., {Randich}, S., {et~al.} 2018, \aap, 617, A106

\bibitem[{{Magrini} {et~al.}(2021){Magrini}, {Vescovi}, {Casali}, {Cristallo},
  {Viscasillas V{\'a}zquez}, {Cescutti}, {Spina}, {Van Der Swaelmen}, \&
  {Randich}}]{2021A&A...646L...2M}
{Magrini}, L., {Vescovi}, D., {Casali}, G., {et~al.} 2021, \aap, 646, L2

\bibitem[{{Majewski} {et~al.}(2017){Majewski}, {Schiavon}, {Frinchaboy},
  {Allende Prieto}, {Barkhouser}, {Bizyaev}, {Blank}, {Brunner}, {Burton},
  {Carrera}, {Chojnowski}, {Cunha}, {Epstein}, {Fitzgerald}, {Garc{\'\i}a
  P{\'e}rez}, {Hearty}, {Henderson}, {Holtzman}, {Johnson}, {Lam}, {Lawler},
  {Maseman}, {M{\'e}sz{\'a}ros}, {Nelson}, {Nguyen}, {Nidever}, {Pinsonneault},
  {Shetrone}, {Smee}, {Smith}, {Stolberg}, {Skrutskie}, {Walker}, {Wilson},
  {Zasowski}, {Anders}, {Basu}, {Beland}, {Blanton}, {Bovy}, {Brownstein},
  {Carlberg}, {Chaplin}, {Chiappini}, {Eisenstein}, {Elsworth}, {Feuillet},
  {Fleming}, {Galbraith-Frew}, {Garc{\'\i}a}, {Garc{\'\i}a-Hern{\'a}ndez},
  {Gillespie}, {Girardi}, {Gunn}, {Hasselquist}, {Hayden}, {Hekker}, {Ivans},
  {Kinemuchi}, {Klaene}, {Mahadevan}, {Mathur}, {Mosser}, {Muna}, {Munn},
  {Nichol}, {O'Connell}, {Parejko}, {Robin}, {Rocha-Pinto}, {Schultheis},
  {Serenelli}, {Shane}, {Silva Aguirre}, {Sobeck}, {Thompson}, {Troup},
  {Weinberg}, \& {Zamora}}]{2017AJ....154...94M}
{Majewski}, S.~R., {Schiavon}, R.~P., {Frinchaboy}, P.~M., {et~al.} 2017, \aj,
  154, 94

\bibitem[{{Martig} {et~al.}(2015){Martig}, {Rix}, {Silva Aguirre}, {Hekker},
  {Mosser}, {Elsworth}, {Bovy}, {Stello}, {Anders}, {Garc{\'\i}a}, {Tayar},
  {Rodrigues}, {Basu}, {Carrera}, {Ceillier}, {Chaplin}, {Chiappini},
  {Frinchaboy}, {Garc{\'\i}a-Hern{\'a}ndez}, {Hearty}, {Holtzman}, {Johnson},
  {Majewski}, {Mathur}, {M{\'e}sz{\'a}ros}, {Miglio}, {Nidever}, {Pan},
  {Pinsonneault}, {Schiavon}, {Schneider}, {Serenelli}, {Shetrone}, \&
  {Zamora}}]{Martig15}
{Martig}, M., {Rix}, H.-W., {Silva Aguirre}, V., {et~al.} 2015, \mnras, 451,
  2230

\bibitem[{{Mashonkina} {et~al.}(2019){Mashonkina}, {Sitnova}, {Yakovleva}, \&
  {Belyaev}}]{2019Mashonkina}
{Mashonkina}, L., {Sitnova}, T., {Yakovleva}, S.~A., \& {Belyaev}, A.~K. 2019,
  \aap, 631, A43

\bibitem[{{Matteucci}(2012)}]{2012ceg..book.....M}
{Matteucci}, F. 2012, {Chemical Evolution of Galaxies} (Astronomy and
  Astrophysics Library)

\bibitem[{{Matteucci}(2016)}]{2016Matteucci}
{Matteucci}, F. 2016, in Journal of Physics Conference Series, Vol. 703,
  Journal of Physics Conference Series, 012004

\bibitem[{{Miglio}(2012)}]{2012ASSP...26...11M}
{Miglio}, A. 2012, in Astrophysics and Space Science Proceedings, Vol.~26, Red
  Giants as Probes of the Structure and Evolution of the Milky Way, 11

\bibitem[{{Miglio} {et~al.}(2016){Miglio}, {Chaplin}, {Brogaard}, {Lund},
  {Mosser}, {Davies}, {Handberg}, {Milone}, {Marino}, {Bossini}, {Elsworth},
  {Grundahl}, {Arentoft}, {Bedin}, {Campante}, {Jessen-Hansen}, {Jones},
  {Kuszlewicz}, {Malavolta}, {Nascimbeni}, \& {Sandquist}}]{2016Miglio}
{Miglio}, A., {Chaplin}, W.~J., {Brogaard}, K., {et~al.} 2016, \mnras, 461, 760

\bibitem[{{Minchev} \& {Famaey}(2010)}]{2010ApJ...722..112M}
{Minchev}, I. \& {Famaey}, B. 2010, \apj, 722, 112

\bibitem[{{Minchev} {et~al.}(2012){Minchev}, {Famaey}, {Quillen}, {Di Matteo},
  {Combes}, {Vlaji{\'c}}, {Erwin}, \& {Bland-Hawthorn}}]{2012Minchev}
{Minchev}, I., {Famaey}, B., {Quillen}, A.~C., {et~al.} 2012, \aap, 548, A126

\bibitem[{{Minchev} {et~al.}(2017){Minchev}, {Steinmetz}, {Chiappini},
  {Martig}, {Anders}, {Matijevic}, \& {de Jong}}]{2017Minchev}
{Minchev}, I., {Steinmetz}, M., {Chiappini}, C., {et~al.} 2017, \apj, 834, 27

\bibitem[{{Mosser} {et~al.}(2014){Mosser}, {Benomar}, {Belkacem}, {Goupil},
  {Lagarde}, {Michel}, {Lebreton}, {Stello}, {Vrard}, {Barban}, {Bedding},
  {Deheuvels}, {Chaplin}, {De Ridder}, {Elsworth}, {Montalban}, {Noels},
  {Ouazzani}, {Samadi}, {White}, \& {Kjeldsen}}]{2014Mosser}
{Mosser}, B., {Benomar}, O., {Belkacem}, K., {et~al.} 2014, \aap, 572, L5

\bibitem[{{Nissen}(2015)}]{2015A&A...579A..52N}
{Nissen}, P.~E. 2015, \aap, 579, A52

\bibitem[{{Nissen}(2016)}]{2016Nissen}
{Nissen}, P.~E. 2016, \aap, 593, A65

\bibitem[{{Nissen} {et~al.}(2020){Nissen}, {Christensen-Dalsgaard},
  {Mosumgaard}, {Silva Aguirre}, {Spitoni}, \& {Verma}}]{2020A&A...640A..81N}
{Nissen}, P.~E., {Christensen-Dalsgaard}, J., {Mosumgaard}, J.~R., {et~al.}
  2020, \aap, 640, A81

\bibitem[{{Nissen} \& {Gustafsson}(2018)}]{2018Nissenrew}
{Nissen}, P.~E. \& {Gustafsson}, B. 2018, \aapr, 26, 6

\bibitem[{{Nissen} {et~al.}(2017){Nissen}, {Silva Aguirre},
  {Christensen-Dalsgaard}, {Collet}, {Grundahl}, \&
  {Slumstrup}}]{2017A&A...608A.112N}
{Nissen}, P.~E., {Silva Aguirre}, V., {Christensen-Dalsgaard}, J., {et~al.}
  2017, \aap, 608, A112

\bibitem[{{Noels} \& {Bragaglia}(2015)}]{2015ASSP...39..167N}
{Noels}, A. \& {Bragaglia}, A. 2015, in Astrophysics and Space Science
  Proceedings, Vol.~39, Asteroseismology of Stellar Populations in the Milky
  Way, 167

\bibitem[{{Ou} {et~al.}(2020){Ou}, {Roederer}, {Sneden}, {Cowan}, {Lawler},
  {Shectman}, \& {Thompson}}]{2020ApJ...900..106O}
{Ou}, X., {Roederer}, I.~U., {Sneden}, C., {et~al.} 2020, \apj, 900, 106

\bibitem[{{Pietrinferni} {et~al.}(2021){Pietrinferni}, {Hidalgo}, {Cassisi},
  {Salaris}, {Savino}, {Mucciarelli}, {Verma}, {Silva Aguirre}, {Aparicio}, \&
  {Ferguson}}]{2021ApJ...908..102P}
{Pietrinferni}, A., {Hidalgo}, S., {Cassisi}, S., {et~al.} 2021, \apj, 908, 102

\bibitem[{{Pignatari} {et~al.}(2010){Pignatari}, {Gallino}, {Heil}, {Wiescher},
  {K{\"a}ppeler}, {Herwig}, \& {Bisterzo}}]{2010ApJ...710.1557P}
{Pignatari}, M., {Gallino}, R., {Heil}, M., {et~al.} 2010, \apj, 710, 1557

\bibitem[{{Pinsonneault} {et~al.}(2014){Pinsonneault}, {Elsworth}, {Epstein},
  {Hekker}, {M{\'e}sz{\'a}ros}, {Chaplin}, {Johnson}, {Garc{\'\i}a},
  {Holtzman}, {Mathur}, {Garc{\'\i}a P{\'e}rez}, {Silva Aguirre}, {Girardi},
  {Basu}, {Shetrone}, {Stello}, {Allende Prieto}, {An}, {Beck}, {Beers},
  {Bizyaev}, {Bloemen}, {Bovy}, {Cunha}, {De Ridder}, {Frinchaboy},
  {Garc{\'\i}a-Hern{\'a}ndez}, {Gilliland}, {Harding}, {Hearty}, {Huber},
  {Ivans}, {Kallinger}, {Majewski}, {Metcalfe}, {Miglio}, {Mosser}, {Muna},
  {Nidever}, {Schneider}, {Serenelli}, {Smith}, {Tayar}, {Zamora}, \&
  {Zasowski}}]{2014ApJS..215...19P}
{Pinsonneault}, M.~H., {Elsworth}, Y., {Epstein}, C., {et~al.} 2014, \apjs,
  215, 19

\bibitem[{{Plez}(2012)}]{2012ascl.soft05004P}
{Plez}, B. 2012, {Turbospectrum: Code for spectral synthesis}, Astrophysics
  Source Code Library, record ascl:1205.004

\bibitem[{{Preston} \& {Sneden}(2000)}]{Preston}
{Preston}, G.~W. \& {Sneden}, C. 2000, \aj, 120, 1014

\bibitem[{{Ratcliffe} {et~al.}(2023{\natexlab{a}}){Ratcliffe}, {Minchev},
  {Anders}, {Khoperskov}, {Guiglion}, {Buck}, {Cunha}, {Queiroz}, {Nitschelm},
  {Meszaros}, {Steinmetz}, {de Jong}, {Nepal}, {Lane}, \&
  {Sobeck}}]{2023MNRAS.525.2208R}
{Ratcliffe}, B., {Minchev}, I., {Anders}, F., {et~al.} 2023{\natexlab{a}},
  \mnras, 525, 2208

\bibitem[{{Ratcliffe} {et~al.}(2023{\natexlab{b}}){Ratcliffe}, {Minchev},
  {Cescutti}, {Spitoni}, {J{\"o}nsson}, {Anders}, {Queiroz}, \&
  {Steinmetz}}]{2023arXiv230711159R}
{Ratcliffe}, B., {Minchev}, I., {Cescutti}, G., {et~al.} 2023{\natexlab{b}},
  arXiv e-prints, arXiv:2307.11159

\bibitem[{{Ratcliffe} {et~al.}(2020){Ratcliffe}, {Ness}, {Johnston}, \&
  {Sen}}]{2020ApJ...900..165R}
{Ratcliffe}, B.~L., {Ness}, M.~K., {Johnston}, K.~V., \& {Sen}, B. 2020, \apj,
  900, 165

\bibitem[{{Recio-Blanco} {et~al.}(2021){Recio-Blanco}, {Fern{\'a}ndez-Alvar},
  {de Laverny}, {Antoja}, {Helmi}, \& {Crida}}]{2021RBlanco}
{Recio-Blanco}, A., {Fern{\'a}ndez-Alvar}, E., {de Laverny}, P., {et~al.} 2021,
  \aap, 648, A108

\bibitem[{{Reddy} \& {Lambert}(2017)}]{2017ApJ...845..151R}
{Reddy}, A. B.~S. \& {Lambert}, D.~L. 2017, \apj, 845, 151

\bibitem[{{Rendle} {et~al.}(2019){Rendle}, {Miglio}, {Chiappini}, {Valentini},
  {Davies}, {Mosser}, {Elsworth}, {Garc{\'\i}a}, {Mathur}, {Jofr{\'e}},
  {Worley}, {Casagrande}, {Girardi}, {Lund}, {Feuillet}, {Gavel}, {Magrini},
  {Khan}, {Rodrigues}, {Johnson}, {Cunha}, {Lane}, {Nitschelm}, \&
  {Chaplin}}]{2019MNRAS.490.4465R}
{Rendle}, B.~M., {Miglio}, A., {Chiappini}, C., {et~al.} 2019, \mnras, 490,
  4465

\bibitem[{{Rix} \& {Bovy}(2013)}]{2013Rix}
{Rix}, H.-W. \& {Bovy}, J. 2013, \aapr, 21, 61

\bibitem[{{Sahlholdt} {et~al.}(2022){Sahlholdt}, {Feltzing}, \&
  {Feuillet}}]{Sahlholdt22}
{Sahlholdt}, C.~L., {Feltzing}, S., \& {Feuillet}, D.~K. 2022, \mnras, 510,
  4669

\bibitem[{{Sahlholdt} \& {Silva Aguirre}(2018)}]{2018Sahlholdt}
{Sahlholdt}, C.~L. \& {Silva Aguirre}, V. 2018, \mnras, 481, L125

\bibitem[{{Sales-Silva} {et~al.}(2022){Sales-Silva}, {Daflon}, {Cunha},
  {Souto}, {Smith}, {Chiappini}, {Donor}, {Frinchaboy},
  {Garc{\'\i}a-Hern{\'a}ndez}, {Hayes}, {Majewski}, {Masseron}, {Schiavon},
  {Weinberg}, {Beaton}, {Fern{\'a}ndez-Trincado}, {J{\"o}nsson}, {Lane},
  {Minniti}, {Manchado}, {Moni Bidin}, {Nitschelm}, {O'Connell}, \&
  {Villanova}}]{2022Ssilva}
{Sales-Silva}, J.~V., {Daflon}, S., {Cunha}, K., {et~al.} 2022, \apj, 926, 154

\bibitem[{{Sandage}(1953)}]{sandage}
{Sandage}, A.~R. 1953, \aj, 58, 61

\bibitem[{{Sellwood} \& {Binney}(2002{\natexlab{a}})}]{2002MNRAS.336..785S}
{Sellwood}, J.~A. \& {Binney}, J.~J. 2002{\natexlab{a}}, \mnras, 336, 785

\bibitem[{{Sellwood} \& {Binney}(2002{\natexlab{b}})}]{2002Sellwood}
{Sellwood}, J.~A. \& {Binney}, J.~J. 2002{\natexlab{b}}, \mnras, 336, 785

\bibitem[{{Sharma} {et~al.}(2022){Sharma}, {Hayden}, {Bland-Hawthorn},
  {Stello}, {Buder}, {Zinn}, {Spina}, {Kallinger}, {Asplund}, {De Silva},
  {D'Orazi}, {Freeman}, {Kos}, {Lewis}, {Lin}, {Lind}, {Martell},
  {Schlesinger}, {Simpson}, {Zucker}, {Zwitter}, {Chen}, {Cotar}, {Kafle},
  {Khanna}, {Tepper-Garcia}, {Wang}, \& {Wittenmyer}}]{2022Sharma}
{Sharma}, S., {Hayden}, M.~R., {Bland-Hawthorn}, J., {et~al.} 2022, \mnras,
  510, 734

\bibitem[{{Sharma} {et~al.}(2018){Sharma}, {Stello}, {Buder}, {Kos},
  {Bland-Hawthorn}, {Asplund}, {Duong}, {Lin}, {Lind}, {Ness}, {Huber},
  {Zwitter}, {Traven}, {Hon}, {Kafle}, {Khanna}, {Saddon}, {Anguiano}, {Casey},
  {Freeman}, {Martell}, {De Silva}, {Simpson}, {Wittenmyer}, \&
  {Zucker}}]{2018MNRAS.473.2004S}
{Sharma}, S., {Stello}, D., {Buder}, S., {et~al.} 2018, \mnras, 473, 2004

\bibitem[{{Silva Aguirre} {et~al.}(2018){Silva Aguirre}, {Bojsen-Hansen},
  {Slumstrup}, {Casagrande}, {Kawata}, {Ciuc{\v{a}}}, {Handberg}, {Lund},
  {Mosumgaard}, {Huber}, {Johnson}, {Pinsonneault}, {Serenelli}, {Stello},
  {Tayar}, {Bird}, {Cassisi}, {Hon}, {Martig}, {Nissen}, {Rix},
  {Sch{\"o}nrich}, {Sahlholdt}, {Trick}, \& {Yu}}]{2018MNRAS.475.5487S}
{Silva Aguirre}, V., {Bojsen-Hansen}, M., {Slumstrup}, D., {et~al.} 2018,
  \mnras, 475, 5487

\bibitem[{{Silva Aguirre} {et~al.}(2012){Silva Aguirre}, {Casagrande}, {Basu},
  {Campante}, {Chaplin}, {Huber}, {Miglio}, {Serenelli}, {Ballot}, {Bedding},
  {Christensen-Dalsgaard}, {Creevey}, {Elsworth}, {Garc{\'\i}a}, {Gilliland},
  {Hekker}, {Kjeldsen}, {Mathur}, {Metcalfe}, {Monteiro}, {Mosser},
  {Pinsonneault}, {Stello}, {Weiss}, {Tenenbaum}, {Twicken}, \&
  {Uddin}}]{2012Aguirre}
{Silva Aguirre}, V., {Casagrande}, L., {Basu}, S., {et~al.} 2012, \apj, 757, 99

\bibitem[{{Silva Aguirre} {et~al.}(2015){Silva Aguirre}, {Davies}, {Basu},
  {Christensen-Dalsgaard}, {Creevey}, {Metcalfe}, {Bedding}, {Casagrande},
  {Handberg}, {Lund}, {Nissen}, {Chaplin}, {Huber}, {Serenelli}, {Stello}, {Van
  Eylen}, {Campante}, {Elsworth}, {Gilliland}, {Hekker}, {Karoff}, {Kawaler},
  {Kjeldsen}, \& {Lundkvist}}]{2015MNRAS.452.2127S}
{Silva Aguirre}, V., {Davies}, G.~R., {Basu}, S., {et~al.} 2015, \mnras, 452,
  2127

\bibitem[{{Silva Aguirre} {et~al.}(2020){Silva Aguirre}, {Stello}, {Stokholm},
  {Mosumgaard}, {Ball}, {Basu}, {Bossini}, {Bugnet}, {Buzasi}, {Campante},
  {Carboneau}, {Chaplin}, {Corsaro}, {Davies}, {Elsworth}, {Garc{\'\i}a},
  {Gaulme}, {Hall}, {Handberg}, {Hon}, {Kallinger}, {Kang}, {Lund}, {Mathur},
  {Mints}, {Mosser}, {{\c{C}}elik Orhan}, {Rodrigues}, {Vrard}, {Y{\i}ld{\i}z},
  {Zinn}, {{\"O}rtel}, {Beck}, {Bell}, {Guo}, {Jiang}, {Kuszlewicz}, {Kuehn},
  {Li}, {Lundkvist}, {Pinsonneault}, {Tayar}, {Cunha}, {Hekker}, {Huber},
  {Miglio}, {F.~G. Monteiro}, {Slumstrup}, {Winther}, {Angelou}, {Benomar},
  {B{\'o}di}, {De Moura}, {Deheuvels}, {Derekas}, {Di Mauro}, {Dupret},
  {Jim{\'e}nez}, {Lebreton}, {Matthews}, {Nardetto}, {do Nascimento},
  {Pereira}, {Rodr{\'\i}guez D{\'\i}az}, {Serenelli}, {Spitoni},
  {Stonkut{\.{e}}}, {Su{\'a}rez}, {Szab{\'o}}, {Van Eylen}, {Ventura}, {Verma},
  {Weiss}, {Wu}, {Barclay}, {Christensen-Dalsgaard}, {Jenkins}, {Kjeldsen},
  {Ricker}, {Seager}, \& {Vanderspek}}]{2020ApJ...889L..34S}
{Silva Aguirre}, V., {Stello}, D., {Stokholm}, A., {et~al.} 2020, \apjl, 889,
  L34

\bibitem[{{Slumstrup} {et~al.}(2017){Slumstrup}, {Grundahl}, {Brogaard},
  {Thygesen}, {Nissen}, {Jessen-Hansen}, {Van Eylen}, \&
  {Pedersen}}]{2017A&A...604L...8S}
{Slumstrup}, D., {Grundahl}, F., {Brogaard}, K., {et~al.} 2017, \aap, 604, L8

\bibitem[{{Slumstrup} {et~al.}(2019){Slumstrup}, {Grundahl}, {Silva Aguirre},
  \& {Brogaard}}]{2019Slumstrup}
{Slumstrup}, D., {Grundahl}, F., {Silva Aguirre}, V., \& {Brogaard}, K. 2019,
  \aap, 622, A111

\bibitem[{{Spina} {et~al.}(2018){Spina}, {Mel{\'e}ndez}, {Karakas}, {dos
  Santos}, {Bedell}, {Asplund}, {Ram{\'\i}rez}, {Yong}, {Alves-Brito}, {Bean},
  \& {Dreizler}}]{2018MNRAS.474.2580S}
{Spina}, L., {Mel{\'e}ndez}, J., {Karakas}, A.~I., {et~al.} 2018, \mnras, 474,
  2580

\bibitem[{{Spina} {et~al.}(2016){Spina}, {Mel{\'e}ndez}, {Karakas},
  {Ram{\'\i}rez}, {Monroe}, {Asplund}, \& {Yong}}]{2016Spina}
{Spina}, L., {Mel{\'e}ndez}, J., {Karakas}, A.~I., {et~al.} 2016, \aap, 593,
  A125

\bibitem[{{Stello} {et~al.}(2016){Stello}, {Vanderburg}, {Casagrande},
  {Gilliland}, {Silva Aguirre}, {Sandquist}, {Leiner}, {Mathieu}, \&
  {Soderblom}}]{2016Stello}
{Stello}, D., {Vanderburg}, A., {Casagrande}, L., {et~al.} 2016, \apj, 832, 133

\bibitem[{{Stokholm} {et~al.}(2023){Stokholm}, {Aguirre B{\o}rsen-Koch},
  {Stello}, {Hon}, \& {Reyes}}]{2023MNRAS.524.1634S}
{Stokholm}, A., {Aguirre B{\o}rsen-Koch}, V., {Stello}, D., {Hon}, M., \&
  {Reyes}, C. 2023, \mnras, 524, 1634

\bibitem[{{Storm} \& {Bergemann}(2023)}]{2023MNRAS.525.3718S}
{Storm}, N. \& {Bergemann}, M. 2023, \mnras, 525, 3718

\bibitem[{{Tautvai{\v{s}}ien{\.{e}}} {et~al.}(2021){Tautvai{\v{s}}ien{\.{e}}},
  {Viscasillas V{\'a}zquez}, {Mikolaitis}, {Stonkut{\.{e}}},
  {Minkevi{\v{c}}i{\={u}}t{\.{e}}}, {Drazdauskas}, \&
  {Bagdonas}}]{2021A&A...649A.126T}
{Tautvai{\v{s}}ien{\.{e}}}, G., {Viscasillas V{\'a}zquez}, C., {Mikolaitis},
  {\v{S}}., {et~al.} 2021, \aap, 649, A126

\bibitem[{{Titarenko} {et~al.}(2019){Titarenko}, {Recio-Blanco}, {de Laverny},
  {Hayden}, \& {Guiglion}}]{2019A&A...622A..59T}
{Titarenko}, A., {Recio-Blanco}, A., {de Laverny}, P., {Hayden}, M., \&
  {Guiglion}, G. 2019, \aap, 622, A59

\bibitem[{{Travaglio} {et~al.}(2004){Travaglio}, {Gallino}, {Arnone}, {Cowan},
  {Jordan}, \& {Sneden}}]{Travaglio2004}
{Travaglio}, C., {Gallino}, R., {Arnone}, E., {et~al.} 2004, \apj, 601, 864

\bibitem[{{Tucci Maia} {et~al.}(2016){Tucci Maia}, {Ram{\'\i}rez},
  {Mel{\'e}ndez}, {Bedell}, {Bean}, \& {Asplund}}]{2016A&A...590A..32T}
{Tucci Maia}, M., {Ram{\'\i}rez}, I., {Mel{\'e}ndez}, J., {et~al.} 2016, \aap,
  590, A32

\bibitem[{{Valentini} {et~al.}(2016){Valentini}, {Chiappini}, {Miglio},
  {Montalb{\'a}n}, {Rodrigues}, {Mosser}, {Anders}, {CoRoT RG Group}, \& {GES
  Consortium}}]{2016AN....337..970V}
{Valentini}, M., {Chiappini}, C., {Miglio}, A., {et~al.} 2016, Astronomische
  Nachrichten, 337, 970

\bibitem[{{Viscasillas V{\'a}zquez} {et~al.}(2022){Viscasillas V{\'a}zquez},
  {Magrini}, {Casali}, {Tautvai{\v{s}}ien{\.{e}}}, {Spina}, {Van der Swaelmen},
  {Randich}, {Bensby}, {Bragaglia}, {Friel}, {Feltzing}, {Sacco}, {Turchi},
  {Jim{\'e}nez-Esteban}, {D'Orazi}, {Delgado-Mena}, {Mikolaitis},
  {Drazdauskas}, {Minkevi{\v{c}}i{\={u}}t{\.{e}}}, {Stonkut{\.{e}}},
  {Bagdonas}, {Montes}, {Guiglion}, {Baratella}, {Tabernero}, {Gilmore},
  {Alfaro}, {Francois}, {Korn}, {Smiljanic}, {Bergemann}, {Franciosini},
  {Gonneau}, {Hourihane}, {Worley}, \& {Zaggia}}]{2022A&A...660A.135V}
{Viscasillas V{\'a}zquez}, C., {Magrini}, L., {Casali}, G., {et~al.} 2022,
  \aap, 660, A135

\bibitem[{{Walsen} {et~al.}(2023){Walsen}, {Jofr{\'e}}, {Buder}, {Yaxley},
  {Das}, {Yates}, {Hua}, {Signor}, {Eldridge}, {Rojas-Arriagada}, {Tissera},
  {Johnston}, {Aguilera-G{\'o}mez}, {Zoccali}, {Gilmore}, \&
  {Foley}}]{Walsen23}
{Walsen}, K., {Jofr{\'e}}, P., {Buder}, S., {et~al.} 2023, arXiv e-prints,
  arXiv:2310.15107

\bibitem[{{Zenati} {et~al.}(2023){Zenati}, {Perets}, {Dessart},
  {Jacobson-Gal{\'a}n}, {Toonen}, \& {Rest}}]{2023Zenati}
{Zenati}, Y., {Perets}, H.~B., {Dessart}, L., {et~al.} 2023, \apj, 944, 22

\end{thebibliography}
\newpage

\begin{appendix}\label{sec:append}
\onecolumn

\section{Target information}
\begin{table}[!h]
    \centering
    \caption{The EPIC (K2) IDs are listed in the first column for the entire sample. Subsequently, the \textit{Gaia} IDs along with  with their corresponding coordinates (RA and DEC) and \textit{G} magnitudes are presented. The signal-to-noise ratios (SNR) are presented in the sixth column. The asteroseismic frequency separation $\Delta \nu$, together with the frequency of maximum power $\nu_{max}$ were obtained from the K2 catalogue. Finally the 2MASS magnitudes are reported in J, H and K bands. The two stars likely parts of binary systems are not listed in the table.}
    \label{tab:infotarg}
    \tiny
    \begin{tabular}{|c|c|c|c|c|c|c|c|c|c|c|}
        \toprule
        EPIC ID & \textit{Gaia} DR3 ID & RA & DEC & $G_{\mathrm{mag}}$ & SNR & $\Delta_{\nu}$ & $\nu_{max}$ & J$_{mag}$ & H$_{mag}$ & K$_{mag}$ \\
        \midrule
        \begin{filecontents*}{targets.csv}
Unnamed: 0,EPICID,GaiaDRID,RA,DEC,Gmag,SNR,DELTANU,NUMAX,JMAG,HMAG,KMAG,Numeric_ID
0,212328705,3603874691200483584,204.108972,-16.70291629,9.56,97,5.63,56.38,7.95,7.41,7.28,3603874691200483584
1,201387934,3791659118172624512,169.8884225,-1.143003424,9.88,94,4.31,35.09,8.29,7.76,7.61,3791659118172624512
2,201610179,3810779934617363584,170.7994691,2.201317576,9.54,94,8.43,93.78,8.05,7.59,7.46,3810779934617363584
3,201615553,3810802165368121984,171.0301093,2.283891994,9.56,93,3.73,33.54,7.94,7.43,7.29,3810802165368121984
4,201436971,3797212025554640640,173.1636164,-0.406484518,9.95,92,5.72,58.86,8.31,7.78,7.64,3797212025554640640
5,206049832,2601007287942620032,338.6186258,-13.42684939,9.64,92,1.8,34.01,8.09,7.55,7.45,2601007287942620032
6,210729904,47764610473398016,64.89874585,18.80338613,10.25,91,14.27,251.11,8.36,7.8,7.66,47764610473398016
7,205989874,2599221543620591232,336.0198833,-15.12118093,9.37,91,3.45,26.76,7.84,7.33,7.22,2599221543620591232
8,210875646,51670728611455360,58.95835214,21.01350234,10.49,90,0.82,4.25,8.29,7.49,7.34,51670728611455360
9,205935544,2595016839357041664,337.9320516,-16.85770001,10.38,90,4.14,30.29,8.88,8.44,8.26,2595016839357041664
10,206166172,2614051064966374784,332.5811764,-10.40084002,10.11,90,6.0,58.19,8.51,7.94,7.81,2614051064966374784
11,210557561,46220449471982208,60.49102134,16.3313616,9.61,90,5.48,56.08,7.68,7.08,6.92,46220449471982208
12,201504843,3796029608173623296,177.2799633,0.596013632,11.3,89,14.27,188.58,9.95,9.52,9.45,3796029608173623296
13,206057733,2600597681206509824,334.3914421,-13.21705184,10.56,89,6.98,74.18,9.07,8.57,8.44,2600597681206509824
14,201834045,3817027222247093760,169.2843704,5.997252977,11.37,89,4.67,44.29,9.68,9.1,8.98,3817027222247093760
15,211006381,52907060716766080,63.16466585,23.07606484,10.23,88,3.61,32.3,7.86,7.13,6.93,52907060716766080
16,210904019,63700485331065472,57.41907273,21.4528989,10.17,88,1.44,9.74,8.26,7.6,7.43,63700485331065472
17,201801438,3813366776239894272,172.7472781,5.393485992,20.66,88,7.44,71.93,8.79,8.35,8.18,3813366776239894272
18,201584294,3796479514587811072,177.0167383,1.800451943,10.23,88,4.82,43.74,8.54,7.94,7.82,3796479514587811072
19,201380381,3794078971466576768,173.3835933,-1.25747885,10.47,88,7.17,73.28,8.97,8.42,8.36,3794078971466576768
20,201695150,3800805366992735360,174.0922748,3.557615343,9.5,87,9.94,109.25,7.97,7.47,7.34,3800805366992735360
21,206012087,2599468860721809408,335.9863649,-14.47835686,9.48,87,5.76,59.82,7.93,7.42,7.3,2599468860721809408
22,210329876,3304805952194320256,59.77071415,11.64807462,9.55,86,4.16,32.95,7.65,7.05,6.9,3304805952194320256
23,205991527,2599236971143099648,335.7188622,-15.06953592,9.37,86,4.65,41.69,7.71,7.17,7.03,2599236971143099648
24,211015044,53665345782804480,63.08209558,23.21443206,10.4,86,2.8,30.12,8.34,7.71,7.56,53665345782804480
25,201798083,3813371964560265856,172.451626,5.332099278,9.83,86,3.63,29.84,8.19,7.62,7.47,3813371964560265856
26,210977032,65359510936065792,59.17248117,22.61729657,20.59,85,7.44,71.93,8.88,8.35,8.18,65359510936065792
27,210724982,57497693561981952,51.56661758,18.73063083,9.86,85,4.73,40.4,8.33,7.82,7.74,57497693561981952
28,210990463,53626622358017920,62.46723912,22.82266847,10.79,85,3.04,34.42,8.81,8.22,8.07,53626622358017920
29,211083040,65991936282836864,60.31149905,24.29785844,10.78,84,5.12,46.84,9.12,8.6,8.51,65991936282836864
30,211091343,66218951074368512,59.40521966,24.42591654,9.81,84,1.8,15.06,7.92,7.26,7.13,66218951074368512
31,205953136,2596385074203302528,340.4845892,-16.26738208,10.99,83,4.81,44.09,9.4,8.78,8.61,2596385074203302528
32,211108599,66103845951674112,60.77734577,24.70042008,9.99,83,6.69,71.92,8.18,7.66,7.52,66103845951674112
33,210480597,40016455113058560,59.70249644,15.07969951,9.94,83,1.92,14.61,7.79,7.15,6.93,40016455113058560
34,210982421,52859541198660480,63.14881434,22.70042056,9.62,83,3.83,38.41,7.63,7.04,6.9,52859541198660480
35,210578532,46279857459569664,60.67986033,16.64887014,9.79,82,1.61,11.15,7.65,7.0,6.82,46279857459569664
36,211120167,66685453242145792,58.61152019,24.89475755,9.95,81,4.38,38.44,8.32,7.76,7.64,66685453242145792
37,201393756,3797665612756053888,170.0989951,-1.055593065,10.97,81,4.97,46.2,9.36,8.85,8.67,3797665612756053888
38,211032114,148965309961550720,64.4060835,23.4813859,9.67,81,5.41,58.97,7.78,7.28,7.09,148965309961550720
39,211029152,53722417308529792,62.11758335,23.43338483,10.72,79,7.68,43.09,8.41,7.64,7.42,53722417308529792
40,211100795,66612404439062784,58.73661193,24.57333701,20.51,79,7.3,75.15,8.09,7.55,7.41,66612404439062784
41,212323768,3605133498870789120,205.278068,-16.84117728,11.28,79,2.32,15.59,9.53,8.89,8.81,3605133498870789120
42,206218229,2608958401983554816,339.5905542,-9.469920682,9.76,79,4.9,44.75,8.03,7.47,7.27,2608958401983554816
43,203478307,6043454739975673344,243.1821768,-26.00487066,10.83,79,5.12,45.23,8.84,8.22,8.1,6043454739975673344
44,203367145,6235380954236279552,239.6732505,-26.37878524,10.2,79,4.69,42.57,8.41,7.85,7.73,6235380954236279552
45,211044578,149027466728165504,63.9129976,23.68320126,9.87,78,4.71,38.77,7.86,7.3,7.13,149027466728165504
46,201531324,3796185154709476608,177.612019,1.006774282,9.29,78,3.64,31.58,7.87,7.35,7.24,3796185154709476608
47,210780208,57845345394803456,52.12542368,19.56914201,10.7,78,4.45,43.41,8.88,8.3,8.15,57845345394803456
48,206509377,2626530320077918464,335.1881504,-4.687467974,9.9,77,4.45,37.76,8.31,7.8,7.67,2626530320077918464
49,201473281,3804057176928023168,168.3857731,0.120790002,9.97,77,16.02,214.69,8.54,8.03,7.91,3804057176928023168
50,206120536,2613677127933423744,330.7351883,-11.56185003,9.23,76,5.24,54.36,7.65,7.1,7.0,2613677127933423744
51,205929054,6825961802359128576,334.1660451,-17.08712916,11.17,76,7.2,72.63,9.73,9.24,9.09,6825961802359128576
52,211039480,53690978147432960,62.96880068,23.60085938,10.39,76,7.24,87.93,8.61,8.12,7.94,53690978147432960
53,210976157,64078992209400064,58.0044248,22.60290508,10.47,74,4.29,37.92,8.56,7.95,7.8,64078992209400064
54,201799986,3813361381760859392,172.623422,5.366302365,9.85,73,1.39,8.59,8.18,7.58,7.49,3813361381760859392
55,206322548,2622263214234941184,336.4938927,-7.887875111,11.66,72,4.63,26.75,10.2,9.71,9.58,2622263214234941184
56,206187923,2608825739033972224,339.0435866,-9.96799251,9.84,71,18.46,221.03,8.28,7.81,7.65,2608825739033972224
57,210977788,52834080632668288,63.34477847,22.62833215,10.87,70,4.24,39.13,8.74,8.13,7.92,52834080632668288
58,210995892,65417372735307904,59.25809294,22.91023579,11.09,70,3.31,27.56,9.25,8.64,8.49,65417372735307904
59,210964487,52802126076038272,63.03918788,22.41391435,9.95,70,3.52,26.71,7.74,7.08,6.86,52802126076038272
60,201741868,3800997304787056256,174.1671701,4.345257611,11.42,70,5.66,56.13,9.8,9.23,9.11,3800997304787056256
61,211017969,148983898578548224,63.7152624,23.25812944,11.64,69,4.34,34.07,9.57,8.98,8.79,148983898578548224
62,203848638,6034822646009063040,253.1051686,-24.7023971,10.09,69,6.48,59.12,8.39,7.85,7.76,6034822646009063040
63,203722894,6049568265144531456,243.3517865,-25.16200543,10.95,69,4.4,35.95,9.09,8.5,8.33,6049568265144531456
64,204640114,6243730920412078976,242.0595416,-21.5012362,10.52,69,2.12,28.31,8.63,8.01,7.85,6243730920412078976
65,203829170,6049773251041967360,242.5593522,-24.77676238,11.13,67,5.11,48.07,9.4,8.82,8.65,6049773251041967360
66,203722894,6049568265144531456,243.3517865,-25.16200543,10.95,67,4.4,35.95,9.09,8.5,8.33,6049568265144531456
67,210744319,57706291534137856,51.40042643,19.02251796,10.71,67,3.16,29.13,9.03,8.46,8.37,57706291534137856
68,206211780,2615277982503739904,334.7283482,-9.571356794,11.42,66,8.33,88.03,9.95,9.42,9.33,2615277982503739904
69,210922109,63742507291336448,57.34670246,21.73908703,9.68,66,4.7,44.72,7.72,7.1,6.94,63742507291336448
70,201640625,3800313030597421696,172.3870992,2.672229331,12.01,65,6.28,59.03,10.44,9.89,9.79,3800313030597421696
71,212322695,3603646439458466176,204.8885046,-16.87068314,11.15,63,4.98,44.57,9.52,8.98,8.84,3603646439458466176
72,201456500,3794842788451115904,177.0677615,-0.12614963,10.86,22,2.84,26.89,9.26,8.61,8.5,3794842788451115904
\end{filecontents*}
\csvreader[head to column names, late after line=\\]{targets.csv}{}{\EPICID & \GaiaDRID & \RA & \DEC & \Gmag &\SNR & \DELTANU & \NUMAX & \JMAG & \HMAG & \KMAG  }
        \bottomrule
    \end{tabular}
\end{table}
\clearpage

\begin{table}[htbp]
    \centering
    \caption{Linear regression fit coefficients for the three metallicity groups as defined in Table \ref{tab:groups}. The slopes and their sigmas are expressed in dex $\mathrm{Gyr}^{-1}$. The $\rho$ values indicate the Pearson correlation coefficients for each relation.}
    \label{tab:coeff_alpha_n2}
    \tiny
    \begin{tabular}{|c|c|c|c|c|c|c|c|c|c|}
        \toprule
        Ratio & Group & Slope & $\sigma_{\text{slope}}$ & $\rho$ & Ratio & Group & Slope & $\sigma_{\text{slope}}$ & $\rho$  \\
        \midrule
        \begin{filecontents*}{coeff_2.csv}
Ratio,Group,Slope,sigmaslope,rho,Ratio,Group,Slope,sigmaslope,rho
Sr$/$Mg,poor,-0.024,0.01,0.653,Nd$/$Si,poor,-0.009,0.007,0.338
Sr$/$Mg,mix,-0.037,0.011,0.785,Nd$/$Si,mix,-0.016,0.005,0.65
Sr$/$Mg,solar,-0.044,0.027,0.762,Nd$/$Si,solar,-0.03,0.009,0.749
Sr$/$Si,poor,-0.019,0.013,0.502,Nd$/$Ca,poor,-0.007,0.005,0.397
Sr$/$Si,mix,-0.036,0.008,0.783,Nd$/$Ca,mix,-0.015,0.003,0.741
Sr$/$Si,solar,-0.038,0.018,0.743,Nd$/$Ca,solar,-0.024,0.005,0.811
Sr$/$Ca,poor,-0.019,0.011,0.479,Nd$/$Ti,poor,-0.01,0.008,0.356
Sr$/$Ca,mix,-0.035,0.004,0.839,Nd$/$Ti,mix,-0.019,0.003,0.754
Sr$/$Ca,solar,-0.034,0.018,0.761,Nd$/$Ti,solar,-0.03,0.006,0.811
Sr$/$Ti,poor,-0.018,0.009,0.52,Sr$/$Eu,poor,-0.021,0.013,0.472
Sr$/$Ti,mix,-0.038,0.006,0.834,Sr$/$Eu,mix,-0.027,0.009,0.681
Sr$/$Ti,solar,-0.04,0.018,0.785,Sr$/$Eu,solar,-0.024,0.009,0.7
Y$/$Mg,poor,-0.015,0.007,0.555,Sr$/$V,poor,-0.024,0.011,0.423
Y$/$Mg,mix,-0.029,0.007,0.742,Sr$/$V,mix,-0.032,0.005,0.787
Y$/$Mg,solar,-0.036,0.018,0.698,Sr$/$V,solar,-0.043,0.017,0.805
Y$/$Si,poor,-0.008,0.005,0.404,Sr$/$Al,poor,-0.024,0.015,0.531
Y$/$Si,mix,-0.027,0.005,0.765,Sr$/$Al,mix,-0.036,0.01,0.785
Y$/$Si,solar,-0.036,0.019,0.658,Sr$/$Al,solar,-0.045,0.019,0.793
Y$/$Ca,poor,-0.008,0.004,0.465,Y$/$Eu,poor,-0.009,0.007,0.369
Y$/$Ca,mix,-0.024,0.003,0.814,Y$/$Eu,mix,-0.021,0.009,0.617
Y$/$Ca,solar,-0.028,0.012,0.712,Y$/$Eu,solar,-0.016,0.008,0.474
Y$/$Ti,poor,-0.009,0.004,0.445,Y$/$V,poor,-0.009,0.006,0.337
Y$/$Ti,mix,-0.029,0.005,0.807,Y$/$V,mix,-0.024,0.005,0.732
Y$/$Ti,solar,-0.033,0.009,0.764,Y$/$V,solar,-0.034,0.013,0.776
Zr$/$Mg,poor,-0.014,0.007,0.597,Y$/$Al,poor,-0.012,0.008,0.461
Zr$/$Mg,mix,-0.034,0.007,0.803,Y$/$Al,mix,-0.029,0.005,0.758
Zr$/$Mg,solar,-0.06,0.021,0.84,Y$/$Al,solar,-0.044,0.019,0.739
Zr$/$Si,poor,-0.007,0.007,0.342,Zr$/$Eu,poor,-0.009,0.01,0.298
Zr$/$Si,mix,-0.032,0.004,0.845,Zr$/$Eu,mix,-0.026,0.008,0.764
Zr$/$Si,solar,-0.056,0.017,0.861,Zr$/$Eu,solar,-0.037,0.007,0.857
Zr$/$Ca,poor,-0.007,0.006,0.345,Zr$/$V,poor,-0.007,0.01,0.191
Zr$/$Ca,mix,-0.029,0.003,0.866,Zr$/$V,mix,-0.029,0.004,0.829
Zr$/$Ca,solar,-0.05,0.016,0.847,Zr$/$V,solar,-0.057,0.013,0.87
Zr$/$Ti,poor,-0.009,0.006,0.394,Zr$/$Al,poor,-0.01,0.009,0.453
Zr$/$Ti,mix,-0.033,0.003,0.876,Zr$/$Al,mix,-0.035,0.007,0.82
Zr$/$Ti,solar,-0.057,0.018,0.844,Zr$/$Al,solar,-0.071,0.021,0.876
La$/$Mg,poor,-0.016,0.008,0.591,La$/$Eu,poor,-0.01,0.006,0.52
La$/$Mg,mix,-0.028,0.006,0.772,La$/$Eu,mix,-0.02,0.007,0.658
La$/$Mg,solar,-0.029,0.009,0.768,La$/$Eu,solar,-0.011,0.003,0.696
La$/$Si,poor,-0.011,0.006,0.532,La$/$V,poor,-0.01,0.005,0.599
La$/$Si,mix,-0.024,0.005,0.782,La$/$V,mix,-0.021,0.003,0.801
La$/$Si,solar,-0.029,0.011,0.748,La$/$V,solar,-0.034,0.008,0.836
La$/$Ca,poor,-0.011,0.005,0.57,La$/$Al,poor,-0.017,0.007,0.595
La$/$Ca,mix,-0.023,0.003,0.861,La$/$Al,mix,-0.026,0.005,0.792
La$/$Ca,solar,-0.021,0.006,0.766,La$/$Al,solar,-0.037,0.013,0.814
La$/$Ti,poor,-0.013,0.006,0.533,Ce$/$Eu,poor,-0.016,0.006,0.642
La$/$Ti,mix,-0.027,0.003,0.86,Ce$/$Eu,mix,-0.024,0.008,0.715
La$/$Ti,solar,-0.029,0.007,0.812,Ce$/$Eu,solar,-0.014,0.004,0.691
Ce$/$Mg,poor,-0.02,0.007,0.708,Ce$/$V,poor,-0.015,0.005,0.673
Ce$/$Mg,mix,-0.032,0.007,0.809,Ce$/$V,mix,-0.024,0.003,0.809
Ce$/$Mg,solar,-0.03,0.013,0.762,Ce$/$V,solar,-0.034,0.01,0.813
Ce$/$Si,poor,-0.015,0.006,0.647,Ce$/$Al,poor,-0.02,0.006,0.692
Ce$/$Si,mix,-0.028,0.007,0.807,Ce$/$Al,mix,-0.03,0.005,0.824
Ce$/$Si,solar,-0.03,0.011,0.763,Ce$/$Al,solar,-0.035,0.012,0.811
Ce$/$Ca,poor,-0.016,0.004,0.706,Nd$/$Eu,poor,-0.01,0.007,0.409
Ce$/$Ca,mix,-0.027,0.003,0.883,Nd$/$Eu,mix,-0.012,0.006,0.5
Ce$/$Ca,solar,-0.025,0.008,0.77,Nd$/$Eu,solar,-0.012,0.004,0.607
Ce$/$Ti,poor,-0.016,0.006,0.647,Nd$/$V,poor,-0.008,0.004,0.451
Ce$/$Ti,mix,-0.031,0.004,0.871,Nd$/$V,mix,-0.015,0.004,0.6
Ce$/$Ti,solar,-0.03,0.008,0.806,Nd$/$V,solar,-0.032,0.007,0.799
Nd$/$Mg,poor,-0.013,0.011,0.448,Nd$/$Al,poor,-0.014,0.006,0.504
Nd$/$Mg,mix,-0.019,0.007,0.634,Nd$/$Al,mix,-0.02,0.006,0.684
Nd$/$Mg,solar,-0.034,0.012,0.718,Nd$/$Al,solar,-0.034,0.011,0.797
\end{filecontents*}
\csvreader[head to column names, late after line=\\]{coeff_2.csv}{}{\Ratio & \Group & \Slope & \sigmaslope & \rho &\Ratio & \Group & \Slope & \sigmaslope & \rho  }
        \bottomrule
    \end{tabular}
\end{table}

\end{appendix}

\end{document}